\documentclass[aps,amssymb,amsmath,pra,twocolumn,showpacs,superscriptaddress,longbibliography]{revtex4-2}

\usepackage{ulem}

\usepackage{graphicx}
\graphicspath{{figs_main/}{figs_SI/}}

\usepackage{dcolumn}
\usepackage{bm}
\usepackage{subfigure}
\usepackage[font=small, justification=Justified]{caption}
\usepackage{color}
\usepackage{soul}

\usepackage{amsfonts,verbatim}

\usepackage[margin=0.6in]{geometry}

\usepackage{tikz}

\usepackage[english]{babel}
\usepackage{dsfont}
\usepackage{latexsym}
\usepackage{float}
\usepackage{afterpage}
\usepackage{enumitem}
\usepackage{mleftright}

\definecolor{lR}{rgb}{1, 0.8, 0.79}

\usepackage{eso-pic, graphicx}

\usepackage{listings}
\usepackage{multirow}
\usepackage{xcolor,colortbl}

\usepackage{bbm}
\usepackage{upgreek}
\usepackage{newtxtext,newtxmath}

\newcommand{\red}[2][]{\textcolor{red}{#1#2}}

\newcommand{\nocontentsline}[3]{}
\newcommand{\tocless}[2]{\bgroup\let\addcontentsline=\nocontentsline#1{#2}\egroup}

\newcommand{\Q}[1]{\red{#1}}

\definecolor{Ablue}{rgb}{0.96,0.24,0.00}

\definecolor{Abluetitle}{rgb}{0.,0.24,0.51}

\definecolor{orange}{rgb}{0.96,0.24,0.00}

\definecolor{darkred}{rgb}{0.55, 0.0, 0.0}

\definecolor{darksalmon}{rgb}{0.91, 0.59, 0.48}
\definecolor{maroon}{cmyk}{0,0.87,0.68,0.32}

\setcitestyle{numbers,square,citesep={,\kern-.24em}}

\definecolor{mustard}{rgb}{1.0, 0.86, 0.35}

\addto\captionsenglish{\renewcommand{\figurename}{Fig.}}

\definecolor{Gray}{gray}{0.85}
\definecolor{LightCyan}{rgb}{0.88,1,1}
\newcolumntype{a}{${>}${\columncolor{Gray}}c}
\newcolumntype{b}{${>}${\columncolor{White}}c}

\usepackage{array}
\newcolumntype{L}[1]{${>}${\raggedright\let\newline\\\arraybackslash\hspace{0pt}}m{#1}}
\newcolumntype{C}[1]{${>}${\centering\let\newline\\\arraybackslash\hspace{0pt}}m{#1}}
\newcolumntype{R}[1]{${>}${\raggedleft\let\newline\\\arraybackslash\hspace{0pt}}m{#1}}

\newcolumntype{P}[1]{{>}{\centering\arraybackslash}p{#1}}
\newcolumntype{M}[1]{{>}{\centering\arraybackslash}m{#1}}

\usepackage[colorlinks=true , citecolor=black,urlcolor=blue]{hyperref}

\newcommand{\app}{\approx}

\newcommand{\Cs}{{}^{13}\R{C}}

\newcommand{\mC}[0]{\mathcal{C}}

\newcommand{\xD}{\Delta}

\newcommand{\sq}[1]{\sqrt{#1}}

\newcommand{\mw}{\R{MW}}

\newcommand{\mF}[0]{\mathcal{F}}

\newcommand{\mA}[0]{\mathcal{A}}

\newcommand{\rt}{\rightarrow}

\newcommand{\beq}{\begin{equation}}
\newcommand{\eeq}{\end{equation}}
                  
\newcommand{\benum}{\begin{enumerate}}
\newcommand{\eenum}{\end{enumerate}}
                    
\newcommand{\bit}{\begin{itemize}}
\newcommand{\eit}{\end{itemize}}

\newcommand{\bea}{\begin{eqnarray}}
\newcommand{\eea}{\end{eqnarray}}

\newcommand{\bev}{\begin{verbatim}}
\newcommand{\eev}{\end{verbatim}}

\newcommand{\zt}{\times}

\newcommand{\T}[1]{\textbf{#1}}
\newcommand{\I}[1]{\textit{#1}}
\newcommand{\R}[1]{\textrm{#1}}

\newcommand{\zl}[1]{\label{eqn:#1}}
\newcommand{\zr}[1]{Eq.\,(\ref{eqn:#1})}
\newcommand{\zfl}[1]{\protect\label{fig:#1}}
\newcommand{\zfr}[1]{\figurename\,\ref{fig:#1}}
\newcommand{\ztl}[1]{\label{table:#1}}
\newcommand{\ztr}[1]{Table \ref{table:#1}}
\newcommand{\zsl}[1]{\label{sec:#1}}
\newcommand{\zsr}[1]{Sec. \!\!\!\ref{sec:#1}}

\newcommand{\ba}{\left\{ \begin{array}{lr}}
\newcommand{\ea}{\end{array}\right.}

\newcommand{\BRd}[1]{\textcolor{red}{#1}} 

\newcommand{\blist}[1]{
 \begin{list}{#1}
 \begin{align}
	 arrow
 \end{align}
 $\checkmark\star
  { \setlength{\itemsep}{3pt}
     \setlength{\parsep}{2pt}
     \setlength{\topsep}{3pt}
     \setlength{\partopsep}{0pt}
     \setlength{\leftmargin}{1em}
     \setlength{\labelwidth}{1em}
     \setlength{\labelsep}{0.5em} } }
\newcommand{\elist}{
  \end{list}  }

\DeclareMathSymbol{\vartheta}{\mathalpha}{letters}{"12}
\DeclareMathSymbol{\theta}{\mathalpha}{letters}{"23}
\DeclareMathSymbol{\phi}{\mathalpha}{letters}{"27}
\DeclareMathSymbol{\varphi}{\mathalpha}{letters}{"1E}

\newcommand{\bef}
{
\begin{figure}[htbp]
\centering
}

\newcommand{\eef}{\end{figure}}

\mleftright
\makeatletter
\@addtoreset{section}{part}
\makeatother

\newcommand{\beginsupplement}{%
        \setcounter{table}{0}
        \renewcommand{\tablename}{Supplementary Table}
        \renewcommand{\thetable}{S\arabic{table}}%
        \setcounter{figure}{0}
        \renewcommand{\thefigure}{S\arabic{figure}}%
		\setcounter{page}{1}
		\renewcommand{\figurename}{Fig.} 
		\renewcommand{\thesection}{\:S\arabic{section}}
		\setcounter{section}{0}
     }

\newcommand{\affA}{Department of Chemistry, University of California, Berkeley, Berkeley, CA 94720, USA.}
\newcommand{\affB}{Chemical Sciences Division,  Lawrence Berkeley National Laboratory,  Berkeley, CA 94720, USA.}
\newcommand{\affC}{Energy Geoscience Division, Lawrence Berkeley National Laboratory, Berkeley, CA 94720, USA.}
\newcommand{\affD}{Adamas Nanotechnologies, Inc., Raleigh, NC 27617, USA.}
\newcommand{\affE}{Advanced Biofuels and Bioproducts Process Development Unit (ABPDU), Biological Systems and Engineering Division, Lawrence Berkeley National Laboratory, Berkeley,CA 94720, USA.}
\newcommand{\affF}{CIFAR Azrieli Global Scholars Program, 661 University Ave, Toronto, ON M5G 1M1, Canada.}

\begin{document}
\title{High-precision chemical quantum sensing in flowing monodisperse microdroplets}
\author{Adrisha Sarkar}\thanks{These authors contributed equally to this work}\affiliation{\affA}\affiliation{\affB}
\author{Zachary Jones}\thanks{These authors contributed equally to this work}\affiliation{\affA}\affiliation{\affE}
\author{Madhur Parashar}\affiliation{\affA}
\author{Emanuel Druga}\affiliation{\affA}
\author{Amala Akkiraju}\affiliation{\affA}
\author{Sophie Conti}\affiliation{\affA}
\author{Pranav Krishnamoorthi}\affiliation{\affA}
\author{Srisai Nachuri}\affiliation{\affA}
\author{Parker Aman}\affiliation{\affA}
\author{Mohammad Hashemi}\affiliation{\affA}
\author{Nicholas Nunn}\affiliation{\affD}
\author{Marco Torelli}\affiliation{\affD}
\author{Benjamin Gilbert}\affiliation{\affC}
\author{Kevin R. Wilson}\affiliation{\affB}
\author{Olga Shenderova}\affiliation{\affD}
\author{Deepti Tanjore}\affiliation{\affE}
\author{Ashok Ajoy}\thanks{ashokaj@berkeley.edu}\affiliation{\affA}\affiliation{\affB}\affiliation{\affF}

\begin{abstract}
We report on a novel flow-based method for high-precision chemical detection that integrates quantum sensing with droplet microfluidics. We deploy nanodiamond (ND) particles hosting fluorescent nitrogen vacancy (NV) defect centers as quantum sensors in rapidly flowing, monodisperse, picoliter-volume microdroplets containing analyte molecules. ND motion within these microcompartments facilitates close sensor-analyte interaction and mitigates particle heterogeneity. Microdroplet flow rates are rapid (upto 4cm/s) and with minimal drift. Pairing this controlled flow with microwave control of NV electronic spins, we introduce a new noise-suppressed mode of Optically Detected Magnetic Resonance (ODMR) that is sensitive to chemical analytes while resilient against experimental variations, achieving detection of analyte-induced signals at an unprecedented level of a few hundredths of a percent of the ND fluorescence. We demonstrate its application to detecting paramagnetic ions in droplets with simultaneously low limit-of-detection and low analyte volumes, in a manner significantly better than existing technologies. This is combined with exceptional measurement stability over ${>}10^3$s and across hundreds of thousands of droplets, while utilizing minimal sensor volumes and incurring low ND costs (${<}\$0.70$ for an hour of operation). Additionally, we demonstrate using these droplets as micro-confinement chambers by co-encapsulating ND quantum sensors with a variety of analytes, including single cells. This versatility suggests wide-ranging applications, including single-cell metabolomics and real-time intracellular measurements from bioreactors. Our work paves the way for portable, high-sensitivity, amplification-free, chemical assays with high throughput; introduces a new chemical imaging tool for probing chemical reactions within microenvironments; and establishes the foundation for developing movable, arrayed quantum sensors through droplet microfluidics.
\end{abstract}

\maketitle
\pagebreak

\section{Introduction}
Quantum sensing~\cite{Degen17} is rapidly reshaping our ability to discern chemical processes with high sensitivity and spatial resolution,  with the potential to impact a range of disciplines from synthesis to bioengineering~\cite{Zhang21}. Specifically, sensors based on nitrogen-vacancy (NV) defects in diamond~\cite{Doherty12,Jelezko06} translate optically addressable electronic spin state information into detectable fluorescence signals in a manner that is sensitive to the local chemical environment.  This has led to diverse applications, including in-cell thermometry~\cite{Neumann13,Kucsko13,Choi20,Petrini20} and reactive oxygen species detection~\cite{van20,Sharmin21}, high-sensitivity lateral flow assays~\cite{Miller20}, and nuclear magnetic resonance (NMR) measurements in picoliter volumes~\cite{Glenn18, Aslam17}, marking an exciting leap in precision measurement technologies.

Traditionally, quantum sensing for chemical analysis has relied on single crystals hosting shallow NV centers~\cite{Staudacher13, Mamin13}. High-throughput analysis faces challenges due to the small (mm-scale) sizes of these crystals, their substantial cost, and the need for precise crystal orientation~\cite{tetienne12}. Only a fraction of the diamond crystal is used for sensing, and analyte molecules are required to flow over its surface, necessitating complex integration of microfluidic structures directly onto the diamond~\cite{Ziem13,Smits19,BharadwaJ19, Allert22}.

Nanodiamond (ND)-based sensing offers a compelling alternative as they are inherently deployable, and can yield spatially selective sensing in or near targeted volumes of interest. NDs are also low-cost, non-reactive, and bio-inert~\cite{Schrand07,Shenderova19}, and for ${<}$40 nm particles, a significant proportion of their NV centers can interact with external analytes. Advances in nanodiamond chemistry have facilitated surface functionalization to control surface charge, hydrophilicity or hydrophobicity, and for targeting to proteins or cellular organelles~\cite{Chang10,Liu07,Zhang15,Reina19,Zhang19, Jones22}. 

Despite these advantages, ND-based sensing is fraught with challenges. Significant heterogeneity in particle size~\cite{Zheng23}, shape, and NV center coherence times leads to large statistical errors in multi-particle analysis. Additionally, fluorescence fluctuations arise from different particle orientations, and spatial variations in the materials in which the particles are targeted~\cite{Chowdhury19,Rodgers21}. Overcoming these significant challenges is essential to achieving high-precision quantum sensing.

In this paper, we directly address these challenges by deploying nanodiamonds within flowing, monodisperse, picoliter-volume microdroplets~\cite{Huebner08,Theberge10} that host analyte molecules. Rapid movement of the ND particles within the droplets effectively averages out their heterogeneity and ensurs close interaction with the analytes. We take advantage of stable and controllable flow afforded by droplet microfluidics to develop a new method for background-free quantum sensing at high-throughput. Additionally, the dynamic nature of the flowing droplets yields exceptional measurement stability, resistant to experimental variations and temperature shifts. We demonstrate this stability over $>10^3$s measurement and across ${\sim}10^5$ droplets, greatly surpassing the typical stability in conventional quantum sensing experiments~\cite{Fang13, Wolf15}. Additionally, the ND sensor volumes required are minuscule, costing only about \$0.63 for an hour of analysis over hundreds of thousands of droplets.

Our work, therefore, introduces a new platform technology fusing quantum sensing and droplet microfluidics, and is marked by several novel aspects. Picoliter-scale droplets can function as microscopic confinement chambers, encapsulating diverse analytes, ranging from single cells to chemical reaction products, and can stably accommodate a broader range of concentrations than bulk solutions~\cite{Chiu09}. The droplets are precisely controllable in terms of diameter, charge, and environmental conditions, and their movement under flow enhances sensor-analyte mixing~\cite{Grigoriev06, Srisa08}. This approach is also amenable to digital control techniques for droplet ``arithmetic," including mixing, collisions, and sorting, further enhancing their application in quantum sensing. 

\begin{figure*}
\centering
\begin{minipage}{\linewidth}
\makebox[\linewidth]{
 \includegraphics[width=\textwidth]{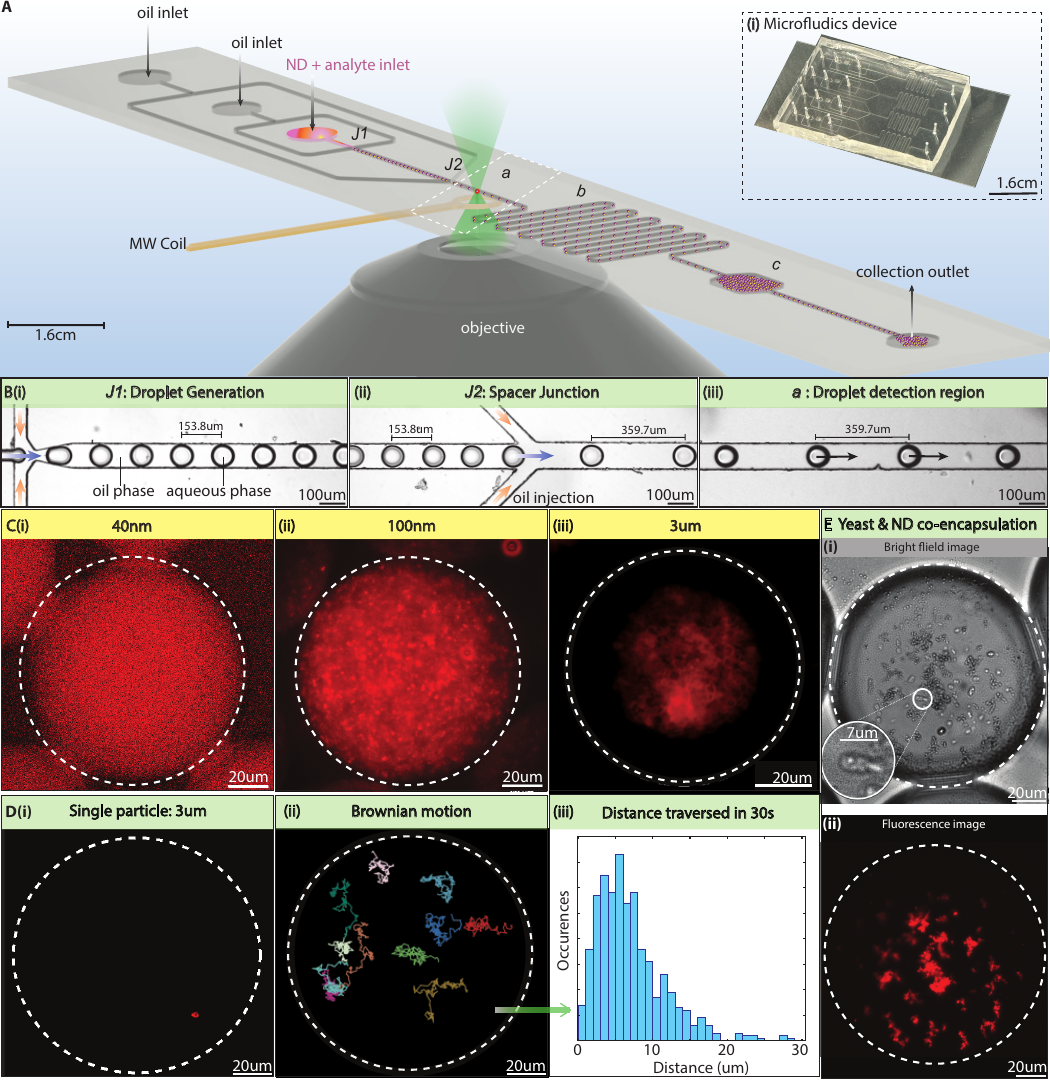}}
\caption{ \T{Nanodiamonds in droplets.} \T{(A)} \T{{Schematic of microfluidic chip}} employed, comprising two inlets for oil and one for water (grey circles). Analyte of interest and NDs are mixed into the latter. Chip features two focusing junctions, $J_1$ and $J_2$. $J_1$ generates ND-filled water droplets, while $J_2$ regulates droplet spacing  (see B). Droplets are analyzed in a region (dashed region $a$) over an objective lens and MW coil, followed by a circuitous region to induce additional mixing within droplets (region $b$) and storage in a collection chamber (region $c$). 
  (i) \I{Inset:} \I{Photograph of chip}; water and oil are delivered via narrow capillaries. See SI~\cite{SI} for fabrication details. 
  \T{(B)} \T{{Brightfield images}} with detailed views of the device regions. (i) \I{Droplet generation} at $J_1$ occurs by pinching water flow by oil (blue and orange arrows respectively) through an orifice. Resulting droplets are stable and monodisperse, and move at a fixed speed and spacing set by fluid flow rates. (ii) \I{Spacer junction} $J_2$ allows adjustment of droplet spacing via oil-flow (arrows). Here interdroplet distance is changed ${\app}154{\rt}360\mu$m. (iii) \I{Analysis region $a$}: Droplets maintain a consistent velocity and separation downstream and are analysed in flow (see A). 
  \T{(C)} \T{{Fluorescence images of droplets}} containing NDs of various sizes, (i) 40 nm, (ii) 100 nm, and (iii) 3 $\mu$m diameter particles. Dashed lines outline the droplet for clarity.
  \T{(D)} \T{{Tracking single NDs in droplets}}. (i) \I{Single $3\mu$m particle} encapsulated within a droplet. 
  (ii) \I{Tracked motion of 100nm particles} within a single droplet~\cite{TrackMate,TrackMate2}, shown for eleven particles tracked via fluorescence over a 30 s. (iii) \I{Histogram of particle displacement} for 200 particle trajectories over 30 s. For moving droplets, the NDs sample larger part of the droplet volume (movie in ref ~\cite{TrackingMovie}). 
  \T{(E)} \T{{Co-encapsulation of cells with NDs}}. (i) Bright-field and (ii) fluorescence images showing yeast cells encapsulated along with 100 nm ND particles. \I{Inset:} zoom into individual cells. See SI~\zsr{Targeting} for ND targeting to these cells.}
\zfl{mfig1}
\end{minipage}
\end{figure*}

\section{Nanodiamond loading in microdroplets}
Our microfluidics platform, schematically depicted in \zfr{mfig1}A and photographed in \zfr{mfig1}A(i), features a device that produces phase-separated, monodisperse, droplets in a water-in-oil emulsion, varying in diameter in range 10-150 $\mu$m and volume 300fL - 500pL. Chips are made from PDMS via soft lithography and bonded to a glass cover slip (\zfr{mfig1}A(i)). Each chip supports multiple devices, each hosting diverse microfluidic structures.

Droplets are formed by constricting an aqueous phase with oil channels (colored arrows in \zfr{mfig1}B(i)) using a fluid-focusing geometry at junction $J_1$, as depicted in \zfr{mfig1}A, through a ${\app}30 \mu$m orifice. This creates a stable stream of uniformly spaced droplets shown in the bright-field image in \zfr{mfig1}B(i). Droplet size and formation rate are controlled by the orifice size and water/oil flow rates. We achieve droplet speeds in excess of 4 cm/s. The fabrication process ensures ${\lesssim}$1 $\mu$m error, guaranteeing high reproducibility across chips (SI \zsr{chips}).

Additionally, the chips host a second junction $J_2$ (\zfr{mfig1}A), which enables oil injection to precisely control droplet spacing. This is demonstrated in \zfr{mfig1}B(ii), showing a spacing change from 154 $\mu$m to 360 $\mu$m. This spacing is maintained constant and droplet flow is highly uniform downstream, as depicted in \zfr{mfig1}B(iii), with consistent speed, uniformity, and stability over several hours (\zsr{stable}). In this region ($a$ in \zfr{mfig1}A) droplets are analyzed over a microscope objective and microwave (MW) coil for imaging and quantum sensing measurements before flowing through a circuitous path ($b$ in \zfr{mfig1}A) that induces intra-droplet mixing and extends the channel for downstream analysis of stationary droplets. Lastly, they are directed into a collection chamber ($c$ in \zfr{mfig1}A), allowing for the simultaneous storage and further examination of over a hundred droplets (SI~\zsr{chamber} and movie in Ref.~\cite{BF_ChipStructures}).

Diamond particles are incorporated into droplets in flow using an aqueous suspension of carboxylated (hydrophilic) NDs, which host ${\sim}$1-3 ppm NV centers, as the dispersed phase. Zeta potential measurements (SI~\zsr{materials}) reveal a surface charge sufficient to confer hydrophilicity and colloidal stability. The fluorescent NDs are thereby completely encapsulated within the droplet, as shown in \zfr{mfig1}C.
An inverted configuration, with NDs in oil droplets surrounded by an aqueous phase, can instead be achieved by coating ND surfaces with polymer chains. We showcase ND loading and droplet control through three movies available in the SI, and on Youtube (Refs.~\cite{BF_ChipStructures, Fl_ChipStructures, 750nmDiamondFlowing}). They illustrate droplet motion in brightfield~\cite{BF_ChipStructures} and fluorescence~\cite{Fl_ChipStructures} through various chip structures in \zfr{mfig1}A, highlighting rapid ND droplet formation, droplet spacing regulation via the dual focusing junctions $J_1,J_2$, and ND mixing within flowing droplets~\cite{750nmDiamondFlowing}.

Fluorescence imaging allows visualization of microdroplets loaded with NDs of varying sizes (SI~\zsr{microscope} describes optical setup), as illustrated in \zfr{mfig1}C with examples of (i) 40 nm, (ii) 100 nm, and (iii) 3 $\mu$m sized particles. The NDs feature a heterogeneous size distribution (${\pm}$30nm) characterized by DLS spectroscopy (see Materials, SI~\zsr{materials}). In \zfr{mfig1}C(i), the 40 nm NDs achieve a well-dispersed distribution within the droplet, occupying 0.01\% of its volume with minimal evident aggregation. Each droplet here contains around 10$^6$ particles and a total diamond mass of ${<}$100 pg. This is significantly lower than that employed in single-crystal diamond microfluidic-channel methods. The 100 nm particles shown in \zfr{mfig1}C(ii) are brighter and remain well dispersed in solution and the 3 $\mu$m particles (\zfr{mfig1}C(iii)), while even brighter, tend to settle at the droplet bottom for stationary droplets, indicating a trade-off between fluorescence intensity and colloidal stability.

At the opposite limit, \zfr{mfig1}D(i) shows the encapsulation of a \I{single} 3$\mu$m ND particle in a droplet. These large particles remain stationary in a static droplet, but can be induced to sample the droplet volume upon motion (movie in Ref.~\cite{750nmDiamondFlowing}). For smaller particles however, Brownian motion is much more pronounced, allowing the NDs to traverse large swathes of the droplet volume. In \zfr{mfig1}D(ii) we measure trajectories of individual 100 nm ND particles within a droplet over 30 s (movie in Ref.~\cite{TrackingMovie}). \zfr{mfig1}D(iii) presents a histogram of total displacement for these particles, based on tracking 200 ND trajectories for the same interval. They traverse distances ${>}$5 $\mu$m, several fold larger than their diameter. The long tail in the distribution points to anomalous diffusion reminiscent of Levy flight processes~\cite{Shlesinger87}. Overall, these large excursions promote interaction with droplet-confined analytes.

Microdroplets can also serve as picoliter-scale containers capable of co-encapsulating entities~\cite{Chiu09}. We demonstrate this in (\zfr{mfig1}E) by loading yeast cells (\I{Rhodosporidium toruloides}) and 100 nm NDs into a droplet. Bright-field and fluorescence images (\zfr{mfig1}E(i) and \zfr{mfig1}E(ii)) highlight the yeast and diamond particles, respectively, with the inset in \zfr{mfig1}F(i) showing a single yeast cell. Diamond aggregation, influenced by ions in the yeast growth medium, can be managed by altering the medium or modifying the diamond surface functionality~\cite{Hemelaar2017}. In SI~\zsr{Targeting} we show that the NDs can be targeted to the yeast cells via surface functionalization with Concanavalin-A, a protein with affinity to the cell surface.

\begin{figure}[t]
  \centering
\begin{minipage}{\linewidth}
\makebox[\linewidth]{
  \includegraphics[width=1\textwidth]{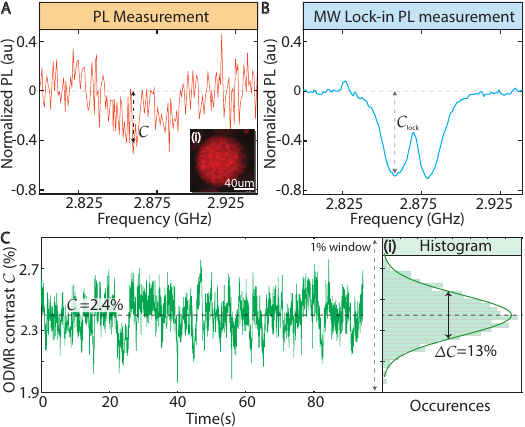}}
  \caption{\T{ODMR of ND particles in droplets}. (A) \I{Conventional ODMR} measurement of 100 nm particles in a single ${\app}100 \mu$m droplet at zero magnetic field. \I{Inset (i):} droplet fluorescence image. Each point is averaged for 1 s with and without MWs. Contrast $\mC$ is marked. Strong fluctuations arise from particle motion. (B) \I{Enhanced ODMR using MW lock-in}, using analog lock-in detector at modulation frequency $f_{\mw}{=}1$kHz. SNR improves 10-fold; contrast $\mC_{\R{lock}}$ is clearer. Strain-mediated dips around 2.87 GHz are visible. (C) \I{ODMR contrast} $\mC$ relavant for chemical sensing obtained via normalizing lock-in signal in (B) at 2.866 GHz to simultaneously measured PL. Here droplets are flowing, and data is sampled every 100 ms over 90 s. (i) \I{Histogram} of $\mC$ data displays variations at the $\Delta\mC{\app}$13\% level, highlighting the challenge for chemical sensing. Solid line is Gaussian fit.}  
\zfl{mfig2}
\end{minipage}
\end{figure}

\section{In-Droplet Chemical sensing by ODMR}
\subsection{Lock-in ODMR and challenges for low LOD sensing}
Chemical sensing utilizes the NV center's electronic spin sensitivity to its environment and the ability to convert this into optical signals via its spin-state-dependent fluorescence. This involves the measurement of NV center optically detected magnetic resonance (ODMR) spectrum, where ND photoluminescence (PL) is monitored as MW excitation is swept in frequency across NV spin transitions (\zfr{mfig2}A). At Earth's field, the PL exhibits a characteristic dip at ${\app}$2.87 GHz, signifying a shift in spin population from the brighter $m_s{=}0$ state to the dimmer $m_s{=}\pm 1$ states on-resonance.

\zfr{mfig2}A shows the first-reported ODMR measurements from ND particles in a single microdroplet, here held static (shown in \zfr{mfig2}A(i)). PL here is at the sub-nW level and is detected using a multi-pixel photon counter. These measurements are inherently noisy due to background, including PDMS autofluorescence (SI~\zsr{autoflour}), and fluctuations stemming from Brownian motion and particle re-orientation.

To enhance measurement signal-to-noise ratio (SNR), we utilize MW lock-in detection by amplitude modulating the applied MWs at $f_{\mw}{=}1$ kHz. This lock-in frequency is chosen from a balance between rates of NV repumping and MW-driven population redistribution. The result, shown in \zfr{mfig2}B, is an order-of-magnitude increase in SNR making the strain splitting near 2.87 GHz clearly visible. 

The ODMR contrast, marked $\mC$ in \zfr{mfig2}A, quantifies the ODMR dip relative to off-resonance PL. At fixed laser and MW powers, $\mC$ serves as a proxy for the NV electronic $T_1$, which is influenced by dipolar interactions with analytes in the droplet. Low limit of detection (LOD) sensing necessitates the ability to detect minute changes in $\mC$. While the lock-in strategy in \zfr{mfig2}B yields an increase in SNR, its contrast $\mC_{\R{lock}}$ remains susceptible to non-analyte specific factors like fluorescence variations from background and particle motion the absence of an off-resonance reference in this measurement. This poses challenges for chemical sensing.

A mitigating strategy involves continuously measuring NV PL and normalizing it to the resonant ODMR lock-in signal in \zfr{mfig2}B, creating a ratio-metric measurement shown in \zfr{mfig2}C. Here, the ODMR signal from flowing droplets with MWs at 2.86 GHz over 90 s is normalized to the total PL measured at each instant. The normalized contrast, still referred to as $\mC$ for convenience, in \zfr{mfig2}C has an average of 2.4\% (dashed line); a 1\% contrast window is shown here for clarity. Nonetheless, noise from PL measurements still affects $\mC$. Sensitive detection of analytes is limited by ability to resolve small fluctuations about this baseline contrast $ \mC$. We will refer to this measurement precision as $\xD\mC$, and by convention, we report it as a percentage of the base $\mC$ level, rather than in absolute units. Indeed $\mC$ itself varies based on sample and experiment conditions, but the percent error $\xD\mC$ allows for a standardized comparison across different experiments. From a histogram of the data (\zfr{mfig2}C(i)), we estimate $\Delta \mC$ ${\app}13\%$ in this case, setting a bound on the quantitative sensing at low LOD.

\begin{figure*}
\centering
\begin{minipage}{\linewidth}
\makebox[\linewidth]{
 \includegraphics[width=0.95\textwidth]{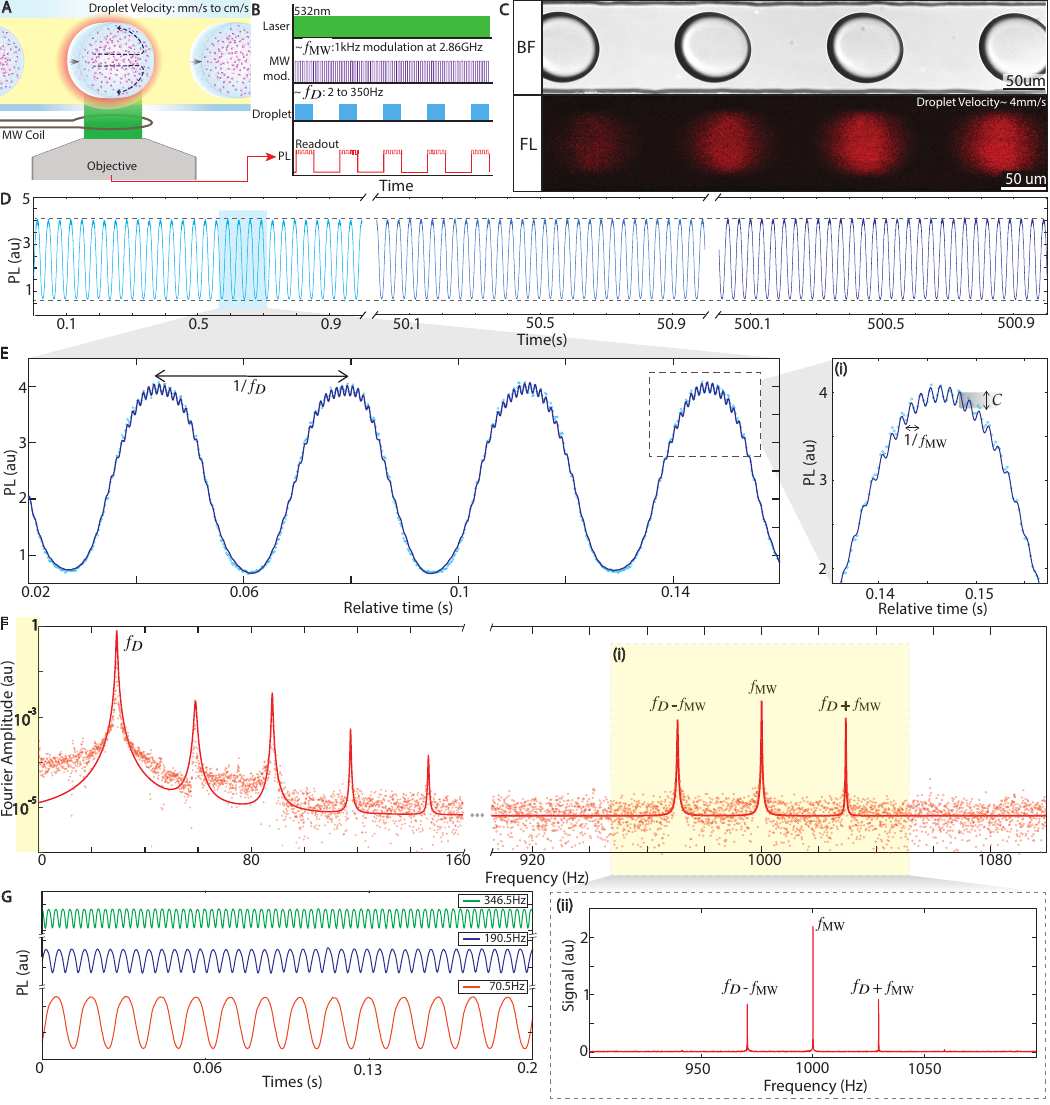}}
  \caption{\T{Droplet double modulation lock-in detection.} \T{(A)} \T{Schematic:} Stream of droplets hosting NDs continuously flow through the analysis region where they are illuminated by light that closely matches the droplet diameter, and  MW excitation at the ODMR resonance ${\app}$2.866 GHz. Mixing of NDs in droplet is illustrated by the arrows. \T{(B)} \T{Protocol:} under continuous 532 nm illumination, two combined modulations are imposed upon the collected PL (red line), mediated by droplet flow at $f_D$, and MWs at $f_{\mw}$. 
  \T{(C)} \T{Bright-field and fluorescence images} of droplets containing 100 nm NDs flowing at $f_D{\app}30$ droplets per second. 
  \T{(D)} \T{Long-time PL} from flowing droplets captured over a 8.3-min period. Three panels show representative 1s windows. PL profile is highly stable with $f_D{=}30$Hz (\zfr{mfig4}); dashed horizontal lines indicate intensity limits for clarity. 
  \T{(E)} \T{Double modulation imprinted into PL.} Panel shows representative 140 ms (${\app}$ 4 droplet) segment of the PL trace corresponding to 100 nm NDs in droplets. PL modulates at $f_D$ due to successive droplets entering and leaving the field of view is evident. (ii) Smaller ODMR modulation at $f_{\mw}$ is observable in the zoom-in. Grey shading reflects ODMR contrast. In main panel and inset, blue points represent the data, and the solid line is a fit to \zr{PL}.  
  \T{(F)} \T{Fourier transform intensity of PL} collected over $t{=}15$s displayed on a logarithmic scale. Points are data and solid lines are Lorentzian fits. Droplet modulation generates a sharp peak $\mF(f_D)$ at $f_D$ and its harmonics, stemming from the square-wave-like droplet profile. Peak at zero frequency is excluded for simplicity. Sharp peaks arise at the MW modulation, $f_{\mw}{=}1$kHz, while a combination with flow leads to peaks at $f_{\mw}\pm f_D$ (yellow shaded regions). Narrow Fourier linewidths are evident for all peaks (see \zfr{mfig4}B). 
  (ii) \I{Linear scale FT}: Data in yellow shaded region in main panel is plotted on a linear scale highlighting narrow linewidth and high SNR in the measurements. 
  \T{(G)} \T{Tunability of droplet modulation.} PL traces are displayed for three example cases in representative 200 ms windows. Highest frequency $f_D{=}346.5$Hz corresponds to sampling ${>}10^6$ droplets per hour.
}
\zfl{mfig3}
\end{minipage}
\end{figure*}

\subsection{Dual lock-in quantum sensing in droplets}
To improve measurement precision and render ODMR contrast immune to background fluctuations, we employ a strategy utilizing droplet flow, as illustrated in \zfr{mfig3}A. Droplets, uniform in size and nanodiamond content, move at a controlled velocity ($v$) and sequentially enter the analysis region, where an optical spot, roughly equal to the droplet diameter, illuminates them. The spacing between droplets is $vf_D^{-1}$, where $f_D$ is the rate at which droplets are analyzed in region $a$ in \zfr{mfig1}A, and $v$ ranges from 1mm/s - 4cm/s in our experiments. \zfr{mfig3}A also illustrates mixing due to flow (see movie in Ref.~\cite{750nmDiamondFlowing}).

Simultaneously with flow, droplets are subject to MWs amplitude modulated at $f_{\mw}{=}1$ kHz (\zfr{mfig3}B) at 2.866 GHz under continuous laser illumination, introducing two distinct modulations to the PL: $f_D$ and $f_{\mw}$ (\zfr{mfig3}B). By arranging $f_{\mw}{\gg}f_D$, each droplet's signal contains multiple MW modulation cycles. Locking into these frequencies allows us to filter out background noise, a method we term \I{``double lock-in"} detection.

Since the method rests on exploiting droplet flow, we first highlight some of its features. \zfr{mfig3}C illustrates bright-field and fluorescence images of diamond-filled droplets in motion, complementing the stationary images in \zfr{mfig1}C (see also SI~\zsr{videos} and movie in Ref.~\cite{FlowRates_Videos}). Flow exhibits remarkable regular modulation at $f_D$ due to droplet monodispersity, as evidenced by \zfr{mfig3}D displaying measured photoluminescence over 500 s and ${>}10^4$ droplets. Individual 1s windows are shown here, and dashed rails highlight droplet stability (see also \zfr{mfig4}).

\zfr{mfig3}E zooms into a representative 140ms window, showing modulation in the PL both from droplet flow and from the MWs. Solid lines are a fit to the data. The distinct time-scales of both droplet and MW modulations are evident. This is clearer in the inset, \zfr{mfig3}E(i), which focuses on a 30 ms window, illustrating oscillations spaced by $f_{\mw}^{-1}$ (1ms), with amplitude at ${\sim}0.5$\% of raw PL (shaded grey region), consistent with the ODMR contrast in \zfr{mfig2}C.

The PL double modulation can be expressed as the functional form, 
\beq
S(t)=[m(t)+g(t)\cos(2\pi f_Dt)]\cdot[1-\mC(t)\cos(2\pi f_{\mw}t+\phi)]+b(t)
\zl{PL}
\eeq 
where $g(t)$ represents the droplet modulation profile influenced by droplet shape and separation. Microfluidic control enables adjustable profiles, from sinusoidal to square-like (SI~\zsr{flow_control}). $\mC$ represents the ODMR contrast, with time dependence included to account for possible long-term drift across numerous droplets. $m(t)$ reflects the baseline ND contribution to PL, becoming more prominent as inter-droplet spacing decreases (and controllable via junction $J_2$, see \zfr{mfig1}B(ii)). Finally, $b(t)$ captures PL noise from non-ND contributions.

\zfr{mfig3}F displays the Fourier transform amplitude $\mF$ of the PL time series from \zfr{mfig3}E, measured over 15s of droplet flow, after subtracting its mean value. Data here is shown on a logarithmic scale for clarity. It features a distinct Fourier peak at $f_D{=}29$ Hz with a narrow linewidth ($\Delta f_D{\app}1$ Hz), reflecting minimal drift in flow rates (see \zfr{mfig4}). Solid lines here are Lorentzian fits.  Square-like modulation leads to secondary harmonics at multiples of $f_D$, while background signal and autofluorescence occur near zero frequency and can be excluded. Expanding \zr{PL} uncovers frequency bands linked to MW modulation and their combinations with flow, at $f_{\mw}$ and $f_{\mw} {\pm} f_D$. This is shown in the yellow shaded region, \zfr{mfig3}F(i). It is evident the peak intensities here are ${\sim}1$\% that of the droplet modulation, a reflection of contrast $\mC$. \zfr{mfig3}F(ii) shows the same frequency window in a linear scale for clarity.

The ODMR contrast $\mC(t)$ can now be calculated from the ratio of FT peak intensities at the MW and droplet frequencies. Within fixed time windows $\Delta t$, this can be expressed as:
\beq
\mC(t_i)= \frac{\mF(f_{\mw},t_i)+ \mF(f_{\mw}+f_D,t_i) + \mF(f_{\mw}-f_D,t_i)}{\mF(f_D,t_i)}\cdot\frac{1}{f_D(t_i)}
\zl{contrast}
\eeq
Here, $\mF(f,t_i)$ denotes the Fourier intensity at frequency $f$ for a time interval bounded by $t_i$ and $t_i-\Delta t$, averaging the PL over several droplets. Dividing the MW-associated FT peak intensities by the droplet frequency FT peak intensity yields a ratiometric ODMR contrast, factoring in the droplet PL. \zr{contrast} also includes a normalization proportional to droplet frequency to counteract minor frequency drifts that impact the baseline ND-dependent PL ($m(t)$), providing a consistent contrast metric irrespective of droplet flow rate.

\zfr{mfig3}F additionally highlights that the noise profile diminishes ${\underset{\sim}{\propto}}1/f$, suggesting that higher droplet rates could lead to lower noise while enabling greater analysis throughput. To demonstrate the versatility and control over droplet modulation in our experiments, \zfr{mfig3}G shows smoothed PL profiles at various $f_D$ rates. At the upper limit ($f_D{=}346.5$Hz), more than a million droplets can be analyzed per hour (\zfr{mfig4}C), the droplets flowing here at a rapid velocity, $v{>}$5cm/s (see movie in Ref.~\cite{FlowRates_Videos}).

\zfr{mfig4}A depicts the result of \zr{contrast} applied to 40 nm particles in droplets flowing at $f_D{=}34$Hz, measured over a long period (\(T{=}10^4\)s). Data here is sampled every 100ms (corresponding to roughly 3 droplets), and the top axis quantifies the droplet count. The ODMR contrast, $\mC{=}5.6\%$, is marked by the dashed horizontal line in \zfr{mfig4}A. Data is displayed on an identical 1\% contrast window to draw a comparison to \zfr{mfig2}C. Right panel (\zfr{mfig4}A(i)) shows this as a histogram, overlaid with the analogous histogram from \zfr{mfig2}C using the analog lock-in for clarity. The histogram linewidth in the case of \zfr{mfig4}A narrows significantly to $\Delta \mC{=}2\%$ highlighting the enhanced measurement precision. \zfr{mfig4}A also illustrates the inherent stability in the measurement of contrast $\mC$, here over more than 2h and 250,000 droplets.
\begin{figure*}
\centering
\begin{minipage}{\linewidth}
\makebox[\linewidth]{
 \includegraphics[width=0.95\textwidth]{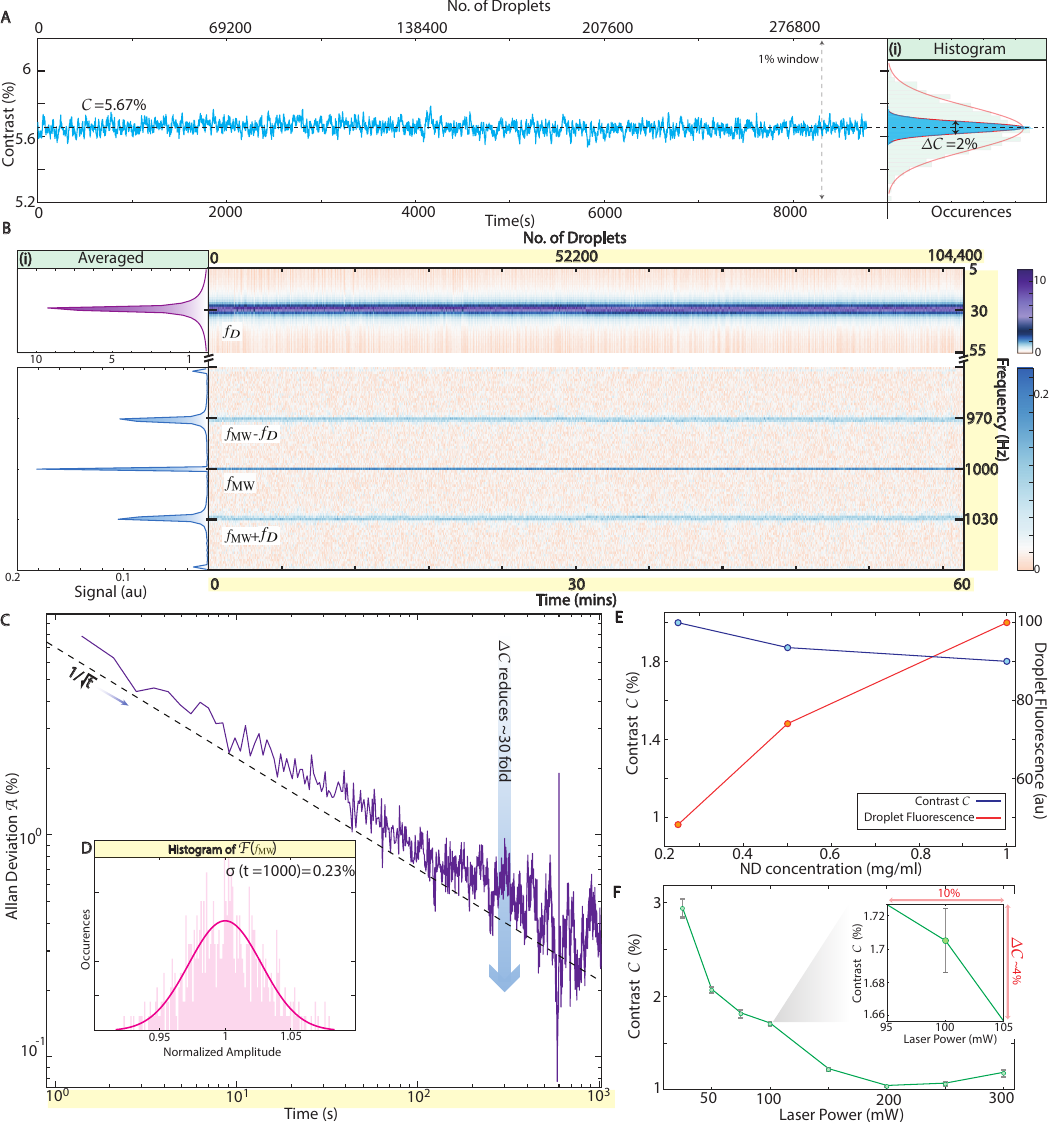}}
  \caption{\T{High stability in-flow droplet measurement.} \T{(A)} \T{Long-time high-precision measurement} of ODMR contrast $\mC$ following \zr{contrast} over $T{\app}2.5\%$ hr period, encompassing ${>}290000$ droplets (upper axis), demonstrating remarkable stability. Shown is a 1\% contrast window similar to \zfr{mfig2}C. Data is sampled every ${\app}$1s. (i) \I{Inset:} Blue bars show histogram of the measured contrast $\mC$ in (A). Solid line is a Gaussian fit; from linewidth we estimate percent error $\xD\mC{=}2 \%$ of the mean contrast (dashed line). Green histogram from \zfr{mfig2}C(ii) is overlaid for reference, highlighting narrowing via double lock-in scheme.
  \T{(B)} \T{Spectrogram of the Fourier peaks} in frequency bands around $f_{D}$ and $f_{\mw}$ measured over 1 hr. Data is presented over successive 700ms windows (corresponding to 20 droplets), for a total of 104k droplets (upper axis). Upper (lower) window spans 50Hz and 120Hz, respectively. Colors indicate FT intensities $\mF(f)$ of PL in the two frequency bands (see colorbars). (i) \I{Left panels:} integrated intensity of spectrogram data plotted against frequency. Narrow peaks are evident, indicating high stability over the entire period.
  \T{(C)}  \T{Allan deviation} $\mA(t)$ shown for 1000 s of data in (B). $\mA(t)$ follows $\propto 1/\sq{t}$ trend (dashed line) for the entire period, highlighting remarkable stability. Percent error $\xD\mC$ reduces over 30-fold as a result. 
\T{(D)} \T{Bounding ND variation per droplet.} Histogram of intensity of  $\mF(f_{\mw})$ peak from spectrogram in (B), measured over 7s bins. Solid line is a Gaussian fit. Extrapolating to 1000s (main panel), we estimate inter droplet ND variation ${<}0.23\%$. This corresponds to ${\lesssim}2300$ particle variation over ${\app}$1M NDs per droplet.
\T{(E-F)} \T{Compensation of experimental variations} via \zr{contrast}.  \T{(E)} Variation with ND concentration of ODMR contrast $\mC$. Red line shows measured droplet fluorescence, while blue line shows corresponding $\mC$. Even a large 400\% change in diamond concentration (D) results only in ${<}0.2$\% change to $\mC$. 
\T{(F)}  Variation laser power (D) (at fixed MW power). \I{Inset:} shows the operational regime for our experiments at ${\app}100$mW. Even 10\% variation in laser power here results only in relative contrast change in $\xD\mC$ of ${\app}4$\%.}
\zfl{mfig4}
\end{minipage}
\end{figure*}

\begin{figure*}
  \centering
  \begin{minipage}{\linewidth}
\makebox[\linewidth]{
  \includegraphics[width=0.95\textwidth]{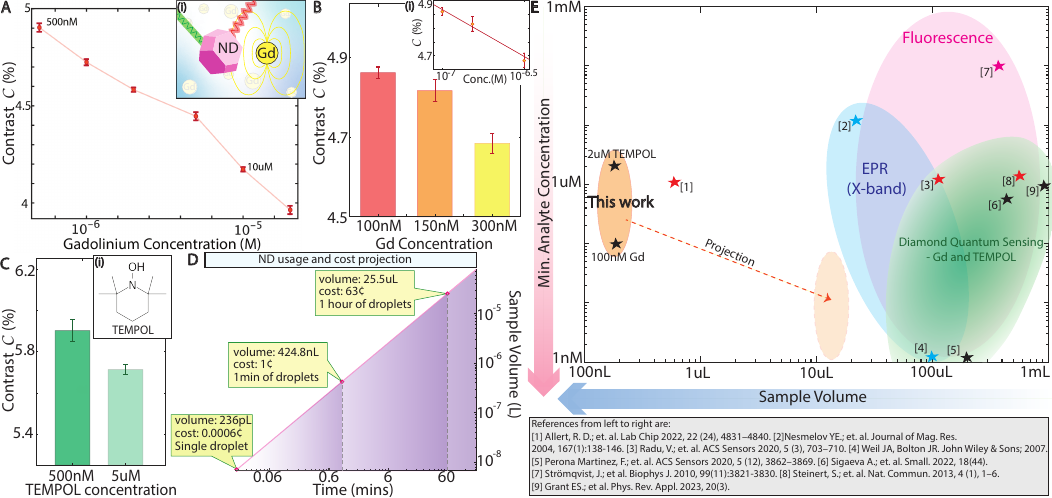}}
  \caption{\T{Sensing paramagnetic species in flowing microdroplets.}
\T{(A)} \T{Gd$^{3+}$ ion detection.} ODMR contrast $\mC$ as function of Gd$^{3+}$ concentration measured in ${\app}50\mu$m droplets flowing at $f_D{\sim}$40Hz. Data points are red circles; error bars reflect $\xD\mC$ for 4 min measurements.
\I{Inset:} Schematic of sensing. Spin noise from Gd$^{3+}$ ions affects NV $T_1$, and converts to measurable change in ODMR contrast $\mC$.
\T{(B)} \T{Gadolinium sensitivity.} Results of separate experiment measuring a lower range of Gd$^{3^+}$concentration. \I{Inset (i):} Data plotted against a log scale in concentration. We estimate a LOD ${\app}100$nM.
\T{(C)} \T{TEMPOL sensing.} Similar measurements for two concentrations of TEMPOL in flowing ${\app}50\mu$m droplets over 1min each. We estimate a LOD ${<}2\mu$M.
\T{(D)} \T{Cost and volume scaling} for in-droplet sensing shown on a logarithmic scale. Marked points correspond to a single droplet, 1 minute, and 1hr of measurements, assuming flow at $f_D{=}30$Hz. 
\T{(E)} \T{Landscape of chemical quantum sensing techniques.} Comparison of related methods (diamond sensing, EPR, fluorescence) for radical/paramagnetic analyte concentration on axes corresponding to sample volume required and lowest detectable concentration reported. Ideal sensing constitues bottom-left region of plot (arrows). Stars are derived from specific references; shaded regions illustrate approximate sensing boundaries. Our work is shown by orange region. Red stars represent single crystal based sensing and black stars represent ND based sensing. Light orange region shows projected improvement from averaging to $10^3$s (\zfr{mfig4}C).}
\zfl{mfig5}
\end{minipage}
\end{figure*}

\subsection{High stability quantum sensing in flow}
\zsl{stable}
To delve deeper into the observed stability in \zfr{mfig4}A, we analyze the data as a time-domain spectrogram in \zfr{mfig4}B. This entails taking a Fourier transform of the PL over small windows $\xD t{=}0.7$s, equivalent to 20 droplets, for an hour, and tracking the resulting spectra over 104,400 droplets (top axis). Colors in \zfr{mfig4}B represent Fourier spectral intensity, and the vertical position indicates frequency. We focus on frequency windows near $f_D$ and $f_{\mw}$, using two distinct color bars for clarity. The resulting horizontal band-like patterns underscore the remarkable stability of the data. The left panel, \zfr{mfig4}B(i), shows the integrated signal across these bands. The narrow linewidths, evident even over this extended period, reflect the system stability. 

Another perspective on the stability is provided through an Allan deviation analysis, applied to the data in \zfr{mfig4}B and depicted in \zfr{mfig4}C. Allan deviation, $\mA(t)$, provides insight into how measurement precision of $\mC$ (ultimately related to analyte LOD) may improve with longer averaging time $t$ or increasing number of droplets. Remarkably, our experimentally measured $\mA(t)$ closely aligns with the theoretically expected ${\propto}1/\sqrt{t}$ trend (dashed line) beyond $10^3$s of averaging, yielding a ${>}$30-fold reduction in $\xD\mC$ (blue arrow in \zfr{mfig4}C). At the lowest point, this corresponds to detecting PL changes to a few-hundredth of a percent. \zfr{mfig4}C marks a notable improvement over previous studies \cite{Fujiwara2020realtime,wolf2015subpicotesla,Hart2021prapplied,Gu2023}, where equivalent ${\propto}1/\sqrt{t}$ scaling is challenging to obtain, and only achievable through highly sophisticated compensation strategies. We attribute this enhanced stability to the immunity of the confined aqueous droplet volume to temperature drifts, their exposure to laser illumination only for short instants ($f_D^{-1}{\sim}33$ms), as well as compensation for laser power fluctuations and ND loading variations by the ratiometric scheme in \zr{contrast}. 

Variability in FT intensities of the band centered at $f_{\mw}$ provides a convenient means to estimate an upper bound on the inter-droplet ND variation. \zfr{mfig4}D shows histogram of the $\mF(f_{\mw},t_i)$ amplitude binned every 7s ($\sim$200 droplets). When extrapolated to $10^3$s following a $t^{-1/2}$ scaling, we obtain an ND variation across droplets lower than $0.23\%$. In absolute terms, this corresponds to a very low droplet-to-droplet variation of ${\lesssim}2300$ NDs over the base level of ${\app}$1M NDs per droplet. 

Now, exploring the impact of potential drift mechanisms, in \zfr{mfig4}E-F, we assess how laser power and ND concentration variations affect ODMR contrast $\mC$. First, in \zfr{mfig4}E we adjust ND concentration over a large range in the flowing droplets, using a 10-way valve (SI \zsr{m_switch}) to load successive samples containing different concentrations while maintaining fixed laser and MW powers. Red points in \zfr{mfig4}E show the resulting change in droplet PL, while blue points show the measured ODMR contrast $\mC$ in percent units. Despite a significant, ${\sim}$500\%-increase in ND concentration, ODMR contrast variation is less than $\xD \mC{\sim}$11\%. Given that inter-droplet ND concentration variation is ${<}0.23$\% (\zfr{mfig4}D), ND  number variations minimally impact contrast.

Increasing laser power, with MW power held constant (\zfr{mfig4}F) affects $\mC$ due to an interplay of NV center repolarization and MW-driven population shifts. However, as \zfr{mfig4}F indicates, at our operational power (${\sim}$100mW), a 10\% laser power variation alters ODMR contrast by only about $\xD\mC{=}4$\%. In reality, laser drift is under 1\%, and this leads to an insignificant effect on $\xD\mC$. Overall, \zfr{mfig4}E-F underscores the method's resilience to common experimental fluctuations.

We comment finally that the double lock-in scheme could be performed through alternate approaches. MW modulation could be replaced by magnetic modulation, with potential advantages of a higher modulation depth (${>}$10\%)~\cite{Singam16,Jones21,Torelli20}. However, it lacks the precise frequency control and long-term stability of MW lock-in.  The latter also benefits from lower $1/f$ noise due to its higher operating frequency. While laser modulation, such as with an optical chopper, is a possible alternate method to $f_D$ droplet modulation, droplet flow, as introduced here, has distinct advantages: (1) immunity to autofluorescence (SI \zsr{autoflour}), (2) averaging effects across droplets to reduce heterogeneity, (3) suitability for high-throughput analysis in a flowing geometry with (4) stable $f_D$ modulation for several hours, and (5) improved thermal stability of flowing versus static droplets, ensuring reliability over long-time analyses.

\subsection{Detection of paramagnetic species in flow}
\zsl{sensing}
Leveraging the enhanced precision above,  we detect chemical analytes in flowing droplets, starting with Gadolinium (Gd$^{3+}$) ions in GdCl$_3$ hexahydrate as a model system. The sensing mechanism, depicted in \zfr{mfig5}A(i) inset, relies on changes in the NV center $T_1$ relaxation time from spin noise of paramagnetic species, affecting the $m_s{=}\pm 1$ population and, thus, ODMR contrast. This effect is concentration-dependent, allowing for quantitative analyte estimation. Enhanced precision via \zfr{mfig3}-\zfr{mfig4} enables the detection of subtle contrast changes, significantly improving limits of detection (LODs).

In \zfr{mfig5}A, we demonstrate this through Gd$^{3+}$ concentration titration in droplets, with each sample averaged for ${\app}$4 min. Samples are automatically loaded using a 10-way valve (SI \zsr{m_switch}) controlled by a customized program, showcasing the potential for automated analysis. The small error bar in the measured points, $\xD\mC{\app}1\%$ (i.e. 1\% of the bare contrast $\mC$), reflects the high measurement precision. \zfr{mfig5}A highlights the dynamic range over which this system can detect gadolinium ions, ranging from a 500 nM to 20 $\mu$M, above which the ionic strength of the solution causes aggregation of our carboxylated diamond particles. We further investigate the sensitivity at the low end in a separate experiment loading 100 nM, 150 nM, and 300 nM samples in \zfr{mfig5}B. We achieve a LOD of 100 nM over 2 min. of averaging.

While Gd$^{3+}$ has spin $S{=}7/2$, most practical applications target single unpaired electronic spins ($S{=}1/2$). We thus employ 4-Hydroxy-2,2,6,6-tetramethylpiperidine-1-oxyl (TEMPOL) as a second model (\zfr{mfig5}C(i)). TEMPOL is a stable radical probe that could serve as a proxy for other paramagnetic analytes, including reactive oxygen species (ROS), crucial for instance for metabolic studies in cells. \zfr{mfig5}C shows the ability to differentiate between 500nM and 5$\mu$M concentrations. We estimate an LOD of $2\mu$M over 1min. of measurement.

\zfr{mfig5}D highlights how our droplet-based sensing requires a very small quantity of NDs. The red line shows sample volume used for varying measurement times -- 1 droplet, 1 min of averaging, up to 1hr. ND costs, at \$50/mg according to Adamas Nanotechnologies, are \I{remarkably} low, costing less than 30$\mbox{\textcent}$ for 1000s of analysis involving ${\sim}3\zt10^4$ droplets. This benefits also from the NDs' native hydrophilicity~\cite{Petit17} (\zsr{materials}), which negates the need for surface treatments.
Our method therefore not only reduces costs dramatically compared to traditional single-crystal diamond approaches requiring costly growth and surface modifications but also enhances portability. Combined with the data from \zfr{mfig4}, this showcases the capability for long-term, stable, and precise analyses at significantly reduced costs.

Finally, \zfr{mfig5}E shows an overview of the sensing technology landscape, identifying the niche filled by droplet-based quantum sensing. We focus attention here on gadolinium and TEMPOL detection, and comparisons to widely used methods for ROS and spin trap detection. We assess electron paramagnetic resonance (EPR)~\cite{Nesmelov04,Roessler18,Abhyankar22}, fluorescence, and NDs-based sensing across reported LODs and analyte sample volume (\zfr{mfig5}E). \ztr{table} in the SI provides more detail contrasting diamond-based quantum sensing methods. 
An ideal chemical sensing platform would occupy the bottom left corner of this plot. Though direct comparisons are challenging due to the diversity of implementations in each technique, we show regions of applicability of each method (shaded regions in \zfr{mfig5}E), and representative references (marked points). Droplet-based sensing as described here occupies the orange region; and is projected to move along the red-dashed arrow with averaging following \zfr{mfig4}C. \zfr{mfig5}E illustrates that our method already provides significant improvements over existing technologies.

\section{Discussion and Outlook}
Our work combines droplet microfluidics with quantum sensing and introduces many new innovations, including (1) deploying ND quantum sensors within droplets, utilizing confinement and flow to (2) facilitate continuous analysis with high precision, capable of detecting contrast changes by $\xD C{\app}2\%$, (3) with high stability across multiple hours and hundreds of thousands of droplets, (4) all while utilizing minuscule sensor volumes and entailing incredibly low ND costs. Looking forward, the platform technology introduced here anticipates many new interesting directions.

First, the use of ND-loaded droplets opens new possibilities for moving beyond traditional single-crystal diamond sensors to those that closely interact with analyte molecules. Mixing within the droplets averages out particle heterogeneity, and averaging over many droplets is simple, rapid, and improves sensitivity (\zfr{mfig4}C). The mobility afforded by droplet confinement introduces versatility in sensor manipulation—such as via droplet sorting, splitting, and collisions~\cite{Davis17} -- and precise placement over target samples. Utilizing electrostatic forces, charged droplets can be dynamically organized~\cite{Lee15}, forming consistent quantum sensor ensembles that could enable the creation of rearrangeable three-dimensional sensor arrays. We also foresee the development of an \I{“ND particle sorter”} capable of isolating high-quality diamond particles from random populations based on desired characteristics, such as NV center $T_2$ times, streamlining the production of quantum materials for sensing.

Our in-flow analysis method, with its precision, stability, and low-volume capabilities, enables high-throughput, high-sensitivity chemical assays~\cite{Kintses12}.  These assays can be conducted serially using single droplets or in a  wide-field setting using a lock-in camera \cite{wojciechowski2018contributed,parashar2022sub,Hart2021prapplied,webb2022high}. They would be ideal for analysis in amplification-free scenarios or in optically turbid media (e.g. blood), eliminating the need for washing steps; and opening applications in bioengineering, trace pathogen detection for diagnostics, and chemical reaction monitoring \cite{Shembekar2016}. Moreover, the assay devices can be rendered compact and portable, thanks to the remarkably low cost of the ND sensors (\zfr{mfig5}C), even lower than the reagents for droplet generation, combined with high stability (\zfr{mfig4}C), and robustness to laser power drifts (\zfr{mfig4}F). The latter suggests feasibility of using low-cost diode lasers, suggesting compact, field-deployable devices.

Utilizing droplets for cell confinement heralds new directions in high-throughput single-cell analysis. This technique is particularly promising for methods akin to flow cytometry \cite{givan2011flow}, focusing on cellular metabolism through detecting intracellular paramagnetic reactive oxygen species (ROS) \cite{eruslanov2010identification}. We anticipate \I{``quantum-enhanced"} flow cytometry, introducing additional dimensions such as cell-localized $T_1$ relaxometry, providing insights towards single-cell metabolomics alongside traditional cell morphology information.

When combined with real-time sampling, this could also enable unprecedented intracellular measurements directly from bioreactors. By integrating cellular lysis buffers and reagents within droplets or through droplet collisions, we imagine a rapid cellular analysis method aimed at precisely controlling bioreactor conditions to achieve optimal outcomes.

More fundamentally, our approach unlocks new possibilities for in-situ chemical imaging and kinetic analysis within droplet microenvironments~\cite{Lee15}, filling a significant unmet need. ND particle tracking (\zfr{mfig1}E and movie in Ref.~\cite{TrackingMovie}) enables new methods for real-space imaging of diffusion in microconfined environments, including turbid settings, opening new perspectives in chemical and biological systems. 

The Allan deviation depicted in \zfr{mfig4}C highlights measurement stability amongst best reported in literature, but achieved with significant ease of operation. This suggests ND-filled flowing droplets could serve as a promising platform for applications in bulk magnetometry~\cite{Barry20} and as acclerometers and rotation sensors~\cite{Ajoy12g,Jarmola21}. Finally, we note other properties of NDs not utilized in this study but fully compatible with being deployed in flowing droplets. These include the ability to hyperpolarize $\Cs$ nuclear spins in NDs optically via NV centers~\cite{Ajoy17}, and exploit their minute-long transverse lifetimes $T_2'$~\cite{Beatrez2021} to construct movable sensors, high-field magnetometers, and NMR sensors within droplets~\cite{Sahin21}.


\I{Acknowledgments} -- We thank M. Mather, A.Smirnov and V. Radu for insightful discussions. We acknowledge facilities in the U.C. Berkeley Biomolecular Nanotechnology Center (BNC) that were used for device fabrication. We gratefully acknowledge Laboratory Directed Research \& Development (LDRD) funding from Lawrence Berkeley National Laboratory  provided by the Department of Energy (DOE) under Contract No. DE-AC02-05CH11231. We also acknowledge funding from DOE BES (Geoscience, DE-AC02-05CH11231), DOE BER (DE-SC0023065), William M. Keck Foundation (8959), DOE SBIR (DE-SC0022441), AFOSR YIP (FA9550-23-1-0106), and the CIFAR Azrieli Foundation (GS23-013)

\vspace{-5mm} 
\bibliography{Droplet.bbl}

\begin{thebibliography}{80}%
\makeatletter
\providecommand \@ifxundefined [1]{%
 \@ifx{#1\undefined}
}%
\providecommand \@ifnum [1]{%
 \ifnum #1\expandafter \@firstoftwo
 \else \expandafter \@secondoftwo
 \fi
}%
\providecommand \@ifx [1]{%
 \ifx #1\expandafter \@firstoftwo
 \else \expandafter \@secondoftwo
 \fi
}%
\providecommand \natexlab [1]{#1}%
\providecommand \enquote  [1]{``#1''}%
\providecommand \bibnamefont  [1]{#1}%
\providecommand \bibfnamefont [1]{#1}%
\providecommand \citenamefont [1]{#1}%
\providecommand \href@noop [0]{\@secondoftwo}%
\providecommand \href [0]{\begingroup \@sanitize@url \@href}%
\providecommand \@href[1]{\@@startlink{#1}\@@href}%
\providecommand \@@href[1]{\endgroup#1\@@endlink}%
\providecommand \@sanitize@url [0]{\catcode `\\12\catcode `\$12\catcode `\&12\catcode `\#12\catcode `\^12\catcode `\_12\catcode `\%12\relax}%
\providecommand \@@startlink[1]{}%
\providecommand \@@endlink[0]{}%
\providecommand \url  [0]{\begingroup\@sanitize@url \@url }%
\providecommand \@url [1]{\endgroup\@href {#1}{\urlprefix }}%
\providecommand \urlprefix  [0]{URL }%
\providecommand \Eprint [0]{\href }%
\providecommand \doibase [0]{https://doi.org/}%
\providecommand \selectlanguage [0]{\@gobble}%
\providecommand \bibinfo  [0]{\@secondoftwo}%
\providecommand \bibfield  [0]{\@secondoftwo}%
\providecommand \translation [1]{[#1]}%
\providecommand \BibitemOpen [0]{}%
\providecommand \bibitemStop [0]{}%
\providecommand \bibitemNoStop [0]{.\EOS\space}%
\providecommand \EOS [0]{\spacefactor3000\relax}%
\providecommand \BibitemShut  [1]{\csname bibitem#1\endcsname}%
\let\auto@bib@innerbib\@empty
\bibitem [{\citenamefont {Degen}\ \emph {et~al.}(2017)\citenamefont {Degen}, \citenamefont {Reinhard},\ and\ \citenamefont {Cappellaro}}]{Degen17}%
  \BibitemOpen
  \bibfield  {author} {\bibinfo {author} {\bibfnamefont {C.~L.}\ \bibnamefont {Degen}}, \bibinfo {author} {\bibfnamefont {F.}~\bibnamefont {Reinhard}},\ and\ \bibinfo {author} {\bibfnamefont {P.}~\bibnamefont {Cappellaro}},\ }\bibfield  {title} {\bibinfo {title} {Quantum sensing},\ }\href@noop {} {\bibfield  {journal} {\bibinfo  {journal} {Reviews of modern physics}\ }\textbf {\bibinfo {volume} {89}},\ \bibinfo {pages} {035002} (\bibinfo {year} {2017})}\BibitemShut {NoStop}%
\bibitem [{\citenamefont {Zhang}\ \emph {et~al.}(2021)\citenamefont {Zhang}, \citenamefont {Pramanik}, \citenamefont {Zhang}, \citenamefont {Gulka}, \citenamefont {Wang}, \citenamefont {Jing}, \citenamefont {Xu}, \citenamefont {Li}, \citenamefont {Wei}, \citenamefont {Cigler} \emph {et~al.}}]{Zhang21}%
  \BibitemOpen
  \bibfield  {author} {\bibinfo {author} {\bibfnamefont {T.}~\bibnamefont {Zhang}}, \bibinfo {author} {\bibfnamefont {G.}~\bibnamefont {Pramanik}}, \bibinfo {author} {\bibfnamefont {K.}~\bibnamefont {Zhang}}, \bibinfo {author} {\bibfnamefont {M.}~\bibnamefont {Gulka}}, \bibinfo {author} {\bibfnamefont {L.}~\bibnamefont {Wang}}, \bibinfo {author} {\bibfnamefont {J.}~\bibnamefont {Jing}}, \bibinfo {author} {\bibfnamefont {F.}~\bibnamefont {Xu}}, \bibinfo {author} {\bibfnamefont {Z.}~\bibnamefont {Li}}, \bibinfo {author} {\bibfnamefont {Q.}~\bibnamefont {Wei}}, \bibinfo {author} {\bibfnamefont {P.}~\bibnamefont {Cigler}}, \emph {et~al.},\ }\bibfield  {title} {\bibinfo {title} {Toward quantitative bio-sensing with nitrogen--vacancy center in diamond},\ }\href@noop {} {\bibfield  {journal} {\bibinfo  {journal} {ACS sensors}\ }\textbf {\bibinfo {volume} {6}},\ \bibinfo {pages} {2077} (\bibinfo {year} {2021})}\BibitemShut {NoStop}%
\bibitem [{\citenamefont {Doherty}\ \emph {et~al.}(2012)\citenamefont {Doherty}, \citenamefont {Dolde}, \citenamefont {Fedder}, \citenamefont {Jelezko}, \citenamefont {Wrachtrup}, \citenamefont {Manson},\ and\ \citenamefont {Hollenberg}}]{Doherty12}%
  \BibitemOpen
  \bibfield  {author} {\bibinfo {author} {\bibfnamefont {M.~W.}\ \bibnamefont {Doherty}}, \bibinfo {author} {\bibfnamefont {F.}~\bibnamefont {Dolde}}, \bibinfo {author} {\bibfnamefont {H.}~\bibnamefont {Fedder}}, \bibinfo {author} {\bibfnamefont {F.}~\bibnamefont {Jelezko}}, \bibinfo {author} {\bibfnamefont {J.}~\bibnamefont {Wrachtrup}}, \bibinfo {author} {\bibfnamefont {N.~B.}\ \bibnamefont {Manson}},\ and\ \bibinfo {author} {\bibfnamefont {L.~C.~L.}\ \bibnamefont {Hollenberg}},\ }\bibfield  {title} {\bibinfo {title} {Theory of the ground-state spin of the nv center in diamond},\ }\href {https://doi.org/10.1103/PhysRevB.85.205203} {\bibfield  {journal} {\bibinfo  {journal} {Phys. Rev. B}\ }\textbf {\bibinfo {volume} {85}},\ \bibinfo {pages} {205203} (\bibinfo {year} {2012})}\BibitemShut {NoStop}%
\bibitem [{\citenamefont {Jelezko}\ and\ \citenamefont {Wrachtrup}(2006)}]{Jelezko06}%
  \BibitemOpen
  \bibfield  {author} {\bibinfo {author} {\bibfnamefont {F.}~\bibnamefont {Jelezko}}\ and\ \bibinfo {author} {\bibfnamefont {J.}~\bibnamefont {Wrachtrup}},\ }\bibfield  {title} {\bibinfo {title} {Single defect centres in diamond: A review},\ }\href {https://doi.org/10.1002/pssa.200671403} {\bibfield  {journal} {\bibinfo  {journal} {Physica Status Solidi (A)}\ }\textbf {\bibinfo {volume} {203}},\ \bibinfo {pages} {3207} (\bibinfo {year} {2006})}\BibitemShut {NoStop}%
\bibitem [{\citenamefont {Neumann}\ \emph {et~al.}(2013)\citenamefont {Neumann}, \citenamefont {Jakobi}, \citenamefont {Dolde}, \citenamefont {Burk}, \citenamefont {Reuter}, \citenamefont {Waldherr}, \citenamefont {Honert}, \citenamefont {Wolf}, \citenamefont {Brunner}, \citenamefont {Shim} \emph {et~al.}}]{Neumann13}%
  \BibitemOpen
  \bibfield  {author} {\bibinfo {author} {\bibfnamefont {P.}~\bibnamefont {Neumann}}, \bibinfo {author} {\bibfnamefont {I.}~\bibnamefont {Jakobi}}, \bibinfo {author} {\bibfnamefont {F.}~\bibnamefont {Dolde}}, \bibinfo {author} {\bibfnamefont {C.}~\bibnamefont {Burk}}, \bibinfo {author} {\bibfnamefont {R.}~\bibnamefont {Reuter}}, \bibinfo {author} {\bibfnamefont {G.}~\bibnamefont {Waldherr}}, \bibinfo {author} {\bibfnamefont {J.}~\bibnamefont {Honert}}, \bibinfo {author} {\bibfnamefont {T.}~\bibnamefont {Wolf}}, \bibinfo {author} {\bibfnamefont {A.}~\bibnamefont {Brunner}}, \bibinfo {author} {\bibfnamefont {J.~H.}\ \bibnamefont {Shim}}, \emph {et~al.},\ }\bibfield  {title} {\bibinfo {title} {High-precision nanoscale temperature sensing using single defects in diamond},\ }\href@noop {} {\bibfield  {journal} {\bibinfo  {journal} {Nano letters}\ }\textbf {\bibinfo {volume} {13}},\ \bibinfo {pages} {2738} (\bibinfo {year} {2013})}\BibitemShut {NoStop}%
\bibitem [{\citenamefont {Kucsko}\ \emph {et~al.}(2013)\citenamefont {Kucsko}, \citenamefont {Maurer}, \citenamefont {Yao}, \citenamefont {Kubo}, \citenamefont {Noh}, \citenamefont {Lo}, \citenamefont {Park},\ and\ \citenamefont {Lukin}}]{Kucsko13}%
  \BibitemOpen
  \bibfield  {author} {\bibinfo {author} {\bibfnamefont {G.}~\bibnamefont {Kucsko}}, \bibinfo {author} {\bibfnamefont {P.}~\bibnamefont {Maurer}}, \bibinfo {author} {\bibfnamefont {N.~Y.}\ \bibnamefont {Yao}}, \bibinfo {author} {\bibfnamefont {M.}~\bibnamefont {Kubo}}, \bibinfo {author} {\bibfnamefont {H.}~\bibnamefont {Noh}}, \bibinfo {author} {\bibfnamefont {P.}~\bibnamefont {Lo}}, \bibinfo {author} {\bibfnamefont {H.}~\bibnamefont {Park}},\ and\ \bibinfo {author} {\bibfnamefont {M.~D.}\ \bibnamefont {Lukin}},\ }\bibfield  {title} {\bibinfo {title} {Nanometre-scale thermometry in a living cell},\ }\href@noop {} {\bibfield  {journal} {\bibinfo  {journal} {Nature}\ }\textbf {\bibinfo {volume} {500}},\ \bibinfo {pages} {54} (\bibinfo {year} {2013})}\BibitemShut {NoStop}%
\bibitem [{\citenamefont {Choi}\ \emph {et~al.}(2020)\citenamefont {Choi}, \citenamefont {Zhou}, \citenamefont {Landig}, \citenamefont {Wu}, \citenamefont {Yu}, \citenamefont {Von~Stetina}, \citenamefont {Kucsko}, \citenamefont {Mango}, \citenamefont {Needleman}, \citenamefont {Samuel} \emph {et~al.}}]{Choi20}%
  \BibitemOpen
  \bibfield  {author} {\bibinfo {author} {\bibfnamefont {J.}~\bibnamefont {Choi}}, \bibinfo {author} {\bibfnamefont {H.}~\bibnamefont {Zhou}}, \bibinfo {author} {\bibfnamefont {R.}~\bibnamefont {Landig}}, \bibinfo {author} {\bibfnamefont {H.-Y.}\ \bibnamefont {Wu}}, \bibinfo {author} {\bibfnamefont {X.}~\bibnamefont {Yu}}, \bibinfo {author} {\bibfnamefont {S.~E.}\ \bibnamefont {Von~Stetina}}, \bibinfo {author} {\bibfnamefont {G.}~\bibnamefont {Kucsko}}, \bibinfo {author} {\bibfnamefont {S.~E.}\ \bibnamefont {Mango}}, \bibinfo {author} {\bibfnamefont {D.~J.}\ \bibnamefont {Needleman}}, \bibinfo {author} {\bibfnamefont {A.~D.}\ \bibnamefont {Samuel}}, \emph {et~al.},\ }\bibfield  {title} {\bibinfo {title} {Probing and manipulating embryogenesis via nanoscale thermometry and temperature control},\ }\href@noop {} {\bibfield  {journal} {\bibinfo  {journal} {Proceedings of the National Academy of Sciences}\ }\textbf {\bibinfo {volume} {117}},\ \bibinfo {pages} {14636} (\bibinfo {year} {2020})}\BibitemShut {NoStop}%
\bibitem [{\citenamefont {Petrini}\ \emph {et~al.}(2020)\citenamefont {Petrini}, \citenamefont {Moreva}, \citenamefont {Bernardi}, \citenamefont {Traina}, \citenamefont {Tomagra}, \citenamefont {Carabelli}, \citenamefont {Degiovanni},\ and\ \citenamefont {Genovese}}]{Petrini20}%
  \BibitemOpen
  \bibfield  {author} {\bibinfo {author} {\bibfnamefont {G.}~\bibnamefont {Petrini}}, \bibinfo {author} {\bibfnamefont {E.}~\bibnamefont {Moreva}}, \bibinfo {author} {\bibfnamefont {E.}~\bibnamefont {Bernardi}}, \bibinfo {author} {\bibfnamefont {P.}~\bibnamefont {Traina}}, \bibinfo {author} {\bibfnamefont {G.}~\bibnamefont {Tomagra}}, \bibinfo {author} {\bibfnamefont {V.}~\bibnamefont {Carabelli}}, \bibinfo {author} {\bibfnamefont {I.~P.}\ \bibnamefont {Degiovanni}},\ and\ \bibinfo {author} {\bibfnamefont {M.}~\bibnamefont {Genovese}},\ }\bibfield  {title} {\bibinfo {title} {Is a quantum biosensing revolution approaching? perspectives in nv-assisted current and thermal biosensing in living cells},\ }\href@noop {} {\bibfield  {journal} {\bibinfo  {journal} {Advanced Quantum Technologies}\ }\textbf {\bibinfo {volume} {3}},\ \bibinfo {pages} {2000066} (\bibinfo {year} {2020})}\BibitemShut {NoStop}%
\bibitem [{\citenamefont {van~der Laan}\ \emph {et~al.}(2020)\citenamefont {van~der Laan}, \citenamefont {Morita}, \citenamefont {Perona-Martinez},\ and\ \citenamefont {Schirhagl}}]{van20}%
  \BibitemOpen
  \bibfield  {author} {\bibinfo {author} {\bibfnamefont {K.~J.}\ \bibnamefont {van~der Laan}}, \bibinfo {author} {\bibfnamefont {A.}~\bibnamefont {Morita}}, \bibinfo {author} {\bibfnamefont {F.~P.}\ \bibnamefont {Perona-Martinez}},\ and\ \bibinfo {author} {\bibfnamefont {R.}~\bibnamefont {Schirhagl}},\ }\bibfield  {title} {\bibinfo {title} {Evaluation of the oxidative stress response of aging yeast cells in response to internalization of fluorescent nanodiamond biosensors},\ }\href@noop {} {\bibfield  {journal} {\bibinfo  {journal} {Nanomaterials}\ }\textbf {\bibinfo {volume} {10}},\ \bibinfo {pages} {372} (\bibinfo {year} {2020})}\BibitemShut {NoStop}%
\bibitem [{\citenamefont {Sharmin}\ \emph {et~al.}(2021)\citenamefont {Sharmin}, \citenamefont {Hamoh}, \citenamefont {Sigaeva}, \citenamefont {Mzyk}, \citenamefont {Damle}, \citenamefont {Morita}, \citenamefont {Vedelaar},\ and\ \citenamefont {Schirhagl}}]{Sharmin21}%
  \BibitemOpen
  \bibfield  {author} {\bibinfo {author} {\bibfnamefont {R.}~\bibnamefont {Sharmin}}, \bibinfo {author} {\bibfnamefont {T.}~\bibnamefont {Hamoh}}, \bibinfo {author} {\bibfnamefont {A.}~\bibnamefont {Sigaeva}}, \bibinfo {author} {\bibfnamefont {A.}~\bibnamefont {Mzyk}}, \bibinfo {author} {\bibfnamefont {V.~G.}\ \bibnamefont {Damle}}, \bibinfo {author} {\bibfnamefont {A.}~\bibnamefont {Morita}}, \bibinfo {author} {\bibfnamefont {T.}~\bibnamefont {Vedelaar}},\ and\ \bibinfo {author} {\bibfnamefont {R.}~\bibnamefont {Schirhagl}},\ }\bibfield  {title} {\bibinfo {title} {Fluorescent nanodiamonds for detecting free-radical generation in real time during shear stress in human umbilical vein endothelial cells},\ }\href@noop {} {\bibfield  {journal} {\bibinfo  {journal} {ACS sensors}\ }\textbf {\bibinfo {volume} {6}},\ \bibinfo {pages} {4349} (\bibinfo {year} {2021})}\BibitemShut {NoStop}%
\bibitem [{\citenamefont {Miller}\ \emph {et~al.}(2020)\citenamefont {Miller}, \citenamefont {Bezinge}, \citenamefont {Gliddon}, \citenamefont {Huang}, \citenamefont {Dold}, \citenamefont {Gray}, \citenamefont {Heaney}, \citenamefont {Dobson}, \citenamefont {Nastouli}, \citenamefont {Morton} \emph {et~al.}}]{Miller20}%
  \BibitemOpen
  \bibfield  {author} {\bibinfo {author} {\bibfnamefont {B.~S.}\ \bibnamefont {Miller}}, \bibinfo {author} {\bibfnamefont {L.}~\bibnamefont {Bezinge}}, \bibinfo {author} {\bibfnamefont {H.~D.}\ \bibnamefont {Gliddon}}, \bibinfo {author} {\bibfnamefont {D.}~\bibnamefont {Huang}}, \bibinfo {author} {\bibfnamefont {G.}~\bibnamefont {Dold}}, \bibinfo {author} {\bibfnamefont {E.~R.}\ \bibnamefont {Gray}}, \bibinfo {author} {\bibfnamefont {J.}~\bibnamefont {Heaney}}, \bibinfo {author} {\bibfnamefont {P.~J.}\ \bibnamefont {Dobson}}, \bibinfo {author} {\bibfnamefont {E.}~\bibnamefont {Nastouli}}, \bibinfo {author} {\bibfnamefont {J.~J.}\ \bibnamefont {Morton}}, \emph {et~al.},\ }\bibfield  {title} {\bibinfo {title} {Spin-enhanced nanodiamond biosensing for ultrasensitive diagnostics},\ }\href@noop {} {\bibfield  {journal} {\bibinfo  {journal} {Nature}\ }\textbf {\bibinfo {volume} {587}},\ \bibinfo {pages} {588} (\bibinfo {year} {2020})}\BibitemShut {NoStop}%
\bibitem [{\citenamefont {Glenn}\ \emph {et~al.}(2018)\citenamefont {Glenn}, \citenamefont {Bucher}, \citenamefont {Lee}, \citenamefont {Lukin}, \citenamefont {Park},\ and\ \citenamefont {Walsworth}}]{Glenn18}%
  \BibitemOpen
  \bibfield  {author} {\bibinfo {author} {\bibfnamefont {D.~R.}\ \bibnamefont {Glenn}}, \bibinfo {author} {\bibfnamefont {D.~B.}\ \bibnamefont {Bucher}}, \bibinfo {author} {\bibfnamefont {J.}~\bibnamefont {Lee}}, \bibinfo {author} {\bibfnamefont {M.~D.}\ \bibnamefont {Lukin}}, \bibinfo {author} {\bibfnamefont {H.}~\bibnamefont {Park}},\ and\ \bibinfo {author} {\bibfnamefont {R.~L.}\ \bibnamefont {Walsworth}},\ }\bibfield  {title} {\bibinfo {title} {High-resolution magnetic resonance spectroscopy using a solid-state spin sensor},\ }\href@noop {} {\bibfield  {journal} {\bibinfo  {journal} {Nature}\ }\textbf {\bibinfo {volume} {555}},\ \bibinfo {pages} {351} (\bibinfo {year} {2018})}\BibitemShut {NoStop}%
\bibitem [{\citenamefont {Aslam}\ \emph {et~al.}(2017)\citenamefont {Aslam}, \citenamefont {Pfender}, \citenamefont {Neumann}, \citenamefont {Reuter}, \citenamefont {Zappe}, \citenamefont {de~Oliveira}, \citenamefont {Denisenko}, \citenamefont {Sumiya}, \citenamefont {Onoda}, \citenamefont {Isoya} \emph {et~al.}}]{Aslam17}%
  \BibitemOpen
  \bibfield  {author} {\bibinfo {author} {\bibfnamefont {N.}~\bibnamefont {Aslam}}, \bibinfo {author} {\bibfnamefont {M.}~\bibnamefont {Pfender}}, \bibinfo {author} {\bibfnamefont {P.}~\bibnamefont {Neumann}}, \bibinfo {author} {\bibfnamefont {R.}~\bibnamefont {Reuter}}, \bibinfo {author} {\bibfnamefont {A.}~\bibnamefont {Zappe}}, \bibinfo {author} {\bibfnamefont {F.~F.}\ \bibnamefont {de~Oliveira}}, \bibinfo {author} {\bibfnamefont {A.}~\bibnamefont {Denisenko}}, \bibinfo {author} {\bibfnamefont {H.}~\bibnamefont {Sumiya}}, \bibinfo {author} {\bibfnamefont {S.}~\bibnamefont {Onoda}}, \bibinfo {author} {\bibfnamefont {J.}~\bibnamefont {Isoya}}, \emph {et~al.},\ }\bibfield  {title} {\bibinfo {title} {Nanoscale nuclear magnetic resonance with chemical resolution},\ }\href@noop {} {\bibfield  {journal} {\bibinfo  {journal} {Science}\ ,\ \bibinfo {pages} {eaam8697}} (\bibinfo {year} {2017})}\BibitemShut {NoStop}%
\bibitem [{\citenamefont {Staudacher}\ \emph {et~al.}(2013)\citenamefont {Staudacher}, \citenamefont {Shi}, \citenamefont {Pezzagna}, \citenamefont {Meijer}, \citenamefont {Du}, \citenamefont {Meriles}, \citenamefont {Reinhard},\ and\ \citenamefont {Wrachtrup}}]{Staudacher13}%
  \BibitemOpen
  \bibfield  {author} {\bibinfo {author} {\bibfnamefont {T.}~\bibnamefont {Staudacher}}, \bibinfo {author} {\bibfnamefont {F.}~\bibnamefont {Shi}}, \bibinfo {author} {\bibfnamefont {S.}~\bibnamefont {Pezzagna}}, \bibinfo {author} {\bibfnamefont {J.}~\bibnamefont {Meijer}}, \bibinfo {author} {\bibfnamefont {J.}~\bibnamefont {Du}}, \bibinfo {author} {\bibfnamefont {C.~A.}\ \bibnamefont {Meriles}}, \bibinfo {author} {\bibfnamefont {F.}~\bibnamefont {Reinhard}},\ and\ \bibinfo {author} {\bibfnamefont {J.}~\bibnamefont {Wrachtrup}},\ }\bibfield  {title} {\bibinfo {title} {Nuclear magnetic resonance spectroscopy on a (5-nanometer)3 sample volume},\ }\href {https://doi.org/10.1126/science.1231675} {\bibfield  {journal} {\bibinfo  {journal} {Science}\ }\textbf {\bibinfo {volume} {339}},\ \bibinfo {pages} {561} (\bibinfo {year} {2013})}\BibitemShut {NoStop}%
\bibitem [{\citenamefont {Mamin}\ \emph {et~al.}(2013)\citenamefont {Mamin}, \citenamefont {Kim}, \citenamefont {Sherwood}, \citenamefont {Rettner}, \citenamefont {Ohno}, \citenamefont {Awschalom},\ and\ \citenamefont {Rugar}}]{Mamin13}%
  \BibitemOpen
  \bibfield  {author} {\bibinfo {author} {\bibfnamefont {H.~J.}\ \bibnamefont {Mamin}}, \bibinfo {author} {\bibfnamefont {M.}~\bibnamefont {Kim}}, \bibinfo {author} {\bibfnamefont {M.~H.}\ \bibnamefont {Sherwood}}, \bibinfo {author} {\bibfnamefont {C.~T.}\ \bibnamefont {Rettner}}, \bibinfo {author} {\bibfnamefont {K.}~\bibnamefont {Ohno}}, \bibinfo {author} {\bibfnamefont {D.~D.}\ \bibnamefont {Awschalom}},\ and\ \bibinfo {author} {\bibfnamefont {D.}~\bibnamefont {Rugar}},\ }\bibfield  {title} {\bibinfo {title} {Nanoscale nuclear magnetic resonance with a nitrogen-vacancy spin sensor},\ }\href {https://doi.org/10.1126/science.1231540} {\bibfield  {journal} {\bibinfo  {journal} {Science}\ }\textbf {\bibinfo {volume} {339}},\ \bibinfo {pages} {557} (\bibinfo {year} {2013})}\BibitemShut {NoStop}%
\bibitem [{\citenamefont {Tetienne}\ \emph {et~al.}(2012)\citenamefont {Tetienne}, \citenamefont {Rondin}, \citenamefont {Spinicelli}, \citenamefont {Chipaux}, \citenamefont {Debuisschert}, \citenamefont {Roch},\ and\ \citenamefont {Jacques}}]{tetienne12}%
  \BibitemOpen
  \bibfield  {author} {\bibinfo {author} {\bibfnamefont {J.}~\bibnamefont {Tetienne}}, \bibinfo {author} {\bibfnamefont {L.}~\bibnamefont {Rondin}}, \bibinfo {author} {\bibfnamefont {P.}~\bibnamefont {Spinicelli}}, \bibinfo {author} {\bibfnamefont {M.}~\bibnamefont {Chipaux}}, \bibinfo {author} {\bibfnamefont {T.}~\bibnamefont {Debuisschert}}, \bibinfo {author} {\bibfnamefont {J.}~\bibnamefont {Roch}},\ and\ \bibinfo {author} {\bibfnamefont {V.}~\bibnamefont {Jacques}},\ }\bibfield  {title} {\bibinfo {title} {Magnetic-field-dependent photodynamics of single nv defects in diamond: an application to qualitative all-optical magnetic imaging},\ }\href@noop {} {\bibfield  {journal} {\bibinfo  {journal} {New Journal of Physics}\ }\textbf {\bibinfo {volume} {14}},\ \bibinfo {pages} {103033} (\bibinfo {year} {2012})}\BibitemShut {NoStop}%
\bibitem [{\citenamefont {Ziem}\ \emph {et~al.}(2013)\citenamefont {Ziem}, \citenamefont {Gotz}, \citenamefont {Zappe}, \citenamefont {Steinert},\ and\ \citenamefont {Wrachtrup}}]{Ziem13}%
  \BibitemOpen
  \bibfield  {author} {\bibinfo {author} {\bibfnamefont {F.~C.}\ \bibnamefont {Ziem}}, \bibinfo {author} {\bibfnamefont {N.~S.}\ \bibnamefont {Gotz}}, \bibinfo {author} {\bibfnamefont {A.}~\bibnamefont {Zappe}}, \bibinfo {author} {\bibfnamefont {S.}~\bibnamefont {Steinert}},\ and\ \bibinfo {author} {\bibfnamefont {J.}~\bibnamefont {Wrachtrup}},\ }\bibfield  {title} {\bibinfo {title} {Highly sensitive detection of physiological spins in a microfluidic device},\ }\href@noop {} {\bibfield  {journal} {\bibinfo  {journal} {Nano letters}\ }\textbf {\bibinfo {volume} {13}},\ \bibinfo {pages} {4093} (\bibinfo {year} {2013})}\BibitemShut {NoStop}%
\bibitem [{\citenamefont {Smits}\ \emph {et~al.}(2019)\citenamefont {Smits}, \citenamefont {Damron}, \citenamefont {Kehayias}, \citenamefont {McDowell}, \citenamefont {Mosavian}, \citenamefont {Fescenko}, \citenamefont {Ristoff}, \citenamefont {Laraoui}, \citenamefont {Jarmola},\ and\ \citenamefont {Acosta}}]{Smits19}%
  \BibitemOpen
  \bibfield  {author} {\bibinfo {author} {\bibfnamefont {J.}~\bibnamefont {Smits}}, \bibinfo {author} {\bibfnamefont {J.~T.}\ \bibnamefont {Damron}}, \bibinfo {author} {\bibfnamefont {P.}~\bibnamefont {Kehayias}}, \bibinfo {author} {\bibfnamefont {A.~F.}\ \bibnamefont {McDowell}}, \bibinfo {author} {\bibfnamefont {N.}~\bibnamefont {Mosavian}}, \bibinfo {author} {\bibfnamefont {I.}~\bibnamefont {Fescenko}}, \bibinfo {author} {\bibfnamefont {N.}~\bibnamefont {Ristoff}}, \bibinfo {author} {\bibfnamefont {A.}~\bibnamefont {Laraoui}}, \bibinfo {author} {\bibfnamefont {A.}~\bibnamefont {Jarmola}},\ and\ \bibinfo {author} {\bibfnamefont {V.~M.}\ \bibnamefont {Acosta}},\ }\bibfield  {title} {\bibinfo {title} {Two-dimensional nuclear magnetic resonance spectroscopy with a microfluidic diamond quantum sensor},\ }\href@noop {} {\bibfield  {journal} {\bibinfo  {journal} {Science advances}\ }\textbf {\bibinfo {volume} {5}},\ \bibinfo {pages} {eaaw7895} (\bibinfo {year} {2019})}\BibitemShut {NoStop}%
\bibitem [{\citenamefont {Bharadwaj}\ \emph {et~al.}(2019)\citenamefont {Bharadwaj}, \citenamefont {Jedrkiewicz}, \citenamefont {Hadden}, \citenamefont {Sotillo}, \citenamefont {Vazquez}, \citenamefont {Dentella}, \citenamefont {Fernandez}, \citenamefont {Chiappini}, \citenamefont {Giakoumaki}, \citenamefont {Le~Phu} \emph {et~al.}}]{BharadwaJ19}%
  \BibitemOpen
  \bibfield  {author} {\bibinfo {author} {\bibfnamefont {V.}~\bibnamefont {Bharadwaj}}, \bibinfo {author} {\bibfnamefont {O.}~\bibnamefont {Jedrkiewicz}}, \bibinfo {author} {\bibfnamefont {J.}~\bibnamefont {Hadden}}, \bibinfo {author} {\bibfnamefont {B.}~\bibnamefont {Sotillo}}, \bibinfo {author} {\bibfnamefont {M.~R.}\ \bibnamefont {Vazquez}}, \bibinfo {author} {\bibfnamefont {P.}~\bibnamefont {Dentella}}, \bibinfo {author} {\bibfnamefont {T.~T.}\ \bibnamefont {Fernandez}}, \bibinfo {author} {\bibfnamefont {A.}~\bibnamefont {Chiappini}}, \bibinfo {author} {\bibfnamefont {A.~N.}\ \bibnamefont {Giakoumaki}}, \bibinfo {author} {\bibfnamefont {T.}~\bibnamefont {Le~Phu}}, \emph {et~al.},\ }\bibfield  {title} {\bibinfo {title} {Femtosecond laser written photonic and microfluidic circuits in diamond},\ }\href@noop {} {\bibfield  {journal} {\bibinfo  {journal} {Journal of Physics: Photonics}\ }\textbf {\bibinfo {volume} {1}},\ \bibinfo {pages} {022001} (\bibinfo {year} {2019})}\BibitemShut {NoStop}%
\bibitem [{\citenamefont {Allert}\ \emph {et~al.}(2022)\citenamefont {Allert}, \citenamefont {Bruckmaier}, \citenamefont {Neuling}, \citenamefont {Freire-Moschovitis}, \citenamefont {Liu}, \citenamefont {Schrepel}, \citenamefont {Sch{\"a}tzle}, \citenamefont {Knittel}, \citenamefont {Hermans},\ and\ \citenamefont {Bucher}}]{Allert22}%
  \BibitemOpen
  \bibfield  {author} {\bibinfo {author} {\bibfnamefont {R.~D.}\ \bibnamefont {Allert}}, \bibinfo {author} {\bibfnamefont {F.}~\bibnamefont {Bruckmaier}}, \bibinfo {author} {\bibfnamefont {N.~R.}\ \bibnamefont {Neuling}}, \bibinfo {author} {\bibfnamefont {F.~A.}\ \bibnamefont {Freire-Moschovitis}}, \bibinfo {author} {\bibfnamefont {K.~S.}\ \bibnamefont {Liu}}, \bibinfo {author} {\bibfnamefont {C.}~\bibnamefont {Schrepel}}, \bibinfo {author} {\bibfnamefont {P.}~\bibnamefont {Sch{\"a}tzle}}, \bibinfo {author} {\bibfnamefont {P.}~\bibnamefont {Knittel}}, \bibinfo {author} {\bibfnamefont {M.}~\bibnamefont {Hermans}},\ and\ \bibinfo {author} {\bibfnamefont {D.~B.}\ \bibnamefont {Bucher}},\ }\bibfield  {title} {\bibinfo {title} {Microfluidic quantum sensing platform for lab-on-a-chip applications},\ }\href@noop {} {\bibfield  {journal} {\bibinfo  {journal} {Lab on a Chip}\ }\textbf {\bibinfo {volume} {22}},\ \bibinfo {pages} {4831} (\bibinfo {year} {2022})}\BibitemShut {NoStop}%
\bibitem [{\citenamefont {Schrand}\ \emph {et~al.}(2007)\citenamefont {Schrand}, \citenamefont {Huang}, \citenamefont {Carlson}, \citenamefont {Schlager}, \citenamefont {{\=O}sawa}, \citenamefont {Hussain},\ and\ \citenamefont {Dai}}]{Schrand07}%
  \BibitemOpen
  \bibfield  {author} {\bibinfo {author} {\bibfnamefont {A.~M.}\ \bibnamefont {Schrand}}, \bibinfo {author} {\bibfnamefont {H.}~\bibnamefont {Huang}}, \bibinfo {author} {\bibfnamefont {C.}~\bibnamefont {Carlson}}, \bibinfo {author} {\bibfnamefont {J.~J.}\ \bibnamefont {Schlager}}, \bibinfo {author} {\bibfnamefont {E.}~\bibnamefont {{\=O}sawa}}, \bibinfo {author} {\bibfnamefont {S.~M.}\ \bibnamefont {Hussain}},\ and\ \bibinfo {author} {\bibfnamefont {L.}~\bibnamefont {Dai}},\ }\bibfield  {title} {\bibinfo {title} {Are diamond nanoparticles cytotoxic?},\ }\href@noop {} {\bibfield  {journal} {\bibinfo  {journal} {The journal of physical chemistry B}\ }\textbf {\bibinfo {volume} {111}},\ \bibinfo {pages} {2} (\bibinfo {year} {2007})}\BibitemShut {NoStop}%
\bibitem [{\citenamefont {Shenderova}\ \emph {et~al.}(2019)\citenamefont {Shenderova}, \citenamefont {Shames}, \citenamefont {Nunn}, \citenamefont {Torelli}, \citenamefont {Vlasov},\ and\ \citenamefont {Zaitsev}}]{Shenderova19}%
  \BibitemOpen
  \bibfield  {author} {\bibinfo {author} {\bibfnamefont {O.~A.}\ \bibnamefont {Shenderova}}, \bibinfo {author} {\bibfnamefont {A.~I.}\ \bibnamefont {Shames}}, \bibinfo {author} {\bibfnamefont {N.~A.}\ \bibnamefont {Nunn}}, \bibinfo {author} {\bibfnamefont {M.~D.}\ \bibnamefont {Torelli}}, \bibinfo {author} {\bibfnamefont {I.}~\bibnamefont {Vlasov}},\ and\ \bibinfo {author} {\bibfnamefont {A.}~\bibnamefont {Zaitsev}},\ }\bibfield  {title} {\bibinfo {title} {Synthesis, properties, and applications of fluorescent diamond particles},\ }\href@noop {} {\bibfield  {journal} {\bibinfo  {journal} {Journal of Vacuum Science \& Technology B, Nanotechnology and Microelectronics: Materials, Processing, Measurement, and Phenomena}\ }\textbf {\bibinfo {volume} {37}},\ \bibinfo {pages} {030802} (\bibinfo {year} {2019})}\BibitemShut {NoStop}%
\bibitem [{\citenamefont {Chang}\ \emph {et~al.}(2010)\citenamefont {Chang}, \citenamefont {Hwang}, \citenamefont {Ho}, \citenamefont {Lin}, \citenamefont {Hwu},\ and\ \citenamefont {Horng}}]{Chang10}%
  \BibitemOpen
  \bibfield  {author} {\bibinfo {author} {\bibfnamefont {I.~P.}\ \bibnamefont {Chang}}, \bibinfo {author} {\bibfnamefont {K.~C.}\ \bibnamefont {Hwang}}, \bibinfo {author} {\bibfnamefont {J.-a.~A.}\ \bibnamefont {Ho}}, \bibinfo {author} {\bibfnamefont {C.-C.}\ \bibnamefont {Lin}}, \bibinfo {author} {\bibfnamefont {R.~J.-R.}\ \bibnamefont {Hwu}},\ and\ \bibinfo {author} {\bibfnamefont {J.-C.}\ \bibnamefont {Horng}},\ }\bibfield  {title} {\bibinfo {title} {Facile surface functionalization of nanodiamonds},\ }\href@noop {} {\bibfield  {journal} {\bibinfo  {journal} {Langmuir}\ }\textbf {\bibinfo {volume} {26}},\ \bibinfo {pages} {3685} (\bibinfo {year} {2010})}\BibitemShut {NoStop}%
\bibitem [{\citenamefont {Liu}\ \emph {et~al.}(2007)\citenamefont {Liu}, \citenamefont {Cheng}, \citenamefont {Chang},\ and\ \citenamefont {Chao}}]{Liu07}%
  \BibitemOpen
  \bibfield  {author} {\bibinfo {author} {\bibfnamefont {K.-K.}\ \bibnamefont {Liu}}, \bibinfo {author} {\bibfnamefont {C.-L.}\ \bibnamefont {Cheng}}, \bibinfo {author} {\bibfnamefont {C.-C.}\ \bibnamefont {Chang}},\ and\ \bibinfo {author} {\bibfnamefont {J.-I.}\ \bibnamefont {Chao}},\ }\bibfield  {title} {\bibinfo {title} {Biocompatible and detectable carboxylated nanodiamond on human cell},\ }\href@noop {} {\bibfield  {journal} {\bibinfo  {journal} {Nanotechnology}\ }\textbf {\bibinfo {volume} {18}},\ \bibinfo {pages} {325102} (\bibinfo {year} {2007})}\BibitemShut {NoStop}%
\bibitem [{\citenamefont {Zhang}\ \emph {et~al.}(2015)\citenamefont {Zhang}, \citenamefont {Neumann}, \citenamefont {Lindlau}, \citenamefont {Wu}, \citenamefont {Pramanik}, \citenamefont {Naydenov}, \citenamefont {Jelezko}, \citenamefont {Schuder}, \citenamefont {Huber}, \citenamefont {Huber} \emph {et~al.}}]{Zhang15}%
  \BibitemOpen
  \bibfield  {author} {\bibinfo {author} {\bibfnamefont {T.}~\bibnamefont {Zhang}}, \bibinfo {author} {\bibfnamefont {A.}~\bibnamefont {Neumann}}, \bibinfo {author} {\bibfnamefont {J.}~\bibnamefont {Lindlau}}, \bibinfo {author} {\bibfnamefont {Y.}~\bibnamefont {Wu}}, \bibinfo {author} {\bibfnamefont {G.}~\bibnamefont {Pramanik}}, \bibinfo {author} {\bibfnamefont {B.}~\bibnamefont {Naydenov}}, \bibinfo {author} {\bibfnamefont {F.}~\bibnamefont {Jelezko}}, \bibinfo {author} {\bibfnamefont {F.}~\bibnamefont {Schuder}}, \bibinfo {author} {\bibfnamefont {S.}~\bibnamefont {Huber}}, \bibinfo {author} {\bibfnamefont {M.}~\bibnamefont {Huber}}, \emph {et~al.},\ }\bibfield  {title} {\bibinfo {title} {Dna-based self-assembly of fluorescent nanodiamonds},\ }\href@noop {} {\bibfield  {journal} {\bibinfo  {journal} {Journal of the American Chemical Society}\ }\textbf {\bibinfo {volume} {137}},\ \bibinfo {pages} {9776} (\bibinfo {year} {2015})}\BibitemShut {NoStop}%
\bibitem [{\citenamefont {Reina}\ \emph {et~al.}(2019)\citenamefont {Reina}, \citenamefont {Zhao}, \citenamefont {Bianco},\ and\ \citenamefont {Komatsu}}]{Reina19}%
  \BibitemOpen
  \bibfield  {author} {\bibinfo {author} {\bibfnamefont {G.}~\bibnamefont {Reina}}, \bibinfo {author} {\bibfnamefont {L.}~\bibnamefont {Zhao}}, \bibinfo {author} {\bibfnamefont {A.}~\bibnamefont {Bianco}},\ and\ \bibinfo {author} {\bibfnamefont {N.}~\bibnamefont {Komatsu}},\ }\bibfield  {title} {\bibinfo {title} {Chemical functionalization of nanodiamonds: Opportunities and challenges ahead},\ }\href@noop {} {\bibfield  {journal} {\bibinfo  {journal} {Angewandte Chemie International Edition}\ }\textbf {\bibinfo {volume} {58}},\ \bibinfo {pages} {17918} (\bibinfo {year} {2019})}\BibitemShut {NoStop}%
\bibitem [{\citenamefont {Zhang}\ \emph {et~al.}(2019)\citenamefont {Zhang}, \citenamefont {Tamijani}, \citenamefont {Taylor}, \citenamefont {Zhi}, \citenamefont {Haynes}, \citenamefont {Mason},\ and\ \citenamefont {Hamers}}]{Zhang19}%
  \BibitemOpen
  \bibfield  {author} {\bibinfo {author} {\bibfnamefont {Y.}~\bibnamefont {Zhang}}, \bibinfo {author} {\bibfnamefont {A.~A.}\ \bibnamefont {Tamijani}}, \bibinfo {author} {\bibfnamefont {M.~E.}\ \bibnamefont {Taylor}}, \bibinfo {author} {\bibfnamefont {B.}~\bibnamefont {Zhi}}, \bibinfo {author} {\bibfnamefont {C.~L.}\ \bibnamefont {Haynes}}, \bibinfo {author} {\bibfnamefont {S.~E.}\ \bibnamefont {Mason}},\ and\ \bibinfo {author} {\bibfnamefont {R.~J.}\ \bibnamefont {Hamers}},\ }\bibfield  {title} {\bibinfo {title} {Molecular surface functionalization of carbon materials via radical-induced grafting of terminal alkenes},\ }\href@noop {} {\bibfield  {journal} {\bibinfo  {journal} {Journal of the American Chemical Society}\ }\textbf {\bibinfo {volume} {141}},\ \bibinfo {pages} {8277} (\bibinfo {year} {2019})}\BibitemShut {NoStop}%
\bibitem [{\citenamefont {Jones}\ \emph {et~al.}(2022)\citenamefont {Jones}, \citenamefont {Niemuth}, \citenamefont {Zhang}, \citenamefont {Protter}, \citenamefont {Kinsley}, \citenamefont {Klaper},\ and\ \citenamefont {Hamers}}]{Jones22}%
  \BibitemOpen
  \bibfield  {author} {\bibinfo {author} {\bibfnamefont {Z.~R.}\ \bibnamefont {Jones}}, \bibinfo {author} {\bibfnamefont {N.~J.}\ \bibnamefont {Niemuth}}, \bibinfo {author} {\bibfnamefont {Y.}~\bibnamefont {Zhang}}, \bibinfo {author} {\bibfnamefont {C.~R.}\ \bibnamefont {Protter}}, \bibinfo {author} {\bibfnamefont {P.~C.}\ \bibnamefont {Kinsley}}, \bibinfo {author} {\bibfnamefont {R.~D.}\ \bibnamefont {Klaper}},\ and\ \bibinfo {author} {\bibfnamefont {R.~J.}\ \bibnamefont {Hamers}},\ }\bibfield  {title} {\bibinfo {title} {Use of magnetic modulation of nitrogen-vacancy center fluorescence in nanodiamonds for quantitative analysis of nanoparticles in organisms},\ }\href@noop {} {\bibfield  {journal} {\bibinfo  {journal} {ACS Measurement Science Au}\ }\textbf {\bibinfo {volume} {2}},\ \bibinfo {pages} {351} (\bibinfo {year} {2022})}\BibitemShut {NoStop}%
\bibitem [{\citenamefont {Zheng}\ \emph {et~al.}(2023)\citenamefont {Zheng}, \citenamefont {Shi}, \citenamefont {Jiang}, \citenamefont {Mao}, \citenamefont {Montes}, \citenamefont {Kateris}, \citenamefont {Reimer}, \citenamefont {Wang},\ and\ \citenamefont {Zheng}}]{Zheng23}%
  \BibitemOpen
  \bibfield  {author} {\bibinfo {author} {\bibfnamefont {Q.}~\bibnamefont {Zheng}}, \bibinfo {author} {\bibfnamefont {X.}~\bibnamefont {Shi}}, \bibinfo {author} {\bibfnamefont {J.}~\bibnamefont {Jiang}}, \bibinfo {author} {\bibfnamefont {H.}~\bibnamefont {Mao}}, \bibinfo {author} {\bibfnamefont {N.}~\bibnamefont {Montes}}, \bibinfo {author} {\bibfnamefont {N.}~\bibnamefont {Kateris}}, \bibinfo {author} {\bibfnamefont {J.~A.}\ \bibnamefont {Reimer}}, \bibinfo {author} {\bibfnamefont {H.}~\bibnamefont {Wang}},\ and\ \bibinfo {author} {\bibfnamefont {H.}~\bibnamefont {Zheng}},\ }\bibfield  {title} {\bibinfo {title} {Unveiling the complexity of nanodiamond structures},\ }\href@noop {} {\bibfield  {journal} {\bibinfo  {journal} {Proceedings of the National Academy of Sciences}\ }\textbf {\bibinfo {volume} {120}},\ \bibinfo {pages} {e2301981120} (\bibinfo {year} {2023})}\BibitemShut {NoStop}%
\bibitem [{\citenamefont {Chowdhury}\ \emph {et~al.}(2019)\citenamefont {Chowdhury}, \citenamefont {Chen}, \citenamefont {Eckert}, \citenamefont {Ren}, \citenamefont {Wu}, \citenamefont {Repina},\ and\ \citenamefont {Waller}}]{Chowdhury19}%
  \BibitemOpen
  \bibfield  {author} {\bibinfo {author} {\bibfnamefont {S.}~\bibnamefont {Chowdhury}}, \bibinfo {author} {\bibfnamefont {M.}~\bibnamefont {Chen}}, \bibinfo {author} {\bibfnamefont {R.}~\bibnamefont {Eckert}}, \bibinfo {author} {\bibfnamefont {D.}~\bibnamefont {Ren}}, \bibinfo {author} {\bibfnamefont {F.}~\bibnamefont {Wu}}, \bibinfo {author} {\bibfnamefont {N.}~\bibnamefont {Repina}},\ and\ \bibinfo {author} {\bibfnamefont {L.}~\bibnamefont {Waller}},\ }\bibfield  {title} {\bibinfo {title} {High-resolution 3d refractive index microscopy of multiple-scattering samples from intensity images},\ }\href@noop {} {\bibfield  {journal} {\bibinfo  {journal} {Optica}\ }\textbf {\bibinfo {volume} {6}},\ \bibinfo {pages} {1211} (\bibinfo {year} {2019})}\BibitemShut {NoStop}%
\bibitem [{\citenamefont {Rodgers}\ \emph {et~al.}(2021)\citenamefont {Rodgers}, \citenamefont {Hughes}, \citenamefont {Xie}, \citenamefont {Maurer}, \citenamefont {Kolkowitz}, \citenamefont {Bleszynski~Jayich},\ and\ \citenamefont {de~Leon}}]{Rodgers21}%
  \BibitemOpen
  \bibfield  {author} {\bibinfo {author} {\bibfnamefont {L.~V.~H.}\ \bibnamefont {Rodgers}}, \bibinfo {author} {\bibfnamefont {L.~B.}\ \bibnamefont {Hughes}}, \bibinfo {author} {\bibfnamefont {M.}~\bibnamefont {Xie}}, \bibinfo {author} {\bibfnamefont {P.~C.}\ \bibnamefont {Maurer}}, \bibinfo {author} {\bibfnamefont {S.}~\bibnamefont {Kolkowitz}}, \bibinfo {author} {\bibfnamefont {A.~C.}\ \bibnamefont {Bleszynski~Jayich}},\ and\ \bibinfo {author} {\bibfnamefont {N.~P.}\ \bibnamefont {de~Leon}},\ }\bibfield  {title} {\bibinfo {title} {Materials challenges for quantum technologies based on color centers in diamond},\ }\href@noop {} {\bibfield  {journal} {\bibinfo  {journal} {MRS Bulletin}\ ,\ \bibinfo {pages} {623–633}} (\bibinfo {year} {2021})}\BibitemShut {NoStop}%
\bibitem [{\citenamefont {Huebner}\ \emph {et~al.}(2008)\citenamefont {Huebner}, \citenamefont {Sharma}, \citenamefont {Srisa-Art}, \citenamefont {Hollfelder}, \citenamefont {Edel},\ and\ \citenamefont {Demello}}]{Huebner08}%
  \BibitemOpen
  \bibfield  {author} {\bibinfo {author} {\bibfnamefont {A.}~\bibnamefont {Huebner}}, \bibinfo {author} {\bibfnamefont {S.}~\bibnamefont {Sharma}}, \bibinfo {author} {\bibfnamefont {M.}~\bibnamefont {Srisa-Art}}, \bibinfo {author} {\bibfnamefont {F.}~\bibnamefont {Hollfelder}}, \bibinfo {author} {\bibfnamefont {J.~B.}\ \bibnamefont {Edel}},\ and\ \bibinfo {author} {\bibfnamefont {A.~J.}\ \bibnamefont {Demello}},\ }\bibfield  {title} {\bibinfo {title} {Microdroplets: a sea of applications?},\ }\href@noop {} {\bibfield  {journal} {\bibinfo  {journal} {Lab on a Chip}\ }\textbf {\bibinfo {volume} {8}},\ \bibinfo {pages} {1244} (\bibinfo {year} {2008})}\BibitemShut {NoStop}%
\bibitem [{\citenamefont {Theberge}\ \emph {et~al.}(2010)\citenamefont {Theberge}, \citenamefont {Courtois}, \citenamefont {Schaerli}, \citenamefont {Fischlechner}, \citenamefont {Abell}, \citenamefont {Hollfelder},\ and\ \citenamefont {Huck}}]{Theberge10}%
  \BibitemOpen
  \bibfield  {author} {\bibinfo {author} {\bibfnamefont {A.~B.}\ \bibnamefont {Theberge}}, \bibinfo {author} {\bibfnamefont {F.}~\bibnamefont {Courtois}}, \bibinfo {author} {\bibfnamefont {Y.}~\bibnamefont {Schaerli}}, \bibinfo {author} {\bibfnamefont {M.}~\bibnamefont {Fischlechner}}, \bibinfo {author} {\bibfnamefont {C.}~\bibnamefont {Abell}}, \bibinfo {author} {\bibfnamefont {F.}~\bibnamefont {Hollfelder}},\ and\ \bibinfo {author} {\bibfnamefont {W.~T.}\ \bibnamefont {Huck}},\ }\bibfield  {title} {\bibinfo {title} {Microdroplets in microfluidics: an evolving platform for discoveries in chemistry and biology},\ }\href@noop {} {\bibfield  {journal} {\bibinfo  {journal} {Angewandte Chemie International Edition}\ }\textbf {\bibinfo {volume} {49}},\ \bibinfo {pages} {5846} (\bibinfo {year} {2010})}\BibitemShut {NoStop}%
\bibitem [{\citenamefont {Fang}\ \emph {et~al.}(2013)\citenamefont {Fang}, \citenamefont {Acosta}, \citenamefont {Santori}, \citenamefont {Huang}, \citenamefont {Itoh}, \citenamefont {Watanabe}, \citenamefont {Shikata},\ and\ \citenamefont {Beausoleil}}]{Fang13}%
  \BibitemOpen
  \bibfield  {author} {\bibinfo {author} {\bibfnamefont {K.}~\bibnamefont {Fang}}, \bibinfo {author} {\bibfnamefont {V.~M.}\ \bibnamefont {Acosta}}, \bibinfo {author} {\bibfnamefont {C.}~\bibnamefont {Santori}}, \bibinfo {author} {\bibfnamefont {Z.}~\bibnamefont {Huang}}, \bibinfo {author} {\bibfnamefont {K.~M.}\ \bibnamefont {Itoh}}, \bibinfo {author} {\bibfnamefont {H.}~\bibnamefont {Watanabe}}, \bibinfo {author} {\bibfnamefont {S.}~\bibnamefont {Shikata}},\ and\ \bibinfo {author} {\bibfnamefont {R.~G.}\ \bibnamefont {Beausoleil}},\ }\bibfield  {title} {\bibinfo {title} {High-sensitivity magnetometry based on quantum beats in diamond nitrogen-vacancy centers},\ }\href@noop {} {\bibfield  {journal} {\bibinfo  {journal} {Physical review letters}\ }\textbf {\bibinfo {volume} {110}},\ \bibinfo {pages} {130802} (\bibinfo {year} {2013})}\BibitemShut {NoStop}%
\bibitem [{\citenamefont {Wolf}\ \emph {et~al.}(2015{\natexlab{a}})\citenamefont {Wolf}, \citenamefont {Neumann}, \citenamefont {Nakamura}, \citenamefont {Sumiya}, \citenamefont {Ohshima}, \citenamefont {Isoya},\ and\ \citenamefont {Wrachtrup}}]{Wolf15}%
  \BibitemOpen
  \bibfield  {author} {\bibinfo {author} {\bibfnamefont {T.}~\bibnamefont {Wolf}}, \bibinfo {author} {\bibfnamefont {P.}~\bibnamefont {Neumann}}, \bibinfo {author} {\bibfnamefont {K.}~\bibnamefont {Nakamura}}, \bibinfo {author} {\bibfnamefont {H.}~\bibnamefont {Sumiya}}, \bibinfo {author} {\bibfnamefont {T.}~\bibnamefont {Ohshima}}, \bibinfo {author} {\bibfnamefont {J.}~\bibnamefont {Isoya}},\ and\ \bibinfo {author} {\bibfnamefont {J.}~\bibnamefont {Wrachtrup}},\ }\bibfield  {title} {\bibinfo {title} {Subpicotesla diamond magnetometry},\ }\href@noop {} {\bibfield  {journal} {\bibinfo  {journal} {Physical Review X}\ }\textbf {\bibinfo {volume} {5}},\ \bibinfo {pages} {041001} (\bibinfo {year} {2015}{\natexlab{a}})}\BibitemShut {NoStop}%
\bibitem [{\citenamefont {Chiu}\ and\ \citenamefont {Lorenz}(2009)}]{Chiu09}%
  \BibitemOpen
  \bibfield  {author} {\bibinfo {author} {\bibfnamefont {D.~T.}\ \bibnamefont {Chiu}}\ and\ \bibinfo {author} {\bibfnamefont {R.~M.}\ \bibnamefont {Lorenz}},\ }\bibfield  {title} {\bibinfo {title} {Chemistry and biology in femtoliter and picoliter volume droplets},\ }\href@noop {} {\bibfield  {journal} {\bibinfo  {journal} {Accounts of chemical research}\ }\textbf {\bibinfo {volume} {42}},\ \bibinfo {pages} {649} (\bibinfo {year} {2009})}\BibitemShut {NoStop}%
\bibitem [{\citenamefont {Grigoriev}\ \emph {et~al.}(2006)\citenamefont {Grigoriev}, \citenamefont {Schatz},\ and\ \citenamefont {Sharma}}]{Grigoriev06}%
  \BibitemOpen
  \bibfield  {author} {\bibinfo {author} {\bibfnamefont {R.~O.}\ \bibnamefont {Grigoriev}}, \bibinfo {author} {\bibfnamefont {M.~F.}\ \bibnamefont {Schatz}},\ and\ \bibinfo {author} {\bibfnamefont {V.}~\bibnamefont {Sharma}},\ }\bibfield  {title} {\bibinfo {title} {Chaotic mixing in microdroplets},\ }\href@noop {} {\bibfield  {journal} {\bibinfo  {journal} {Lab on a Chip}\ }\textbf {\bibinfo {volume} {6}},\ \bibinfo {pages} {1369} (\bibinfo {year} {2006})}\BibitemShut {NoStop}%
\bibitem [{\citenamefont {Srisa-Art}\ \emph {et~al.}(2008)\citenamefont {Srisa-Art}, \citenamefont {DeMello},\ and\ \citenamefont {Edel}}]{Srisa08}%
  \BibitemOpen
  \bibfield  {author} {\bibinfo {author} {\bibfnamefont {M.}~\bibnamefont {Srisa-Art}}, \bibinfo {author} {\bibfnamefont {A.~J.}\ \bibnamefont {DeMello}},\ and\ \bibinfo {author} {\bibfnamefont {J.~B.}\ \bibnamefont {Edel}},\ }\bibfield  {title} {\bibinfo {title} {Fluorescence lifetime imaging of mixing dynamics in continuous-flow microdroplet reactors},\ }\href@noop {} {\bibfield  {journal} {\bibinfo  {journal} {Physical review letters}\ }\textbf {\bibinfo {volume} {101}},\ \bibinfo {pages} {014502} (\bibinfo {year} {2008})}\BibitemShut {NoStop}%
\bibitem [{SI()}]{SI}%
  \BibitemOpen
  \href@noop {} {}\bibinfo {note} {See Supplemental Material.}\BibitemShut {Stop}%
\bibitem [{\citenamefont {Ershov}\ \emph {et~al.}(2022)\citenamefont {Ershov}, \citenamefont {Minh-Son}, \citenamefont {W.}, \citenamefont {Rigaud}, \citenamefont {Le~Blanc}, \citenamefont {Charles-Orszag}, \citenamefont {Conway}, \citenamefont {Laine}, \citenamefont {Roy}, \citenamefont {Bonazzi} \emph {et~al.}}]{TrackMate}%
  \BibitemOpen
  \bibfield  {author} {\bibinfo {author} {\bibfnamefont {D.}~\bibnamefont {Ershov}}, \bibinfo {author} {\bibfnamefont {P.}~\bibnamefont {Minh-Son}}, \bibinfo {author} {\bibfnamefont {P.~J.}\ \bibnamefont {W.}}, \bibinfo {author} {\bibfnamefont {S.~U.}\ \bibnamefont {Rigaud}}, \bibinfo {author} {\bibfnamefont {L.}~\bibnamefont {Le~Blanc}}, \bibinfo {author} {\bibfnamefont {A.}~\bibnamefont {Charles-Orszag}}, \bibinfo {author} {\bibfnamefont {J.~R.~W.}\ \bibnamefont {Conway}}, \bibinfo {author} {\bibfnamefont {R.~F.}\ \bibnamefont {Laine}}, \bibinfo {author} {\bibfnamefont {N.~H.}\ \bibnamefont {Roy}}, \bibinfo {author} {\bibfnamefont {D.}~\bibnamefont {Bonazzi}}, \emph {et~al.},\ }\bibfield  {title} {\bibinfo {title} {Trackmate 7: integrating state-of-the-art segmentation algorithms into tracking pipelines},\ }\href@noop {} {\bibfield  {journal} {\bibinfo  {journal} {Nature Methods}\ }\textbf {\bibinfo {volume} {19}},\ \bibinfo {pages} {829} (\bibinfo {year} {2022})}\BibitemShut {NoStop}%
\bibitem [{\citenamefont {Tinevez}\ \emph {et~al.}(2017)\citenamefont {Tinevez}, \citenamefont {Perry}, \citenamefont {Schindelin}, \citenamefont {Hoopes}, \citenamefont {Reynolds}, \citenamefont {Laplantine}, \citenamefont {Bednarek}, \citenamefont {Shorte},\ and\ \citenamefont {Eliceiri}}]{TrackMate2}%
  \BibitemOpen
  \bibfield  {author} {\bibinfo {author} {\bibfnamefont {J.-Y.}\ \bibnamefont {Tinevez}}, \bibinfo {author} {\bibfnamefont {N.}~\bibnamefont {Perry}}, \bibinfo {author} {\bibfnamefont {J.}~\bibnamefont {Schindelin}}, \bibinfo {author} {\bibfnamefont {G.~M.}\ \bibnamefont {Hoopes}}, \bibinfo {author} {\bibfnamefont {G.~D.}\ \bibnamefont {Reynolds}}, \bibinfo {author} {\bibfnamefont {E.}~\bibnamefont {Laplantine}}, \bibinfo {author} {\bibfnamefont {S.~Y.}\ \bibnamefont {Bednarek}}, \bibinfo {author} {\bibfnamefont {S.~L.}\ \bibnamefont {Shorte}},\ and\ \bibinfo {author} {\bibfnamefont {K.~W.}\ \bibnamefont {Eliceiri}},\ }\bibfield  {title} {\bibinfo {title} {Trackmate: An open and extensible platform for single-particle tracking},\ }\href@noop {} {\bibfield  {journal} {\bibinfo  {journal} {Methods}\ }\textbf {\bibinfo {volume} {115}},\ \bibinfo {pages} {80} (\bibinfo {year} {2017})}\BibitemShut {NoStop}%
\bibitem [{Tra(2024)}]{TrackingMovie}%
  \BibitemOpen
  \href {https://youtube.com/watch/B2CJ8moNFok?feature=share} {\bibinfo {title} {Movie illustrating particle tracking 100 nm diamonds in a stationary droplet}} (\bibinfo {year} {2024})\BibitemShut {NoStop}%
\bibitem [{BF_(2024)}]{BF_ChipStructures}%
  \BibitemOpen
  \href {https://www.youtube.com/watch?v=00bFIx5ihxA} {\bibinfo {title} {Brightfield movie of microfluidic device structures}} (\bibinfo {year} {2024})\BibitemShut {NoStop}%
\bibitem [{Fl_(2024)}]{Fl_ChipStructures}%
  \BibitemOpen
  \href {https://youtu.be/P0BiZ5IuXwI} {\bibinfo {title} {Fluorescence movie of microfluidic device structures}} (\bibinfo {year} {2024})\BibitemShut {NoStop}%
\bibitem [{750(2024)}]{750nmDiamondFlowing}%
  \BibitemOpen
  \href {https://youtube.com/shorts/AdvOiFWc3Gk?feature=share} {\bibinfo {title} {Fluorescence movie of 750 nm diamond in droplets}} (\bibinfo {year} {2024})\BibitemShut {NoStop}%
\bibitem [{\citenamefont {Shlesinger}\ \emph {et~al.}(1987)\citenamefont {Shlesinger}, \citenamefont {West},\ and\ \citenamefont {Klafter}}]{Shlesinger87}%
  \BibitemOpen
  \bibfield  {author} {\bibinfo {author} {\bibfnamefont {M.~F.}\ \bibnamefont {Shlesinger}}, \bibinfo {author} {\bibfnamefont {B.}~\bibnamefont {West}},\ and\ \bibinfo {author} {\bibfnamefont {J.}~\bibnamefont {Klafter}},\ }\bibfield  {title} {\bibinfo {title} {Levy dynamics of enhanced diffusion: Application to turbulence},\ }\href@noop {} {\bibfield  {journal} {\bibinfo  {journal} {Physical Review Letters}\ }\textbf {\bibinfo {volume} {58}},\ \bibinfo {pages} {1100} (\bibinfo {year} {1987})}\BibitemShut {NoStop}%
\bibitem [{\citenamefont {Hemelaar}\ \emph {et~al.}(2017)\citenamefont {Hemelaar}, \citenamefont {Nagl}, \citenamefont {Bigot}, \citenamefont {Rodriguez-Garcia}, \citenamefont {de~Vries}, \citenamefont {Chipaux},\ and\ \citenamefont {Schirhagl}}]{Hemelaar2017}%
  \BibitemOpen
  \bibfield  {author} {\bibinfo {author} {\bibfnamefont {S.~R.}\ \bibnamefont {Hemelaar}}, \bibinfo {author} {\bibfnamefont {A.}~\bibnamefont {Nagl}}, \bibinfo {author} {\bibfnamefont {F.}~\bibnamefont {Bigot}}, \bibinfo {author} {\bibfnamefont {M.~M.}\ \bibnamefont {Rodriguez-Garcia}}, \bibinfo {author} {\bibfnamefont {M.~P.}\ \bibnamefont {de~Vries}}, \bibinfo {author} {\bibfnamefont {M.}~\bibnamefont {Chipaux}},\ and\ \bibinfo {author} {\bibfnamefont {R.}~\bibnamefont {Schirhagl}},\ }\bibfield  {title} {\bibinfo {title} {The interaction of fluorescent nanodiamond probes with cellular media},\ }\href {https://doi.org/10.1007/s00604-017-2086-6} {\bibfield  {journal} {\bibinfo  {journal} {Microchimica Acta}\ }\textbf {\bibinfo {volume} {184}},\ \bibinfo {pages} {1001} (\bibinfo {year} {2017})}\BibitemShut {NoStop}%
\bibitem [{Flo(2024)}]{FlowRates_Videos}%
  \BibitemOpen
  \href {https://youtube.com/watch/WL4YMHS5Ois?feature=sharefeature=share} {\bibinfo {title} {Fluorescence movie of various droplet flow profiles and flow rates}} (\bibinfo {year} {2024})\BibitemShut {NoStop}%
\bibitem [{\citenamefont {Fujiwara}\ \emph {et~al.}(2020)\citenamefont {Fujiwara}, \citenamefont {Dohms}, \citenamefont {Suto}, \citenamefont {Nishimura}, \citenamefont {Oshimi}, \citenamefont {Teki}, \citenamefont {Cai}, \citenamefont {Benson},\ and\ \citenamefont {Shikano}}]{Fujiwara2020realtime}%
  \BibitemOpen
  \bibfield  {author} {\bibinfo {author} {\bibfnamefont {M.}~\bibnamefont {Fujiwara}}, \bibinfo {author} {\bibfnamefont {A.}~\bibnamefont {Dohms}}, \bibinfo {author} {\bibfnamefont {K.}~\bibnamefont {Suto}}, \bibinfo {author} {\bibfnamefont {Y.}~\bibnamefont {Nishimura}}, \bibinfo {author} {\bibfnamefont {K.}~\bibnamefont {Oshimi}}, \bibinfo {author} {\bibfnamefont {Y.}~\bibnamefont {Teki}}, \bibinfo {author} {\bibfnamefont {K.}~\bibnamefont {Cai}}, \bibinfo {author} {\bibfnamefont {O.}~\bibnamefont {Benson}},\ and\ \bibinfo {author} {\bibfnamefont {Y.}~\bibnamefont {Shikano}},\ }\bibfield  {title} {\bibinfo {title} {Real-time estimation of the optically detected magnetic resonance shift in diamond quantum thermometry toward biological applications},\ }\href {https://doi.org/10.1103/PhysRevResearch.2.043415} {\bibfield  {journal} {\bibinfo  {journal} {Phys. Rev. Res.}\ }\textbf {\bibinfo {volume} {2}},\ \bibinfo {pages} {043415} (\bibinfo {year} {2020})}\BibitemShut {NoStop}%
\bibitem [{\citenamefont {Wolf}\ \emph {et~al.}(2015{\natexlab{b}})\citenamefont {Wolf}, \citenamefont {Neumann}, \citenamefont {Nakamura}, \citenamefont {Sumiya}, \citenamefont {Ohshima}, \citenamefont {Isoya},\ and\ \citenamefont {Wrachtrup}}]{wolf2015subpicotesla}%
  \BibitemOpen
  \bibfield  {author} {\bibinfo {author} {\bibfnamefont {T.}~\bibnamefont {Wolf}}, \bibinfo {author} {\bibfnamefont {P.}~\bibnamefont {Neumann}}, \bibinfo {author} {\bibfnamefont {K.}~\bibnamefont {Nakamura}}, \bibinfo {author} {\bibfnamefont {H.}~\bibnamefont {Sumiya}}, \bibinfo {author} {\bibfnamefont {T.}~\bibnamefont {Ohshima}}, \bibinfo {author} {\bibfnamefont {J.}~\bibnamefont {Isoya}},\ and\ \bibinfo {author} {\bibfnamefont {J.}~\bibnamefont {Wrachtrup}},\ }\bibfield  {title} {\bibinfo {title} {Subpicotesla diamond magnetometry},\ }\href@noop {} {\bibfield  {journal} {\bibinfo  {journal} {Physical Review X}\ }\textbf {\bibinfo {volume} {5}},\ \bibinfo {pages} {041001} (\bibinfo {year} {2015}{\natexlab{b}})}\BibitemShut {NoStop}%
\bibitem [{\citenamefont {Hart}\ \emph {et~al.}(2021)\citenamefont {Hart}, \citenamefont {Schloss}, \citenamefont {Turner}, \citenamefont {Scheidegger}, \citenamefont {Bauch},\ and\ \citenamefont {Walsworth}}]{Hart2021prapplied}%
  \BibitemOpen
  \bibfield  {author} {\bibinfo {author} {\bibfnamefont {C.~A.}\ \bibnamefont {Hart}}, \bibinfo {author} {\bibfnamefont {J.~M.}\ \bibnamefont {Schloss}}, \bibinfo {author} {\bibfnamefont {M.~J.}\ \bibnamefont {Turner}}, \bibinfo {author} {\bibfnamefont {P.~J.}\ \bibnamefont {Scheidegger}}, \bibinfo {author} {\bibfnamefont {E.}~\bibnamefont {Bauch}},\ and\ \bibinfo {author} {\bibfnamefont {R.~L.}\ \bibnamefont {Walsworth}},\ }\bibfield  {title} {\bibinfo {title} {$\mathrm{N}$-$v$--diamond magnetic microscopy using a double quantum 4-ramsey protocol},\ }\href {https://doi.org/10.1103/PhysRevApplied.15.044020} {\bibfield  {journal} {\bibinfo  {journal} {Phys. Rev. Appl.}\ }\textbf {\bibinfo {volume} {15}},\ \bibinfo {pages} {044020} (\bibinfo {year} {2021})}\BibitemShut {NoStop}%
\bibitem [{\citenamefont {Gu}\ \emph {et~al.}(2023)\citenamefont {Gu}, \citenamefont {Shanahan}, \citenamefont {Hart}, \citenamefont {Belser}, \citenamefont {Shofer}, \citenamefont {Atat{\"u}re},\ and\ \citenamefont {Knowles}}]{Gu2023}%
  \BibitemOpen
  \bibfield  {author} {\bibinfo {author} {\bibfnamefont {Q.}~\bibnamefont {Gu}}, \bibinfo {author} {\bibfnamefont {L.}~\bibnamefont {Shanahan}}, \bibinfo {author} {\bibfnamefont {J.~W.}\ \bibnamefont {Hart}}, \bibinfo {author} {\bibfnamefont {S.}~\bibnamefont {Belser}}, \bibinfo {author} {\bibfnamefont {N.}~\bibnamefont {Shofer}}, \bibinfo {author} {\bibfnamefont {M.}~\bibnamefont {Atat{\"u}re}},\ and\ \bibinfo {author} {\bibfnamefont {H.~S.}\ \bibnamefont {Knowles}},\ }\bibfield  {title} {\bibinfo {title} {Simultaneous nanorheometry and nanothermometry using intracellular diamond quantum sensors},\ }\href {https://doi.org/10.1021/acsnano.3c05285} {\bibfield  {journal} {\bibinfo  {journal} {ACS Nano}\ }\textbf {\bibinfo {volume} {17}},\ \bibinfo {pages} {20034} (\bibinfo {year} {2023})}\BibitemShut {NoStop}%
\bibitem [{\citenamefont {Singam}\ \emph {et~al.}(2016)\citenamefont {Singam}, \citenamefont {Motylewski}, \citenamefont {Monaco}, \citenamefont {Gjorgievska}, \citenamefont {Bourgeois}, \citenamefont {Nesladek}, \citenamefont {Giugliano},\ and\ \citenamefont {Goovaerts}}]{Singam16}%
  \BibitemOpen
  \bibfield  {author} {\bibinfo {author} {\bibfnamefont {S.~K.}\ \bibnamefont {Singam}}, \bibinfo {author} {\bibfnamefont {J.}~\bibnamefont {Motylewski}}, \bibinfo {author} {\bibfnamefont {A.}~\bibnamefont {Monaco}}, \bibinfo {author} {\bibfnamefont {E.}~\bibnamefont {Gjorgievska}}, \bibinfo {author} {\bibfnamefont {E.}~\bibnamefont {Bourgeois}}, \bibinfo {author} {\bibfnamefont {M.}~\bibnamefont {Nesladek}}, \bibinfo {author} {\bibfnamefont {M.}~\bibnamefont {Giugliano}},\ and\ \bibinfo {author} {\bibfnamefont {E.}~\bibnamefont {Goovaerts}},\ }\bibfield  {title} {\bibinfo {title} {Contrast induced by a static magnetic field for improved detection in nanodiamond fluorescence microscopy},\ }\href@noop {} {\bibfield  {journal} {\bibinfo  {journal} {Physical Review Applied}\ }\textbf {\bibinfo {volume} {6}} (\bibinfo {year} {2016})}\BibitemShut {NoStop}%
\bibitem [{\citenamefont {Jones}\ \emph {et~al.}(2020)\citenamefont {Jones}, \citenamefont {Niemuth}, \citenamefont {Robinson}, \citenamefont {Shenderova}, \citenamefont {Klaper},\ and\ \citenamefont {Hamers}}]{Jones21}%
  \BibitemOpen
  \bibfield  {author} {\bibinfo {author} {\bibfnamefont {Z.~R.}\ \bibnamefont {Jones}}, \bibinfo {author} {\bibfnamefont {N.~J.}\ \bibnamefont {Niemuth}}, \bibinfo {author} {\bibfnamefont {M.~E.}\ \bibnamefont {Robinson}}, \bibinfo {author} {\bibfnamefont {O.~A.}\ \bibnamefont {Shenderova}}, \bibinfo {author} {\bibfnamefont {R.~D.}\ \bibnamefont {Klaper}},\ and\ \bibinfo {author} {\bibfnamefont {R.~J.}\ \bibnamefont {Hamers}},\ }\bibfield  {title} {\bibinfo {title} {Selective imaging of diamond nanoparticles within complex matrices using magnetically induced fluorescence contrast},\ }\href@noop {} {\bibfield  {journal} {\bibinfo  {journal} {Environmental Science: Nano}\ }\textbf {\bibinfo {volume} {7}},\ \bibinfo {pages} {525} (\bibinfo {year} {2020})}\BibitemShut {NoStop}%
\bibitem [{\citenamefont {Torelli}\ \emph {et~al.}(2020)\citenamefont {Torelli}, \citenamefont {Nunn}, \citenamefont {Jones}, \citenamefont {Vedelaar}, \citenamefont {Padamati}, \citenamefont {Schirhagl}, \citenamefont {Hamers}, \citenamefont {Shames}, \citenamefont {Danilov}, \citenamefont {Zaitsev},\ and\ \citenamefont {Shenderova}}]{Torelli20}%
  \BibitemOpen
  \bibfield  {author} {\bibinfo {author} {\bibfnamefont {M.~D.}\ \bibnamefont {Torelli}}, \bibinfo {author} {\bibfnamefont {N.~A.}\ \bibnamefont {Nunn}}, \bibinfo {author} {\bibfnamefont {Z.~R.}\ \bibnamefont {Jones}}, \bibinfo {author} {\bibfnamefont {T.}~\bibnamefont {Vedelaar}}, \bibinfo {author} {\bibfnamefont {S.~K.}\ \bibnamefont {Padamati}}, \bibinfo {author} {\bibfnamefont {R.}~\bibnamefont {Schirhagl}}, \bibinfo {author} {\bibfnamefont {R.~J.}\ \bibnamefont {Hamers}}, \bibinfo {author} {\bibfnamefont {A.~I.}\ \bibnamefont {Shames}}, \bibinfo {author} {\bibfnamefont {E.~O.}\ \bibnamefont {Danilov}}, \bibinfo {author} {\bibfnamefont {A.}~\bibnamefont {Zaitsev}},\ and\ \bibinfo {author} {\bibfnamefont {O.~A.}\ \bibnamefont {Shenderova}},\ }\bibfield  {title} {\bibinfo {title} {High temperature treatment of diamond particles toward enhancement of their quantum properties},\ }\href@noop {} {\bibfield  {journal} {\bibinfo  {journal} {Frontiers in Physics}\ }\textbf {\bibinfo {volume} {8}} (\bibinfo {year}
  {2020})}\BibitemShut {NoStop}%
\bibitem [{\citenamefont {Petit}\ \emph {et~al.}(2017)\citenamefont {Petit}, \citenamefont {Puskar}, \citenamefont {Dolenko}, \citenamefont {Choudhury}, \citenamefont {Ritter}, \citenamefont {Burikov}, \citenamefont {Laptinskiy}, \citenamefont {Brzustowski}, \citenamefont {Schade}, \citenamefont {Yuzawa} \emph {et~al.}}]{Petit17}%
  \BibitemOpen
  \bibfield  {author} {\bibinfo {author} {\bibfnamefont {T.}~\bibnamefont {Petit}}, \bibinfo {author} {\bibfnamefont {L.}~\bibnamefont {Puskar}}, \bibinfo {author} {\bibfnamefont {T.}~\bibnamefont {Dolenko}}, \bibinfo {author} {\bibfnamefont {S.}~\bibnamefont {Choudhury}}, \bibinfo {author} {\bibfnamefont {E.}~\bibnamefont {Ritter}}, \bibinfo {author} {\bibfnamefont {S.}~\bibnamefont {Burikov}}, \bibinfo {author} {\bibfnamefont {K.}~\bibnamefont {Laptinskiy}}, \bibinfo {author} {\bibfnamefont {Q.}~\bibnamefont {Brzustowski}}, \bibinfo {author} {\bibfnamefont {U.}~\bibnamefont {Schade}}, \bibinfo {author} {\bibfnamefont {H.}~\bibnamefont {Yuzawa}}, \emph {et~al.},\ }\bibfield  {title} {\bibinfo {title} {Unusual water hydrogen bond network around hydrogenated nanodiamonds},\ }\href@noop {} {\bibfield  {journal} {\bibinfo  {journal} {The Journal of Physical Chemistry C}\ }\textbf {\bibinfo {volume} {121}},\ \bibinfo {pages} {5185} (\bibinfo {year} {2017})}\BibitemShut {NoStop}%
\bibitem [{\citenamefont {Nesmelov}\ \emph {et~al.}(2004)\citenamefont {Nesmelov}, \citenamefont {Gopinath},\ and\ \citenamefont {Thomas}}]{Nesmelov04}%
  \BibitemOpen
  \bibfield  {author} {\bibinfo {author} {\bibfnamefont {Y.~E.}\ \bibnamefont {Nesmelov}}, \bibinfo {author} {\bibfnamefont {A.}~\bibnamefont {Gopinath}},\ and\ \bibinfo {author} {\bibfnamefont {D.~D.}\ \bibnamefont {Thomas}},\ }\bibfield  {title} {\bibinfo {title} {Aqueous sample in an epr cavity: sensitivity considerations},\ }\href@noop {} {\bibfield  {journal} {\bibinfo  {journal} {Journal of Magnetic Resonance}\ }\textbf {\bibinfo {volume} {167}},\ \bibinfo {pages} {138} (\bibinfo {year} {2004})}\BibitemShut {NoStop}%
\bibitem [{\citenamefont {Roessler}\ and\ \citenamefont {Salvadori}(2018)}]{Roessler18}%
  \BibitemOpen
  \bibfield  {author} {\bibinfo {author} {\bibfnamefont {M.~M.}\ \bibnamefont {Roessler}}\ and\ \bibinfo {author} {\bibfnamefont {E.}~\bibnamefont {Salvadori}},\ }\bibfield  {title} {\bibinfo {title} {Principles and applications of epr spectroscopy in the chemical sciences},\ }\href@noop {} {\bibfield  {journal} {\bibinfo  {journal} {Chemical Society Reviews}\ }\textbf {\bibinfo {volume} {47}},\ \bibinfo {pages} {2534} (\bibinfo {year} {2018})}\BibitemShut {NoStop}%
\bibitem [{\citenamefont {Abhyankar}\ \emph {et~al.}(2022)\citenamefont {Abhyankar}, \citenamefont {Agrawal}, \citenamefont {Campbell}, \citenamefont {Maly}, \citenamefont {Shrestha},\ and\ \citenamefont {Szalai}}]{Abhyankar22}%
  \BibitemOpen
  \bibfield  {author} {\bibinfo {author} {\bibfnamefont {N.}~\bibnamefont {Abhyankar}}, \bibinfo {author} {\bibfnamefont {A.}~\bibnamefont {Agrawal}}, \bibinfo {author} {\bibfnamefont {J.}~\bibnamefont {Campbell}}, \bibinfo {author} {\bibfnamefont {T.}~\bibnamefont {Maly}}, \bibinfo {author} {\bibfnamefont {P.}~\bibnamefont {Shrestha}},\ and\ \bibinfo {author} {\bibfnamefont {V.}~\bibnamefont {Szalai}},\ }\bibfield  {title} {\bibinfo {title} {Recent advances in microresonators and supporting instrumentation for electron paramagnetic resonance spectroscopy},\ }\href@noop {} {\bibfield  {journal} {\bibinfo  {journal} {Review of Scientific Instruments}\ }\textbf {\bibinfo {volume} {93}} (\bibinfo {year} {2022})}\BibitemShut {NoStop}%
\bibitem [{\citenamefont {Davis}\ \emph {et~al.}(2017)\citenamefont {Davis}, \citenamefont {Jacobs}, \citenamefont {Houle},\ and\ \citenamefont {Wilson}}]{Davis17}%
  \BibitemOpen
  \bibfield  {author} {\bibinfo {author} {\bibfnamefont {R.~D.}\ \bibnamefont {Davis}}, \bibinfo {author} {\bibfnamefont {M.~I.}\ \bibnamefont {Jacobs}}, \bibinfo {author} {\bibfnamefont {F.~A.}\ \bibnamefont {Houle}},\ and\ \bibinfo {author} {\bibfnamefont {K.~R.}\ \bibnamefont {Wilson}},\ }\bibfield  {title} {\bibinfo {title} {Colliding-droplet microreactor: rapid on-demand inertial mixing and metal-catalyzed aqueous phase oxidation processes},\ }\href@noop {} {\bibfield  {journal} {\bibinfo  {journal} {Analytical chemistry}\ }\textbf {\bibinfo {volume} {89}},\ \bibinfo {pages} {12494} (\bibinfo {year} {2017})}\BibitemShut {NoStop}%
\bibitem [{\citenamefont {Lee}\ \emph {et~al.}(2015)\citenamefont {Lee}, \citenamefont {Banerjee}, \citenamefont {Nam},\ and\ \citenamefont {Zare}}]{Lee15}%
  \BibitemOpen
  \bibfield  {author} {\bibinfo {author} {\bibfnamefont {J.~K.}\ \bibnamefont {Lee}}, \bibinfo {author} {\bibfnamefont {S.}~\bibnamefont {Banerjee}}, \bibinfo {author} {\bibfnamefont {H.~G.}\ \bibnamefont {Nam}},\ and\ \bibinfo {author} {\bibfnamefont {R.~N.}\ \bibnamefont {Zare}},\ }\bibfield  {title} {\bibinfo {title} {Acceleration of reaction in charged microdroplets},\ }\href@noop {} {\bibfield  {journal} {\bibinfo  {journal} {Quarterly reviews of biophysics}\ }\textbf {\bibinfo {volume} {48}},\ \bibinfo {pages} {437} (\bibinfo {year} {2015})}\BibitemShut {NoStop}%
\bibitem [{\citenamefont {Kintses}\ \emph {et~al.}(2012)\citenamefont {Kintses}, \citenamefont {Hein}, \citenamefont {Mohamed}, \citenamefont {Fischlechner}, \citenamefont {Courtois}, \citenamefont {Laine},\ and\ \citenamefont {Hollfelder}}]{Kintses12}%
  \BibitemOpen
  \bibfield  {author} {\bibinfo {author} {\bibfnamefont {B.}~\bibnamefont {Kintses}}, \bibinfo {author} {\bibfnamefont {C.}~\bibnamefont {Hein}}, \bibinfo {author} {\bibfnamefont {M.~F.}\ \bibnamefont {Mohamed}}, \bibinfo {author} {\bibfnamefont {M.}~\bibnamefont {Fischlechner}}, \bibinfo {author} {\bibfnamefont {F.}~\bibnamefont {Courtois}}, \bibinfo {author} {\bibfnamefont {C.}~\bibnamefont {Laine}},\ and\ \bibinfo {author} {\bibfnamefont {F.}~\bibnamefont {Hollfelder}},\ }\bibfield  {title} {\bibinfo {title} {Picoliter cell lysate assays in microfluidic droplet compartments for directed enzyme evolution},\ }\href@noop {} {\bibfield  {journal} {\bibinfo  {journal} {Chemistry \& biology}\ }\textbf {\bibinfo {volume} {19}},\ \bibinfo {pages} {1001} (\bibinfo {year} {2012})}\BibitemShut {NoStop}%
\bibitem [{\citenamefont {Wojciechowski}\ \emph {et~al.}(2018)\citenamefont {Wojciechowski}, \citenamefont {Karadas}, \citenamefont {Huck}, \citenamefont {Osterkamp}, \citenamefont {Jankuhn}, \citenamefont {Meijer}, \citenamefont {Jelezko},\ and\ \citenamefont {Andersen}}]{wojciechowski2018contributed}%
  \BibitemOpen
  \bibfield  {author} {\bibinfo {author} {\bibfnamefont {A.~M.}\ \bibnamefont {Wojciechowski}}, \bibinfo {author} {\bibfnamefont {M.}~\bibnamefont {Karadas}}, \bibinfo {author} {\bibfnamefont {A.}~\bibnamefont {Huck}}, \bibinfo {author} {\bibfnamefont {C.}~\bibnamefont {Osterkamp}}, \bibinfo {author} {\bibfnamefont {S.}~\bibnamefont {Jankuhn}}, \bibinfo {author} {\bibfnamefont {J.}~\bibnamefont {Meijer}}, \bibinfo {author} {\bibfnamefont {F.}~\bibnamefont {Jelezko}},\ and\ \bibinfo {author} {\bibfnamefont {U.~L.}\ \bibnamefont {Andersen}},\ }\bibfield  {title} {\bibinfo {title} {Contributed review: Camera-limits for wide-field magnetic resonance imaging with a nitrogen-vacancy spin sensor},\ }\href@noop {} {\bibfield  {journal} {\bibinfo  {journal} {Review of Scientific Instruments}\ }\textbf {\bibinfo {volume} {89}} (\bibinfo {year} {2018})}\BibitemShut {NoStop}%
\bibitem [{\citenamefont {Parashar}\ \emph {et~al.}(2022)\citenamefont {Parashar}, \citenamefont {Bathla}, \citenamefont {Shishir}, \citenamefont {Gokhale}, \citenamefont {Bandyopadhyay},\ and\ \citenamefont {Saha}}]{parashar2022sub}%
  \BibitemOpen
  \bibfield  {author} {\bibinfo {author} {\bibfnamefont {M.}~\bibnamefont {Parashar}}, \bibinfo {author} {\bibfnamefont {A.}~\bibnamefont {Bathla}}, \bibinfo {author} {\bibfnamefont {D.}~\bibnamefont {Shishir}}, \bibinfo {author} {\bibfnamefont {A.}~\bibnamefont {Gokhale}}, \bibinfo {author} {\bibfnamefont {S.}~\bibnamefont {Bandyopadhyay}},\ and\ \bibinfo {author} {\bibfnamefont {K.}~\bibnamefont {Saha}},\ }\bibfield  {title} {\bibinfo {title} {Sub-second temporal magnetic field microscopy using quantum defects in diamond},\ }\href@noop {} {\bibfield  {journal} {\bibinfo  {journal} {Scientific reports}\ }\textbf {\bibinfo {volume} {12}},\ \bibinfo {pages} {8743} (\bibinfo {year} {2022})}\BibitemShut {NoStop}%
\bibitem [{\citenamefont {Webb}\ \emph {et~al.}(2022)\citenamefont {Webb}, \citenamefont {Troise}, \citenamefont {Hansen}, \citenamefont {Frellsen}, \citenamefont {Osterkamp}, \citenamefont {Jelezko}, \citenamefont {Jankuhn}, \citenamefont {Meijer}, \citenamefont {Berg-S{\o}rensen}, \citenamefont {Perrier} \emph {et~al.}}]{webb2022high}%
  \BibitemOpen
  \bibfield  {author} {\bibinfo {author} {\bibfnamefont {J.~L.}\ \bibnamefont {Webb}}, \bibinfo {author} {\bibfnamefont {L.}~\bibnamefont {Troise}}, \bibinfo {author} {\bibfnamefont {N.~W.}\ \bibnamefont {Hansen}}, \bibinfo {author} {\bibfnamefont {L.~F.}\ \bibnamefont {Frellsen}}, \bibinfo {author} {\bibfnamefont {C.}~\bibnamefont {Osterkamp}}, \bibinfo {author} {\bibfnamefont {F.}~\bibnamefont {Jelezko}}, \bibinfo {author} {\bibfnamefont {S.}~\bibnamefont {Jankuhn}}, \bibinfo {author} {\bibfnamefont {J.}~\bibnamefont {Meijer}}, \bibinfo {author} {\bibfnamefont {K.}~\bibnamefont {Berg-S{\o}rensen}}, \bibinfo {author} {\bibfnamefont {J.-F.}\ \bibnamefont {Perrier}}, \emph {et~al.},\ }\bibfield  {title} {\bibinfo {title} {High-speed wide-field imaging of microcircuitry using nitrogen vacancies in diamond},\ }\href@noop {} {\bibfield  {journal} {\bibinfo  {journal} {Physical Review Applied}\ }\textbf {\bibinfo {volume} {17}},\ \bibinfo {pages} {064051} (\bibinfo {year} {2022})}\BibitemShut {NoStop}%
\bibitem [{\citenamefont {Shembekar}\ \emph {et~al.}(2016)\citenamefont {Shembekar}, \citenamefont {Chaipan}, \citenamefont {Utharala},\ and\ \citenamefont {Merten}}]{Shembekar2016}%
  \BibitemOpen
  \bibfield  {author} {\bibinfo {author} {\bibfnamefont {N.}~\bibnamefont {Shembekar}}, \bibinfo {author} {\bibfnamefont {C.}~\bibnamefont {Chaipan}}, \bibinfo {author} {\bibfnamefont {R.}~\bibnamefont {Utharala}},\ and\ \bibinfo {author} {\bibfnamefont {C.~A.}\ \bibnamefont {Merten}},\ }\bibfield  {title} {\bibinfo {title} {Droplet-based microfluidics in drug discovery{,} transcriptomics and high-throughput molecular genetics},\ }\href {https://doi.org/10.1039/C6LC00249H} {\bibfield  {journal} {\bibinfo  {journal} {Lab Chip}\ }\textbf {\bibinfo {volume} {16}},\ \bibinfo {pages} {1314} (\bibinfo {year} {2016})}\BibitemShut {NoStop}%
\bibitem [{\citenamefont {Givan}(2011)}]{givan2011flow}%
  \BibitemOpen
  \bibfield  {author} {\bibinfo {author} {\bibfnamefont {A.~L.}\ \bibnamefont {Givan}},\ }\bibfield  {title} {\bibinfo {title} {Flow cytometry: an introduction},\ }\href@noop {} {\bibfield  {journal} {\bibinfo  {journal} {Flow cytometry protocols}\ ,\ \bibinfo {pages} {1}} (\bibinfo {year} {2011})}\BibitemShut {NoStop}%
\bibitem [{\citenamefont {Eruslanov}\ and\ \citenamefont {Kusmartsev}(2010)}]{eruslanov2010identification}%
  \BibitemOpen
  \bibfield  {author} {\bibinfo {author} {\bibfnamefont {E.}~\bibnamefont {Eruslanov}}\ and\ \bibinfo {author} {\bibfnamefont {S.}~\bibnamefont {Kusmartsev}},\ }\bibfield  {title} {\bibinfo {title} {Identification of ros using oxidized dcfda and flow-cytometry},\ }\href@noop {} {\bibfield  {journal} {\bibinfo  {journal} {Advanced protocols in oxidative stress II}\ ,\ \bibinfo {pages} {57}} (\bibinfo {year} {2010})}\BibitemShut {NoStop}%
\bibitem [{\citenamefont {Barry}\ \emph {et~al.}(2020)\citenamefont {Barry}, \citenamefont {Schloss}, \citenamefont {Bauch}, \citenamefont {Turner}, \citenamefont {Hart}, \citenamefont {Pham},\ and\ \citenamefont {Walsworth}}]{Barry20}%
  \BibitemOpen
  \bibfield  {author} {\bibinfo {author} {\bibfnamefont {J.~F.}\ \bibnamefont {Barry}}, \bibinfo {author} {\bibfnamefont {J.~M.}\ \bibnamefont {Schloss}}, \bibinfo {author} {\bibfnamefont {E.}~\bibnamefont {Bauch}}, \bibinfo {author} {\bibfnamefont {M.~J.}\ \bibnamefont {Turner}}, \bibinfo {author} {\bibfnamefont {C.~A.}\ \bibnamefont {Hart}}, \bibinfo {author} {\bibfnamefont {L.~M.}\ \bibnamefont {Pham}},\ and\ \bibinfo {author} {\bibfnamefont {R.~L.}\ \bibnamefont {Walsworth}},\ }\bibfield  {title} {\bibinfo {title} {Sensitivity optimization for nv-diamond magnetometry},\ }\href@noop {} {\bibfield  {journal} {\bibinfo  {journal} {Reviews of Modern Physics}\ }\textbf {\bibinfo {volume} {92}},\ \bibinfo {pages} {015004} (\bibinfo {year} {2020})}\BibitemShut {NoStop}%
\bibitem [{\citenamefont {Ajoy}\ and\ \citenamefont {Cappellaro}(2012)}]{Ajoy12g}%
  \BibitemOpen
  \bibfield  {author} {\bibinfo {author} {\bibfnamefont {A.}~\bibnamefont {Ajoy}}\ and\ \bibinfo {author} {\bibfnamefont {P.}~\bibnamefont {Cappellaro}},\ }\bibfield  {title} {\bibinfo {title} {Stable three-axis nuclear-spin gyroscope in diamond},\ }\href {https://doi.org/10.1103/PhysRevA.86.062104} {\bibfield  {journal} {\bibinfo  {journal} {Phys. Rev. A}\ }\textbf {\bibinfo {volume} {86}},\ \bibinfo {pages} {062104} (\bibinfo {year} {2012})}\BibitemShut {NoStop}%
\bibitem [{\citenamefont {Jarmola}\ \emph {et~al.}(2021)\citenamefont {Jarmola}, \citenamefont {Lourette}, \citenamefont {Acosta}, \citenamefont {Birdwell}, \citenamefont {Bl{\"u}mler}, \citenamefont {Budker}, \citenamefont {Ivanov},\ and\ \citenamefont {Malinovsky}}]{Jarmola21}%
  \BibitemOpen
  \bibfield  {author} {\bibinfo {author} {\bibfnamefont {A.}~\bibnamefont {Jarmola}}, \bibinfo {author} {\bibfnamefont {S.}~\bibnamefont {Lourette}}, \bibinfo {author} {\bibfnamefont {V.~M.}\ \bibnamefont {Acosta}}, \bibinfo {author} {\bibfnamefont {A.~G.}\ \bibnamefont {Birdwell}}, \bibinfo {author} {\bibfnamefont {P.}~\bibnamefont {Bl{\"u}mler}}, \bibinfo {author} {\bibfnamefont {D.}~\bibnamefont {Budker}}, \bibinfo {author} {\bibfnamefont {T.}~\bibnamefont {Ivanov}},\ and\ \bibinfo {author} {\bibfnamefont {V.~S.}\ \bibnamefont {Malinovsky}},\ }\bibfield  {title} {\bibinfo {title} {Demonstration of diamond nuclear spin gyroscope},\ }\href@noop {} {\bibfield  {journal} {\bibinfo  {journal} {Science advances}\ }\textbf {\bibinfo {volume} {7}},\ \bibinfo {pages} {eabl3840} (\bibinfo {year} {2021})}\BibitemShut {NoStop}%
\bibitem [{\citenamefont {Ajoy}\ \emph {et~al.}(2018)\citenamefont {Ajoy}, \citenamefont {Liu}, \citenamefont {Nazaryan}, \citenamefont {Lv}, \citenamefont {Zangara}, \citenamefont {Safvati}, \citenamefont {Wang}, \citenamefont {Arnold}, \citenamefont {Li}, \citenamefont {Lin} \emph {et~al.}}]{Ajoy17}%
  \BibitemOpen
  \bibfield  {author} {\bibinfo {author} {\bibfnamefont {A.}~\bibnamefont {Ajoy}}, \bibinfo {author} {\bibfnamefont {K.}~\bibnamefont {Liu}}, \bibinfo {author} {\bibfnamefont {R.}~\bibnamefont {Nazaryan}}, \bibinfo {author} {\bibfnamefont {X.}~\bibnamefont {Lv}}, \bibinfo {author} {\bibfnamefont {P.~R.}\ \bibnamefont {Zangara}}, \bibinfo {author} {\bibfnamefont {B.}~\bibnamefont {Safvati}}, \bibinfo {author} {\bibfnamefont {G.}~\bibnamefont {Wang}}, \bibinfo {author} {\bibfnamefont {D.}~\bibnamefont {Arnold}}, \bibinfo {author} {\bibfnamefont {G.}~\bibnamefont {Li}}, \bibinfo {author} {\bibfnamefont {A.}~\bibnamefont {Lin}}, \emph {et~al.},\ }\bibfield  {title} {\bibinfo {title} {Orientation-independent room temperature optical 13c hyperpolarization in powdered diamond},\ }\href {http://advances.sciencemag.org/content/4/5/eaar5492} {\bibfield  {journal} {\bibinfo  {journal} {Sci. Adv.}\ }\textbf {\bibinfo {volume} {4}},\ \bibinfo {pages} {eaar5492} (\bibinfo {year} {2018})}\BibitemShut {NoStop}%
\bibitem [{\citenamefont {Beatrez}\ \emph {et~al.}(2021)\citenamefont {Beatrez}, \citenamefont {Janes}, \citenamefont {Akkiraju}, \citenamefont {Pillai}, \citenamefont {Oddo}, \citenamefont {Reshetikhin}, \citenamefont {Druga}, \citenamefont {McAllister}, \citenamefont {Elo}, \citenamefont {Gilbert}, \citenamefont {Suter},\ and\ \citenamefont {Ajoy}}]{Beatrez2021}%
  \BibitemOpen
  \bibfield  {author} {\bibinfo {author} {\bibfnamefont {W.}~\bibnamefont {Beatrez}}, \bibinfo {author} {\bibfnamefont {O.}~\bibnamefont {Janes}}, \bibinfo {author} {\bibfnamefont {A.}~\bibnamefont {Akkiraju}}, \bibinfo {author} {\bibfnamefont {A.}~\bibnamefont {Pillai}}, \bibinfo {author} {\bibfnamefont {A.}~\bibnamefont {Oddo}}, \bibinfo {author} {\bibfnamefont {P.}~\bibnamefont {Reshetikhin}}, \bibinfo {author} {\bibfnamefont {E.}~\bibnamefont {Druga}}, \bibinfo {author} {\bibfnamefont {M.}~\bibnamefont {McAllister}}, \bibinfo {author} {\bibfnamefont {M.}~\bibnamefont {Elo}}, \bibinfo {author} {\bibfnamefont {B.}~\bibnamefont {Gilbert}}, \bibinfo {author} {\bibfnamefont {D.}~\bibnamefont {Suter}},\ and\ \bibinfo {author} {\bibfnamefont {A.}~\bibnamefont {Ajoy}},\ }\bibfield  {title} {\bibinfo {title} {Floquet prethermalization with lifetime exceeding 90 s in a bulk hyperpolarized solid},\ }\href {https://doi.org/10.1103/PhysRevLett.127.170603} {\bibfield  {journal} {\bibinfo  {journal} {Phys. Rev. Lett.}\
  }\textbf {\bibinfo {volume} {127}},\ \bibinfo {pages} {170603} (\bibinfo {year} {2021})}\BibitemShut {NoStop}%
\bibitem [{\citenamefont {Sahin}\ \emph {et~al.}(2022)\citenamefont {Sahin}, \citenamefont {de~Leon~Sanchez}, \citenamefont {Conti}, \citenamefont {Akkiraju}, \citenamefont {Reshetikhin}, \citenamefont {Druga}, \citenamefont {Aggarwal}, \citenamefont {Gilbert}, \citenamefont {Bhave},\ and\ \citenamefont {Ajoy}}]{Sahin21}%
  \BibitemOpen
  \bibfield  {author} {\bibinfo {author} {\bibfnamefont {O.}~\bibnamefont {Sahin}}, \bibinfo {author} {\bibfnamefont {E.}~\bibnamefont {de~Leon~Sanchez}}, \bibinfo {author} {\bibfnamefont {S.}~\bibnamefont {Conti}}, \bibinfo {author} {\bibfnamefont {A.}~\bibnamefont {Akkiraju}}, \bibinfo {author} {\bibfnamefont {P.}~\bibnamefont {Reshetikhin}}, \bibinfo {author} {\bibfnamefont {E.}~\bibnamefont {Druga}}, \bibinfo {author} {\bibfnamefont {A.}~\bibnamefont {Aggarwal}}, \bibinfo {author} {\bibfnamefont {B.}~\bibnamefont {Gilbert}}, \bibinfo {author} {\bibfnamefont {S.}~\bibnamefont {Bhave}},\ and\ \bibinfo {author} {\bibfnamefont {A.}~\bibnamefont {Ajoy}},\ }\bibfield  {title} {\bibinfo {title} {High field magnetometry with hyperpolarized nuclear spins},\ }\href@noop {} {\bibfield  {journal} {\bibinfo  {journal} {Nature communications}\ }\textbf {\bibinfo {volume} {13}},\ \bibinfo {pages} {5486} (\bibinfo {year} {2022})}\BibitemShut {NoStop}%
\bibitem [{\citenamefont {Kumar}\ and\ \citenamefont {Dixit}(2017)}]{kumar17}%
  \BibitemOpen
  \bibfield  {author} {\bibinfo {author} {\bibfnamefont {A.}~\bibnamefont {Kumar}}\ and\ \bibinfo {author} {\bibfnamefont {C.~K.}\ \bibnamefont {Dixit}},\ }\bibinfo {title} {Methods for characterization of nanoparticles},\ in\ \href {https://doi.org/10.1016/B978-0-08-100557-6.00003-1} {\emph {\bibinfo {booktitle} {Advances in Nanomedicine for the Delivery of Therapeutic Nucleic Acids}}}\ (\bibinfo  {publisher} {Woodhead Publishing},\ \bibinfo {year} {2017})\ pp.\ \bibinfo {pages} {43--58}\BibitemShut {NoStop}%
\bibitem [{\citenamefont {Blankenship}\ \emph {et~al.}(2023)\citenamefont {Blankenship}, \citenamefont {Jones}, \citenamefont {Zhao}, \citenamefont {Singh}, \citenamefont {Sarkar}, \citenamefont {Li}, \citenamefont {Suh}, \citenamefont {Chen}, \citenamefont {Grigoropoulos},\ and\ \citenamefont {Ajoy}}]{Blankenship23}%
  \BibitemOpen
  \bibfield  {author} {\bibinfo {author} {\bibfnamefont {B.~W.}\ \bibnamefont {Blankenship}}, \bibinfo {author} {\bibfnamefont {Z.~R.}\ \bibnamefont {Jones}}, \bibinfo {author} {\bibfnamefont {N.}~\bibnamefont {Zhao}}, \bibinfo {author} {\bibfnamefont {H.}~\bibnamefont {Singh}}, \bibinfo {author} {\bibfnamefont {A.}~\bibnamefont {Sarkar}}, \bibinfo {author} {\bibfnamefont {R.}~\bibnamefont {Li}}, \bibinfo {author} {\bibfnamefont {L.}~\bibnamefont {Suh}}, \bibinfo {author} {\bibfnamefont {A.}~\bibnamefont {Chen}}, \bibinfo {author} {\bibfnamefont {C.~P.}\ \bibnamefont {Grigoropoulos}},\ and\ \bibinfo {author} {\bibfnamefont {A.}~\bibnamefont {Ajoy}},\ }\bibfield  {title} {\bibinfo {title} {Complex three-dimensional microscale structures for quantum sensing applications},\ }\href@noop {} {\bibfield  {journal} {\bibinfo  {journal} {Nano Letters}\ }\textbf {\bibinfo {volume} {23}},\ \bibinfo {pages} {9272} (\bibinfo {year} {2023})}\BibitemShut {NoStop}%
\bibitem [{\citenamefont {Allan}\ \emph {et~al.}(1987)\citenamefont {Allan} \emph {et~al.}}]{allan1987time}%
  \BibitemOpen
  \bibfield  {author} {\bibinfo {author} {\bibfnamefont {D.~W.}\ \bibnamefont {Allan}} \emph {et~al.},\ }\bibfield  {title} {\bibinfo {title} {Time and frequency(time-domain) characterization, estimation, and prediction of precision clocks and oscillators},\ }\href@noop {} {\bibfield  {journal} {\bibinfo  {journal} {IEEE transactions on ultrasonics, ferroelectrics, and frequency control}\ }\textbf {\bibinfo {volume} {34}},\ \bibinfo {pages} {647} (\bibinfo {year} {1987})}\BibitemShut {NoStop}%
\bibitem [{\citenamefont {Hopcroft}(2024)}]{hopcroftCode}%
  \BibitemOpen
  \bibfield  {author} {\bibinfo {author} {\bibfnamefont {M.}~\bibnamefont {Hopcroft}},\ }\bibfield  {title} {\bibinfo {title} {allan (https://www.mathworks.com/matlabcentral/fileexchange/13246-allan), matlab central file exchange},\ }\href@noop {} {\bibfield  {journal} {\bibinfo  {journal} {MATLAB Central file exchange}\ } (\bibinfo {year} {February 23, 2024})}\BibitemShut {NoStop}%
\bibitem [{\citenamefont {Barton}\ \emph {et~al.}(2020)\citenamefont {Barton}, \citenamefont {Gulka}, \citenamefont {Tarabek}, \citenamefont {Mindarava}, \citenamefont {Wang}, \citenamefont {Schimer}, \citenamefont {Raabova}, \citenamefont {Bednar}, \citenamefont {Plenio}, \citenamefont {Jelezko} \emph {et~al.}}]{Barton20}%
  \BibitemOpen
  \bibfield  {author} {\bibinfo {author} {\bibfnamefont {J.}~\bibnamefont {Barton}}, \bibinfo {author} {\bibfnamefont {M.}~\bibnamefont {Gulka}}, \bibinfo {author} {\bibfnamefont {J.}~\bibnamefont {Tarabek}}, \bibinfo {author} {\bibfnamefont {Y.}~\bibnamefont {Mindarava}}, \bibinfo {author} {\bibfnamefont {Z.}~\bibnamefont {Wang}}, \bibinfo {author} {\bibfnamefont {J.}~\bibnamefont {Schimer}}, \bibinfo {author} {\bibfnamefont {H.}~\bibnamefont {Raabova}}, \bibinfo {author} {\bibfnamefont {J.}~\bibnamefont {Bednar}}, \bibinfo {author} {\bibfnamefont {M.~B.}\ \bibnamefont {Plenio}}, \bibinfo {author} {\bibfnamefont {F.}~\bibnamefont {Jelezko}}, \emph {et~al.},\ }\bibfield  {title} {\bibinfo {title} {Nanoscale dynamic readout of a chemical redox process using radicals coupled with nitrogen-vacancy centers in nanodiamonds},\ }\href@noop {} {\bibfield  {journal} {\bibinfo  {journal} {ACS nano}\ }\textbf {\bibinfo {volume} {14}},\ \bibinfo {pages} {12938} (\bibinfo {year} {2020})}\BibitemShut {NoStop}%
\bibitem [{\citenamefont {Perona~Martinez}\ \emph {et~al.}(2020)\citenamefont {Perona~Martinez}, \citenamefont {Nusantara}, \citenamefont {Chipaux}, \citenamefont {Padamati},\ and\ \citenamefont {Schirhagl}}]{Perona20}%
  \BibitemOpen
  \bibfield  {author} {\bibinfo {author} {\bibfnamefont {F.}~\bibnamefont {Perona~Martinez}}, \bibinfo {author} {\bibfnamefont {A.~C.}\ \bibnamefont {Nusantara}}, \bibinfo {author} {\bibfnamefont {M.}~\bibnamefont {Chipaux}}, \bibinfo {author} {\bibfnamefont {S.~K.}\ \bibnamefont {Padamati}},\ and\ \bibinfo {author} {\bibfnamefont {R.}~\bibnamefont {Schirhagl}},\ }\bibfield  {title} {\bibinfo {title} {Nanodiamond relaxometry-based detection of free-radical species when produced in chemical reactions in biologically relevant conditions},\ }\href@noop {} {\bibfield  {journal} {\bibinfo  {journal} {ACS sensors}\ }\textbf {\bibinfo {volume} {5}},\ \bibinfo {pages} {3862} (\bibinfo {year} {2020})}\BibitemShut {NoStop}%
\end{thebibliography}%
\vspace{-1mm}

\clearpage
\onecolumngrid
\begin{center}
\textbf{\large{\textit{Supplementary Information:} \\ \smallskip High-precision chemical quantum sensing in flowing monodisperse microdroplets}} \\\smallskip

\hfill \break
\smallskip
Adrisha Sarkar,$^{1,2,\BRd{\ast}}$  Zachary Jones,$^{1,3,\BRd{\ast}}$  Madhur Parashar,$^{1}$ Emanuel Druga,$^{1}$ Amala Akkiraju,$^{1}$ Sophie Conti,$^{1}$ \\Pranav Krishnamoorthi,$^{1}$ Srisai Nachuri,$^{1}$ Parker Aman,$^{1}$ Mohammad Hashemi,$^{1}$
Nicholas Nunn,$^{4}$  \\Marco Torelli,$^{4}$ Benjamin Gilbert,$^{5}$ Kevin R. Wilson,$^{2}$ Olga Shenderova,$^{4}$ Deepti Tanjore,$^{3}$ and Ashok Ajoy,$^{1,2,5}$\\
\smallskip
\I{${}^{1}$ {\small Department of Chemistry, University of California, Berkeley, Berkeley, CA 94720, USA.}} \\
\I{${}^{2}$ {\small Chemical Sciences Division,  Lawrence Berkeley National Laboratory,  Berkeley, CA 94720, USA.}} \\
\I{${}^{3}$ {\small Advanced Biofuels and Bioproducts Process Development Unit (AB-PDU), Biological Systems and Engineering Division, Lawrence Berkeley National Laboratory, Berkeley,CA 94720, USA.}}\\
\I{${}^{4}$ {\small Energy Geoscience Division, Lawrence Berkeley National Laboratory, Berkeley, CA 94720, USA.}}\\
\I{${}^{5}$ {\small Adamas Nanotechnologies, Inc., Raleigh, NC 27617, USA.}}\\
\I{${}^{6}$ {\small CIFAR Azrieli Global Scholars Program, 661 University Ave, Toronto, ON M5G 1M1, Canada.}}\\
{${}^{\BRd{\ast}}$ {\small These authors contributed equally to this work.}}

\end{center}

\twocolumngrid

\beginsupplement
\tableofcontents

\section{Materials}
\zsl{materials}
Diamond particles of all sizes are from Adamas Nanotechnologies (adamasnano.com). Unless otherwise stated particles are carboxly-terminated and implanted with approximately 2 ppm NV centers. 

Particle sizes and charges are determined by the manufacturer using dynamic light scattering (DLS) with zeta potential. Results of particle characterization are shown in \zfr{DLS} showing hydrodynamic diameter (A) and Zeta-potential (B) for the 40 nm and 100 nm particles used in our experiments. The DLS volume average (A) shown gives an indication of the average size and population heterogeneity, and is scewed toward higher values as large particles scatter more than small. Number mean (d$_n$)given as the inset is indicative of the average size of the population of particles. The particles represent a distribution of sizes with number mean 37 nm and 107 nm for 40 nm. The width of the peak is indicative of variation in sizes that demonstrates particle-to-particle variability, especially in the smaller sized particles which have a FWHM$\approx30$ nm.

Both sets of particles have robust colloidal stability owing to their zeta potential ($\zeta$). A measure of the surface charge of nanoparticles in solution, $\zeta$ gives information about the colloidal stability of the NDs. In general, nanoparticles with zeta potential less than -30 mV (or greater than +30 mV) will remain suspended in aqueous solution.~\cite{kumar17} This is largely due to electrostatic repulsion between particles ensuring they stay separated by enough distance to avoid aggregation. \zfr{DLS}B shows that the NDs have negative surface charge: -37 mV for the 40 nm particles and -56 mV for the 100 nm particles, which is sufficient to remain well dispersed in our aqueous samples.

\begin{figure}[t]
  \centering
  {\includegraphics[width=0.40 \textwidth]{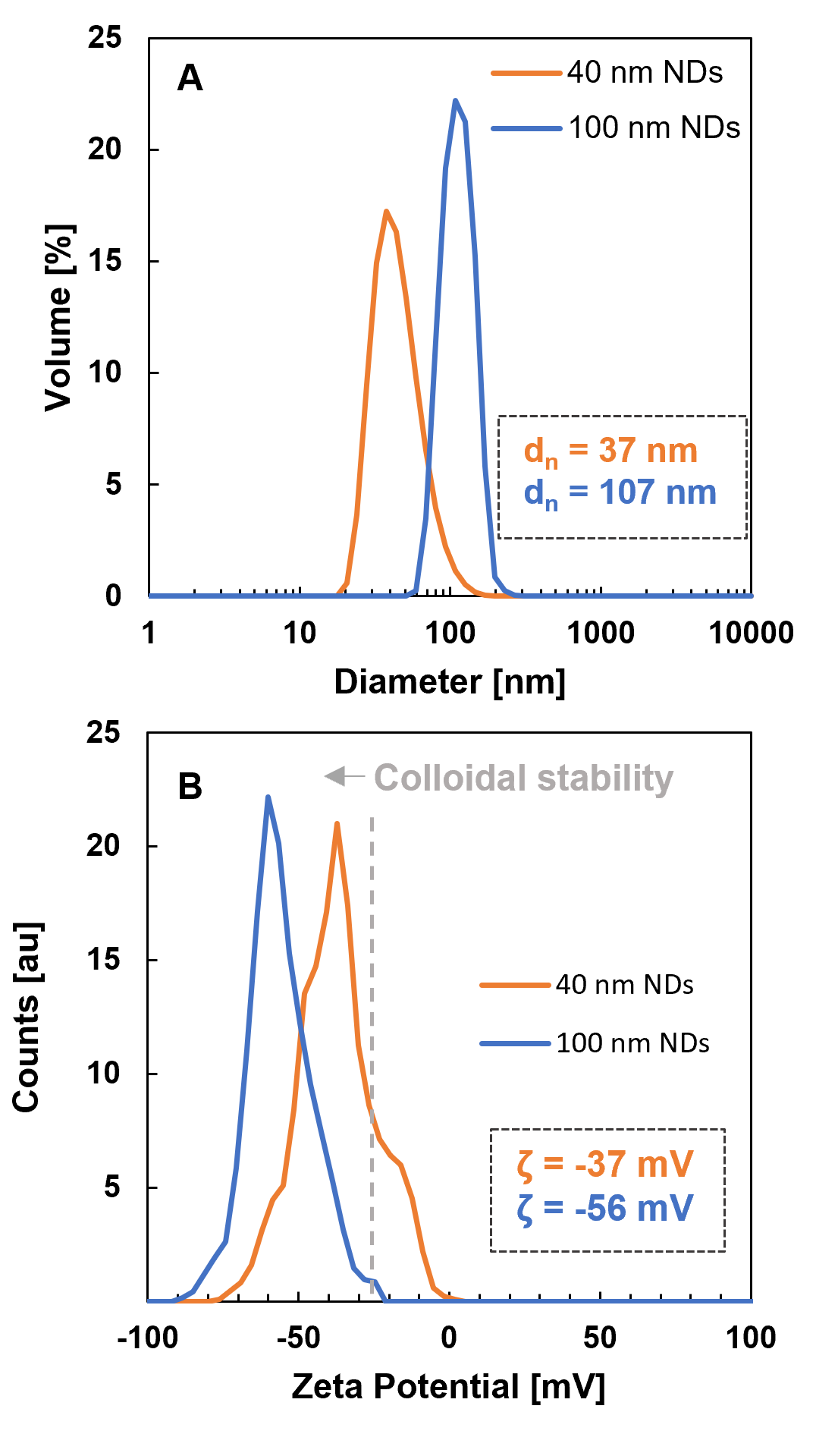}}
  \caption{\T{Nanodiamond DLS and Zeta potential measurements.} Size and surface charge characterization for the 40 nm and 100 nm particles used in our experiments. Dynamic Light Scattering (DLS) plots (A) show histograms of the measurement of hydrodynamic diameter for the population of particles measured. Volume average plot are shown, indicating the proportion of the volume of particles comprised of a given diameter, while number mean (average sizes of all particles) is given as an inset. Zeta potential measurements (B) shows histograms of the measured surface charge from a population of particles, along with the reported value (inset). Dashed line represents roughly the deliniation between colloidally stable and not. 
}
\zfl{DLS}
\end{figure}

\section{Microfluidics details}

\zsl{microfluidics}
\subsection{Device fabrication}
\zsl{chips}
\noindent{\T{\textit{Silicon Master wafer}}}

A 4-inch silicon wafer is cleaned thoroughly with an acetone wash followed by an IPA wash, then dried with nitrogen and placed on a 150$^{\circ}$C hotplate for 10 min. This wafer is then centered in a spin coater and vacuum is used to hold the wafer in place. A layer of photoresist (SU8-2025) is spin-coated to the wafer, under conditions recommended by the SU8 datasheet and spin coater specifications, to produce a 40um film thickness. About 4ml of the photoresist is poured onto the wafer and spun under the appropriate conditions (500 rpm for 10s followed by 2000 rpm for 30s in our apparatus). After removing the wafer from the spin coater it is pre-baked on a hotplate at 65$^{\circ}$C for 3 min and 95$^{\circ}$C for 6 min. A mylar photomask with the microfluidic device design is used to UV expose and cure the photoresist. After cooling to room temperature, the wafer is placed in a mask aligner with the photomask on top of it, ink side down. The mask aligner is closed and clamped down ensuring the best possible contact for the wafer and photomask. Based on the UV power delivered by the mask aligner we calculate the exposure time:

	Exposure time = (Exposure Dosage) / (Power Intensity) * (Scaling Factor) 

For 40 $\mu$m channel height for $160$ mJ/cm$^2$ energy, we expose for 20 s at 10 mW/cm$^2$.
Post exposure, the wafer is baked on the hotplate for 1 min at 65 $^{\circ}$C and then 6 min at 95 $^{\circ}$C. Once the wafer has cooled to room temperature it is submerged in an SU8 developer bath with agitation for 7.5 min then rinsed with isopropyl alcohol. If white streaks are noticed, the wafer is underdeveloped and is re-submerged in developer in 20 s increments till no streaks are visible. Once no SU8 residue is present on the wafer, it is rinsed with isopropyl alcohol and deionized water then dried. The wafer is finally hard baked at 150 $^{\circ}$C for 10 min. Stylus profilometry and light microscopy are used to assess feature sizes and integrity.

\noindent{\T{\textit{Chip fabrication}}}

To create the PDMS on glass microfluidic device we begin by creating a PDMS cast from the master wafer. A Sylgard 216 A+B Clear Silicon compound consisting of a PDMS pre-polymer (A) and curing agent (B) which are vigorously mixed together at a 10:1 ratio. 10 mg of PDMS pre-polymer and curing agent per square inch provided an adequate thickness ($^{\sim}$5mm) for the microfluidics device. The total volume of PDMS prepared can be adjusted to suit the device thickness desired. The mixed PDMS is then left to degas in a vacuum chamber for 30-60 min until all gas bubbles disappear. This uncured PDMS is poured onto the master mold. The master mold is laid flat, pattern side up in a petri dish and the PDMS is poured on top then baked in a 65 $^{\circ}$C oven for at least 6 hr and for up to 12 hr. Once removed from the oven and cooled, the desired PDMS portions for each chip are cut out carefully using an Xacto knife. The inlet ports for the microfluidics devices are created by punching vertical holes with a biopsy punch compatible with the inlet/outlet tubing diameter. To remove particulate matter and dust off the PDMS surface, scotch tape is used to clean and protect the PDMS. 

\noindent{\T{\textit{Bonding to glass}}}

A glass coverslip, is cleaned and prepare by wiping with a lint-free wipe, rinsing for 1 minute with Milli-Q water, and manually scrubbing all surfaces and edges. This is followed by an acetone and IPA wash and scrub. To remove chemical residues a final Milli-Q water wash is done without touching the main face of the glass. The coverslip is dried using compressed air and checked for marks or spots. If any such issues are noted the glass must be re-cleaned or a new coverslip must be prepped. It is important that the coverslip chosen for bonding to the PDMS is spotless. 

The final PDMS-on-glass device is assembled using oxygen plasma bonding. Once the bonding chamber is initialized, the glass coverslip and PDMS piece are inserted into the chamber, both bonding face side up. The vacuum pumps are engaged to remove air until stabilized (${\lesssim}$0.32 Torr). Oxygen gas is then released into the chamber until pressure re-stabilizes (${\app}$0.4 Torr). The radio-frequency (RF) power is turned on to deliver 24 W UV light for 24 s. The RF, oxygen and vacuum are disengaged to allow access to the bonding chamber. Immediately, the PDMS is touched to the glass and gently pressed to facilitate the bonding via the reactive oxidative species deposited. The device is left for ${{>}}$12 hrs in a 65 $^{\circ}$C oven before use. 

\subsection{Microdroplet production}
\zsl{droplet_production}
To generate microdroplets, pressure controlling pumps (Fluigent FlowEZ) pressurize reagent vessels filled with a dispersed phase (aqueous in our case) and a continuous phase (2\% SPAN80 (Sigma) surfactant in mineral oil) and fitted with pressure caps to deliver and fluid to the devices. After inserting inlet tubing and outlet tubing into the device, pressure applied to the water and oil channels at a ratio of approximately 1:2 (w:o) creates droplet emulsions. We adjust set points to optimize for desire flow rate, droplet size and droplet spacing. 

\subsection{Device Features}

\begin{figure*}[t]
  \centering
  {\includegraphics[width=0.9 \textwidth]{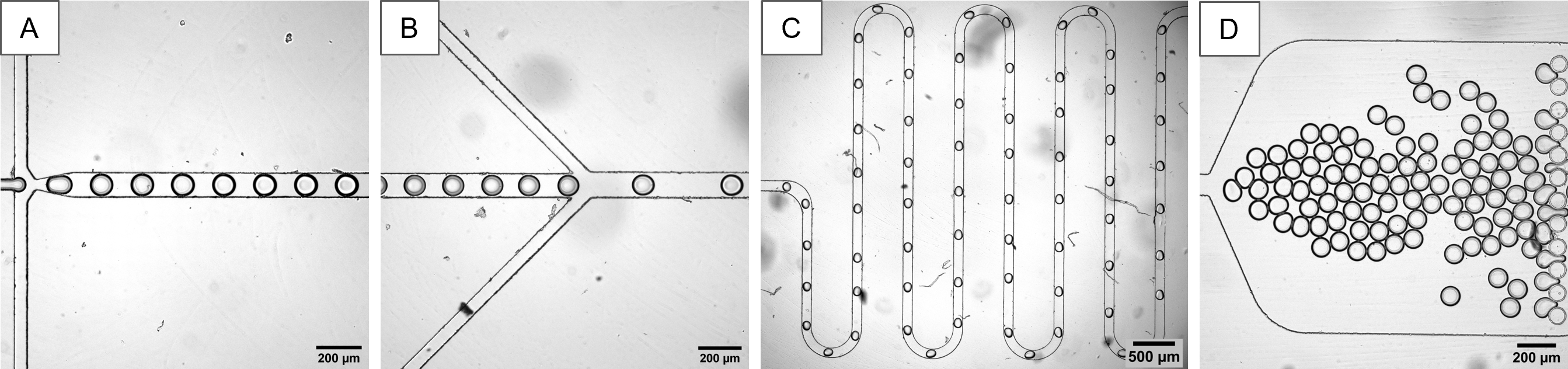}}
  \caption{\T{Detail of microfluidic devices} used for dual lock in measurements. Our devices include a fluid focusing junction (a) to generate aqueous droplets in an oil continuous phase and a down-chip second oil addition (b) which enables control over the spacing of droplets. As needed, droplet contents can be allowed to mix by following the hairpin turns shown is c, or can be stored by filling the chamber shown in d. 
}
\zfl{ChipImages}
\end{figure*}

We design varying photolithography masks to form structures like those shown in \zfr{ChipImages} to produce different droplet outcomes.

\noindent{\T{\textit{Fluid focusing junction}}}

As discussed in the main text, we use a fluid focusing junction (see \zfr{ChipImages}A) to form droplets at a constant size, and rate. We have flexibility in the speed and size of droplet production with control of the relative flow rates of the aqueous and oil phases entering the junction. Microfluidic device design affords an additional layer of control over these factors. 

\noindent{\T{\textit{Droplet separator junction}}}

We control the inter-droplet spacing via the addition of spacer oil in a geometry shown in \zfr{ChipImages}B. The same oil as the continuous phase (mineral oil in our case) so it has no effect on the already formed droplets. Changes to the pressure applied to this channel do however impact the overall resistance in the device with in turn impacts the size and rate of droplets produced upstream, so some fine-tuning is required to achieve the desired profile. 

\noindent{\T{\textit{Extended length for Droplet analysis and some mixing}}}

This region is mainly to extend the length along which the stable droplets travel. Our devices enables some enhanced droplet mixing by including sharp loops and turns in the channel so the droplets will change direction quickly causing mixing of their contents. \zfr{ChipImages}C shows an example of such a channel where a droplet moving in one direction abruptly makes a 180 degree turn causing the momentum of its contents to induce mixing.

\noindent{\T{\textit{Droplet collection chamber}}}
\zsl{chamber}

Here, we describe the on-chip collection chamber used for storing individual droplets post-analysis. Within the chamber, droplets can self-assemble into a regular lattice, advantageous for high-throughput optical analysis. One collection chamber geometry is shown in \zfr{ChipImages}D, which includes an optional barrier structure to further limit flow and collect a higher density of droplets.

The main advantage of the collection chamber is enabling delayed analysis of stored particles, achievable through wide-field microscopy or ensemble methods like NMR. Each droplet, containing NDs, remains a separate, confined environment. \zfr{chamber} provides further detailed images of the collection chamber in brightfield (top) and fluorescence (bottom) under varying optical magnification. Diamond-filled droplets assemble in such a way to be readily analyzed in a wide-field configuration.

\begin{figure}[t]
  \centering
  {\includegraphics[width=0.49 \textwidth]{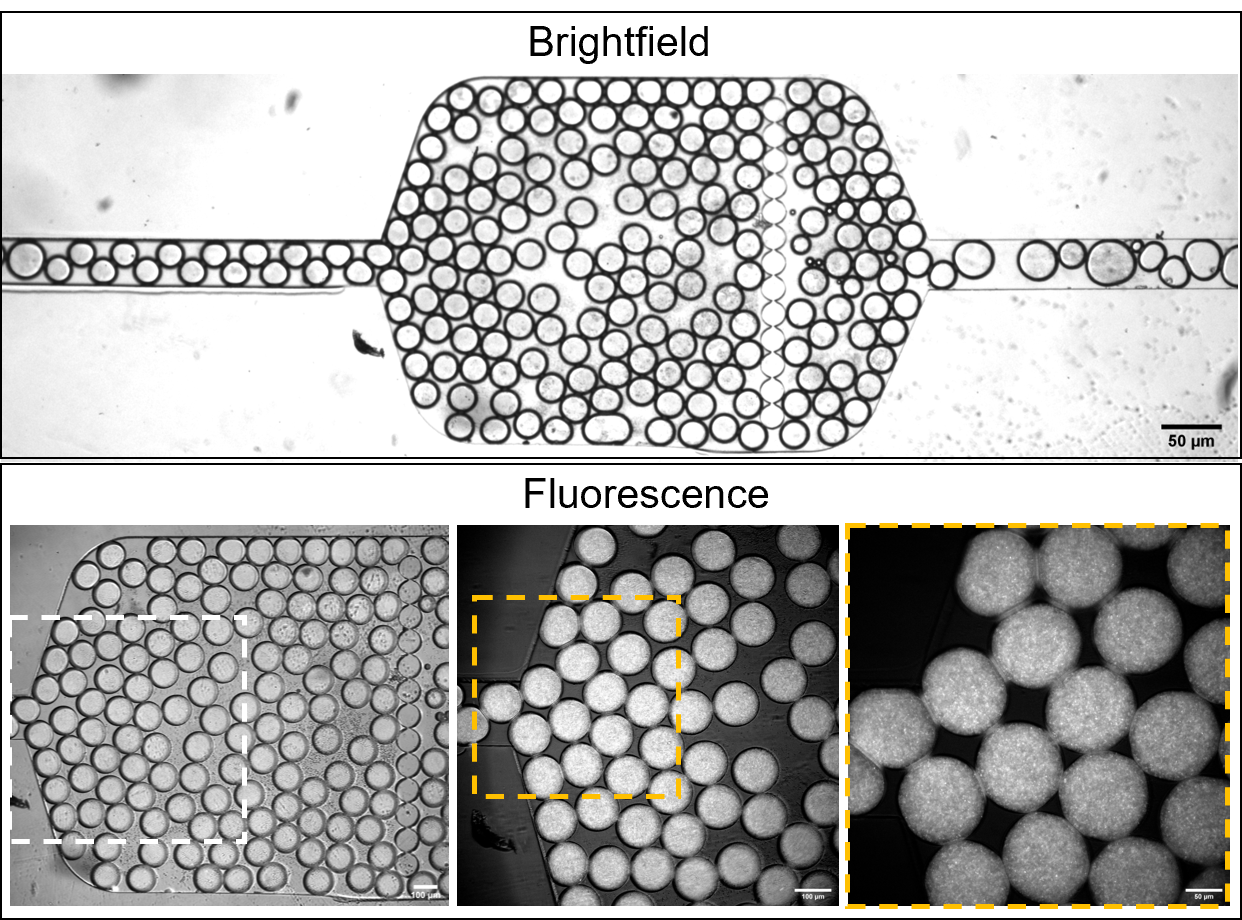}}
  \caption{\T{Droplet collection chamber.} Panels show droplet collection chamber in brightfiled (top) and fluorescence (bottom) at 10x, 20x and 40x magnification (left to right). Fluorescence from 100 nm NDs is visible using mcherry filter set (ex/em:$\approx{540/600 nm}$).}
\zfl{chamber}
\end{figure}

\subsection{Microfluidic valve and automation of sample loading}
\zsl{m_switch}
In order to deliver samples to our microfluidic devices consistently and with high repeatability, we use either a custom LabView program or, most commonly, a graphical user interface designed by Fluigent inc. (Oxygen) to control timing, triggering, and pump applied pressures. We use a 10-position fluidic valve (m-switch, Fluigent Inc.) as well as a 10-port manifold which delivers equal pressure to the ten channels of the switch from a single fluidic pressure controller (Flow-EZ, Fluigent). To load a set of samples, we fill centrifuge tubes with the aqueous phase, generally diamond and analyte, and attach them to varying ports on the fluidic valve. Two positions on the switch are reserved for deionized water and for a fluorinated oil in order to form a plug to separate subsequent samples. By pulling oil-water-oil-analyte repeatedly over each sample we effectively “queue up” a run of samples which will arrive at the microfluidic device for analysis.

In our experiment, each sample requires some settling time for the droplet rate to equilibrate after the disturbance of phases changes and momentary pressure interruptions. This delay is approximately 10 min in our experiment and could  be shortened with fluidic optimization.

\zfr{m_switch} shows a typical example of loading three individual ND samples onto the chip. Each sample is introduced for 150 s, separated by 100 s intervals of water droplets. The figure presents the raw data from these experiments, predominantly consisting of fluorescence from ND particles. During water droplet intervals, the signal detected is at background levels.

The data display brief changes in FL associated with the introduction of new ND sets onto the chip, resolving in under 5 s due to initial particle loading instability. The FL then stabilizes, maintaining a steady state for extended periods, up to 9 h as necessary. When the M-switch is closed, it induces a sudden pressure change, causing a ripple in the ND flow and associated FL, approximately 10 s before the ND flow ceases and transitions to water droplets.

\begin{figure}[t]
  \centering
  {\includegraphics[width=0.45 \textwidth]{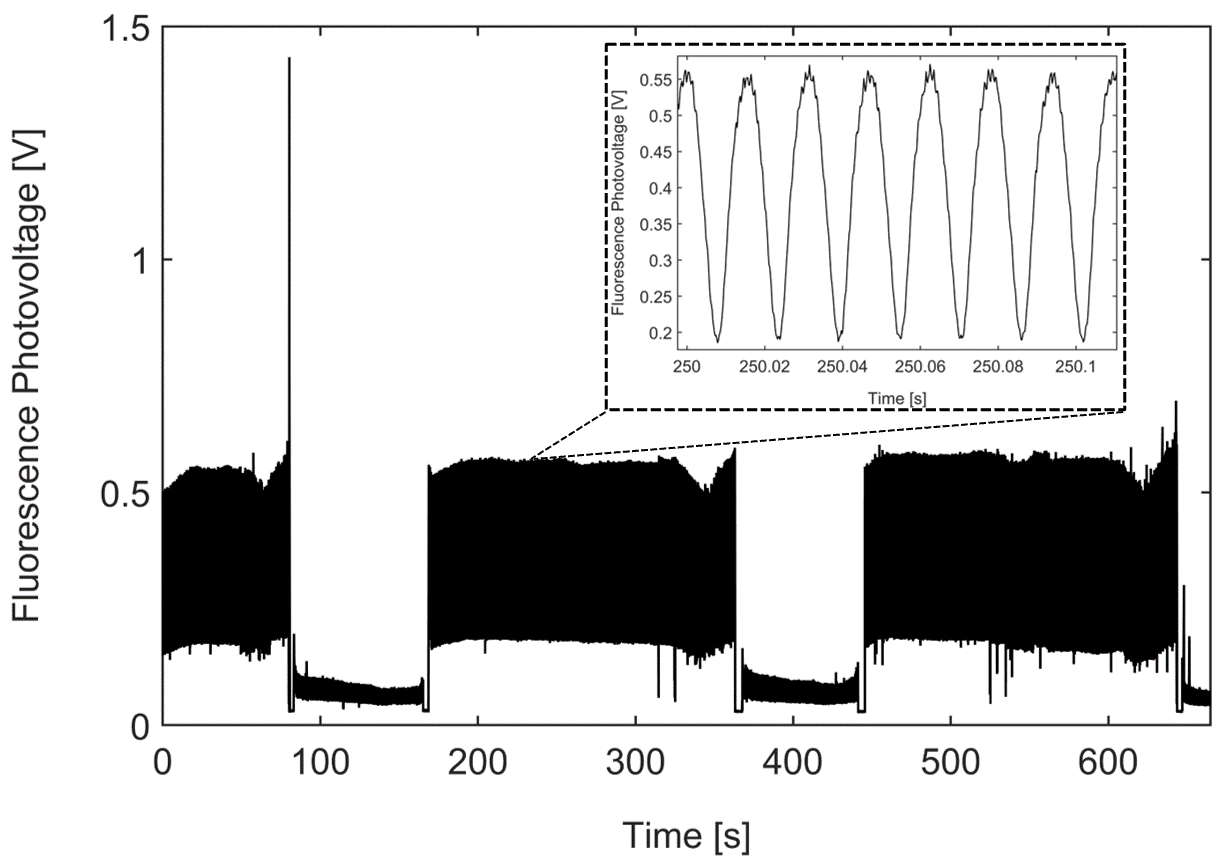}}
  \caption{\T{Automated sample loading} Three successive ND samples loaded onto the chip. Each sample is loaded for 200 s, punctuated by 75 s of pure water droplets, as visible in raw fluorescence traces. A 10-way valve controls introduction of diamond sample, water, or small spacer plugs of oil to maintain sample separation. Switch operation induces transient ND concentration/fluorescence fluctuations due to line pressure changes, lasting about 2s (opening) and 45s (closing). However, a steady-state region is marked by stable fluorescence, and is maintainable for several hours (\zfr{mfig4}). Inset shows the droplet profile over a small time window. }
\zfl{m_switch}
\end{figure}

\begin{figure*}[t]
  \centering
  {\includegraphics[width=0.95\textwidth]{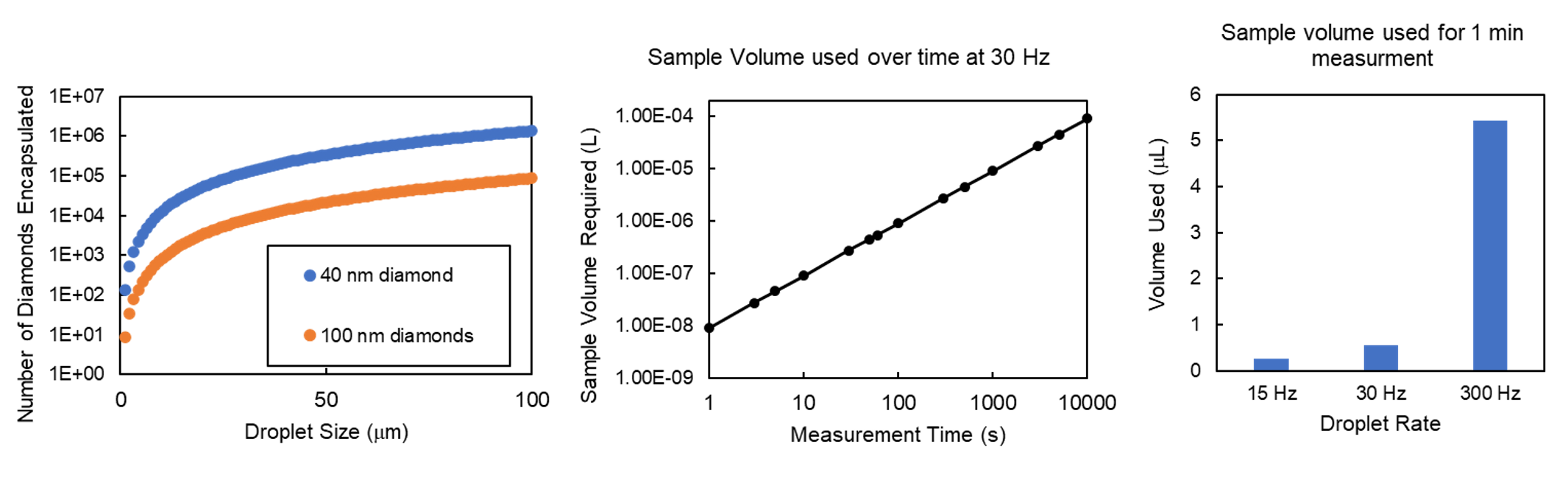}}
  \caption{\T{Microdroplet and diamond loading parameters.} (A) Calculation reflecting the number of diamond particles encapsulated within a single microdroplet for a given microdroplet size. Values are based on initial concentration of diamond of 0.5 $mg/mL$ for either 40 nm (blue) or 100 nm (orange) diameter diamond particles. (B) Calculated sample volume require to perform dual lock-in experiments with droplet rate of 30 Hz for a given measurement time. (C) Calculated sample volume required to perform dual lock-in measurements for 1 minute at a given droplet rate
}
\zfl{stats}
\end{figure*}

\section{Control of droplet modulation profile}
\zsl{flow_control}
Here we provide further details on controlling droplet modulation profiles in our experiments. The droplet chips feature the ability to control flow profiles by adjusting pressures in various channels. This control is primarily achieved through two methods: injecting spacer oil from junction $J_2$ (see \zfr{mfig1} of main paper and \zfr{ChipImages}) and varying the optical spot size for analysis. By altering the amount of spacer oil injected from junction $J_2$, the shape of the droplet modulation profile can be varied significantly. Additionally, the optical spot size employed constitutes another layer of control. It can be adjusted to be smaller than the droplet diameter, resulting in regions without fluorescence (FL) when the spacer oil passes through the analysis region, or it can be enlarged to cover an area larger than the droplet diameter.

\zfr{flow_control} illustrates two examples of droplet modulation control, along with their Fourier Transform (FT) profiles. In \zfr{flow_control}A, the droplet modulation appears nearly sinusoidal, resulting in dominant peaks at $f_D$ in the FT intensity profile (\zfr{flow_control}B). The second example (\zfr{flow_control}C) shows square-like modulation, producing odd harmonics in the $f_D$ profile (\zfr{flow_control}D).

The normalization procedure in \zr{contrast} can incorporate the additional $f_D$ harmonics when necessary. The presence of harmonics may assist in analysis, as they help filter out external noise that does not align with the harmonic pattern, thereby improving result accuracy.

\begin{figure}[t]
  \centering
  {\includegraphics[width=0.49 \textwidth]{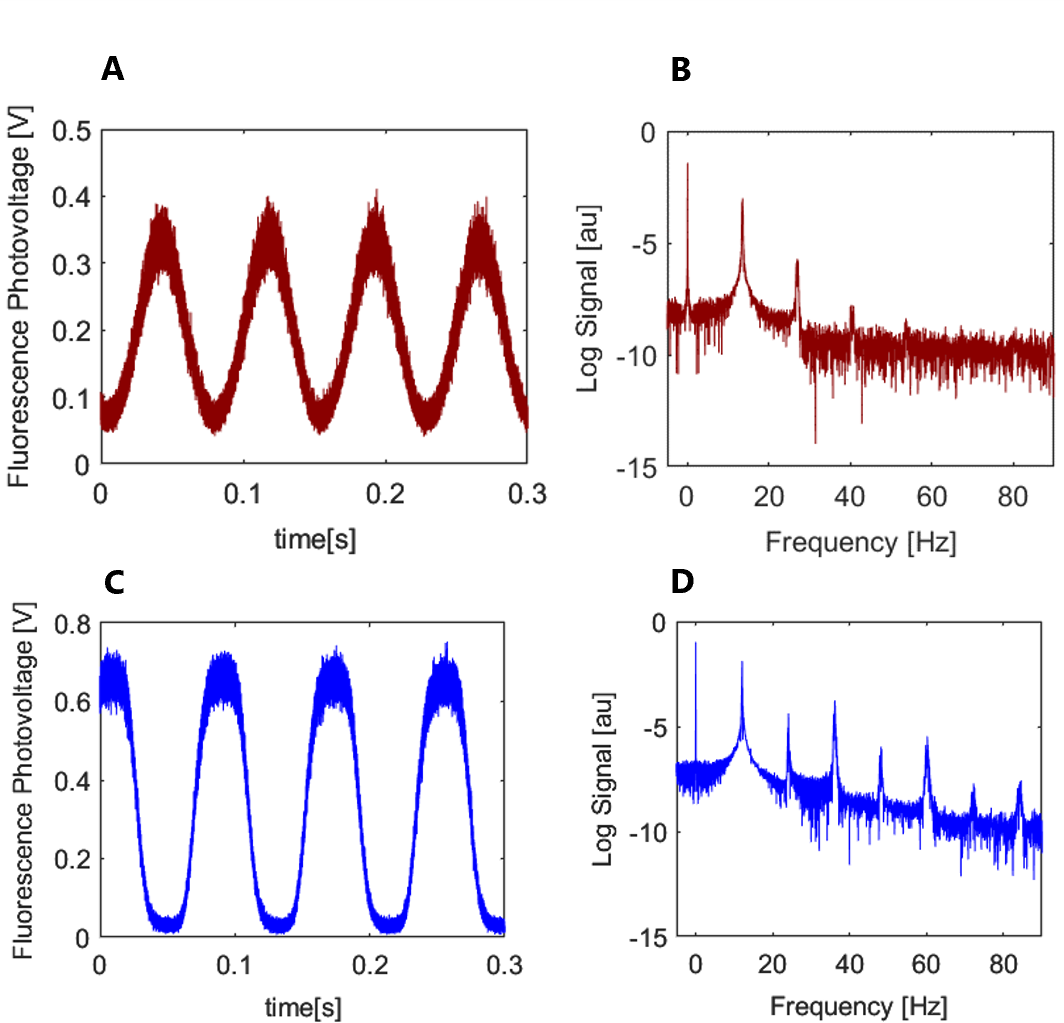}}
  \caption{\T{Control of droplet flow modulation profiles} achieved by varying the flow rates at the two junctions, $J_1$ and $J_2$, and adjusting the optical spot size. A Panel illustrates sinusoidal-like profile. B panel: Fourier intensity spectrum, indicating a single dominant peak at $f_D{=}13.5$Hz with lesser peaks at the first hand second harmonics. C Panel displays a more square-like profile, with the FT (D) revealing several harmonics of $f_D$, predominantly the odd harmonics.}
\zfl{flow_control}
\end{figure}

\section{Imaging and sensing instrumentation}
\subsection{Microscope and optics}
\zsl{microscope}
Fluorescence images and ODMR measurements were performed on a custom-built microscope system that has been reported previously and is briefly described here ~\cite{Blankenship23}. A schematic of the setup is shown in \zfr{opticsDiagram}. Around an enclosure and stage (MadCityLabs RM21 Versa), excitation comes from either a 532 nm laser (Coherent Verdi G15) for fluorescence or a goose-neck lamp above the sample for brightfield images. The laser source is reflected by a dichroic mirror (Thorlabs DMLP605R) before being brought to a focus at the back plane of a 20x objective (Olympus LMPLFLN 20x) using a 150 mm focal length lens. The objective impinges the excitation onto a sample held in place on a piezo-position stage (MadCityLabs RM21). Red emission from samples passes through the dichroic and is directed through color filters (600 nm long pass (Thorlabs FEL0600-1), 680/25 nm band pass (Semrock FF01-680/42-25), 735/25 BP (Semrock FF01-735/28) or a combination), before passing through a 200 mm focal length tube lens. A 30:70 beam splitter projects the resulting image onto both an SCMOS camera (Teledyne Kinetix) and toward a focusing lens to a multi-pixel photon counter (Hamamatsu C14452) for detection. 

Some widefield images were collected using a Nikon Ti2U microscope with the same Kinetix camera at either 4x, 10x, 20x, or 40x magnification. Diamond fluorescence is collected using CYT5-HQY filter cube (ex/em $620/710$ nm) or mcherry filter cube (ex/em:$\approx{540/600 nm}$).

\begin{figure}[t]
  \centering
  {\includegraphics[width=0.49\textwidth]{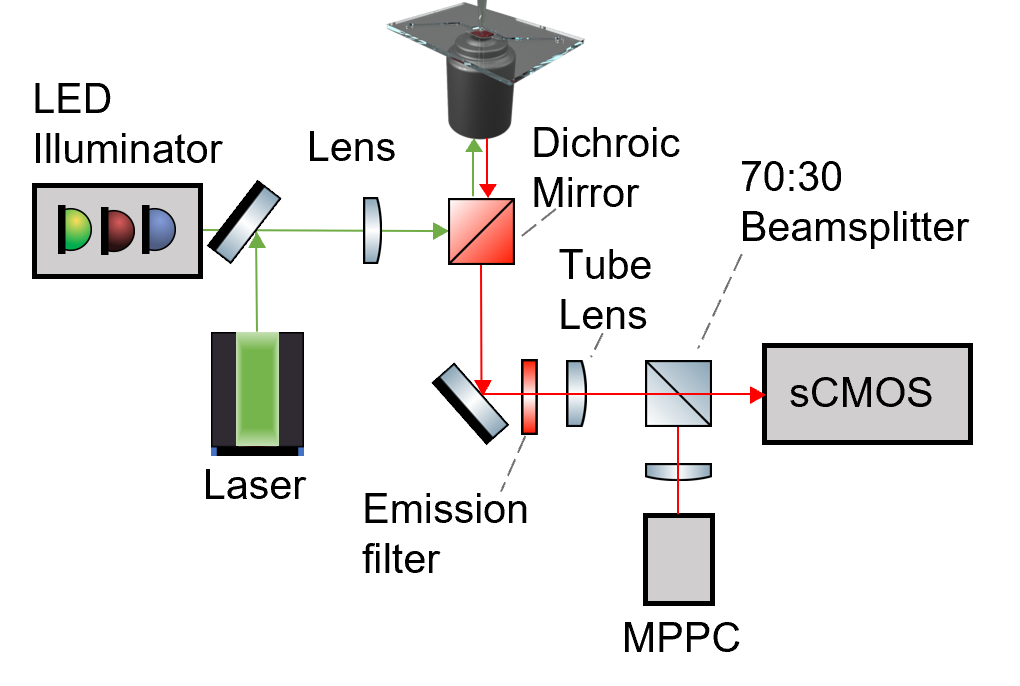}}
  \caption{\T{Detail of microfluidic devices} used for dual lock in measurements. Our devices include a fluid focusing junction (a) to generate aqueous droplets in an oil continuous phase and a down-chip second oil addition (b) which enables control over the spacing of droplets. As needed, droplet contents can be allowed to mix by following the hairpin turns shown is c, or can be stored by filling the chamber shown in d. 
}
\zfl{opticsDiagram}
\end{figure}

\subsection{Optically detected magnetic resonance measurements}

Microwave excitation is generated using a microwave synthesizer (HP 8664A) outputting 0 dBm into a microwave switch which controls whether the MW are passed to the sample or ‘off’ (Minicircuits ZYSW-2-50DR). A 16 W amplifier (Minicircuits ZHL-16W-43-S+) increases the power input into a 100 W amplifier (empower BBM4A6AK5), the output of which passes through a circulator (Pasternack, PE83CR1004) before being delivered to the sample using a circular loop antenna placed between the objective and the sample such that the laser passes through the center of the loop. Microwaves are modulated at a chosen rate (generally 1 KHz) by using the reference output of a lock-in amplifier (Stanford Research SR830) to drive the microwave switch. 

The fluorescence signal detected by the MPPC operating in Geiger mode to generate a photovoltage which is directed both to a data acquisition card (NI DAQ 6215) and to the input of the lock-in amplifier. Lock-in settings like sensitivity, Lock-in time constant and dynamic reserve are chosen to maximize SNR and minimize measurement time and are dependent on signal magnitude and the timescale of signal changes that need to be detected. 

For measurements using the Lock-in amplifier a customized LabView program scans microwave frequency while measuring Lock-in and fluorescence signal or measures signal over time at a set MW frequency as droplets pass through the field of view. For dual lock-in measurements fluorescence photovoltage is measured directly using the analog channels of data acquisition card.

\section{Targeting nanodiamond to yeast cells}
\zsl{Targeting}
As a demonstration of the potential future impacts of our measurement platform we show that diamond particles can be targeted directly to yeast cells. While effective diamond internalization requires removal of the yeast cell wall, with appropriate functionalization, ND particles can be targeted to the exterior of the yeast via the cell wall. ~\zfr{targeting} shows images of the result of such targeting. Here, NDs are coated with Concanavalin A, a protein that interacts with sugar groups on the cell surface. The brightfield image shows the location of 5 \textit{Rhodosporidium toruloides} cells that have been incubated in a solution of Con-A coated NDs. The fluorescence image shows ND emission localized to the cells indicating successful targeting.

\begin{figure}[t]
  \centering
  {\includegraphics[width=0.45 \textwidth]{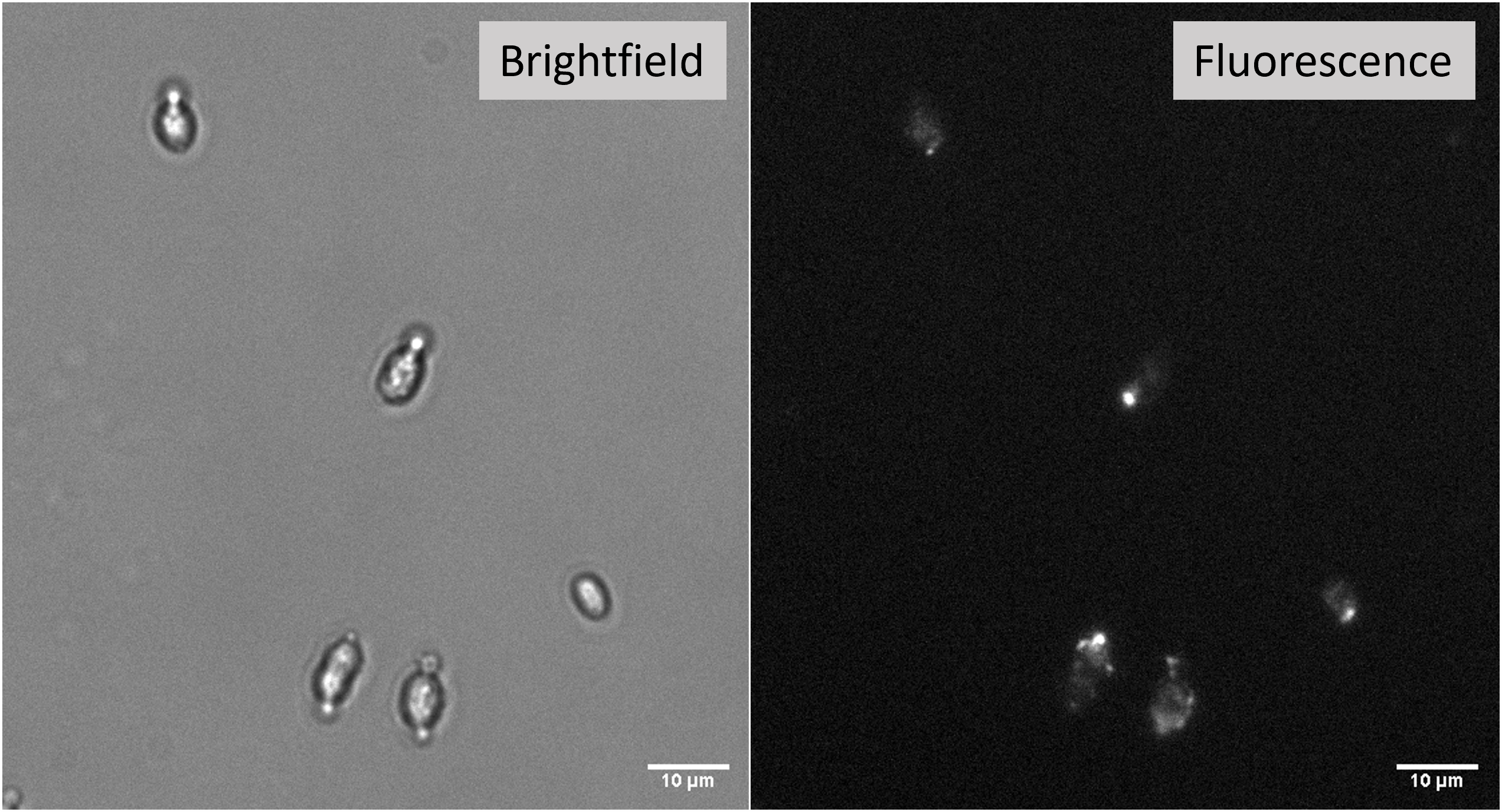}}
  \caption{\T{Targeting diamond to yeast cells} brightfield (left) and fluorescence (right) images of 100 nm NDs targeted to \textit{R. toruloides} using Concanavalin A. Images collected using CYT5-HQY filter set with emission centered near the NV center maximum at 700 nm.}
\zfl{targeting}
\end{figure}

\begin{figure}[t]
  \centering
  {\includegraphics[width=0.45 \textwidth]{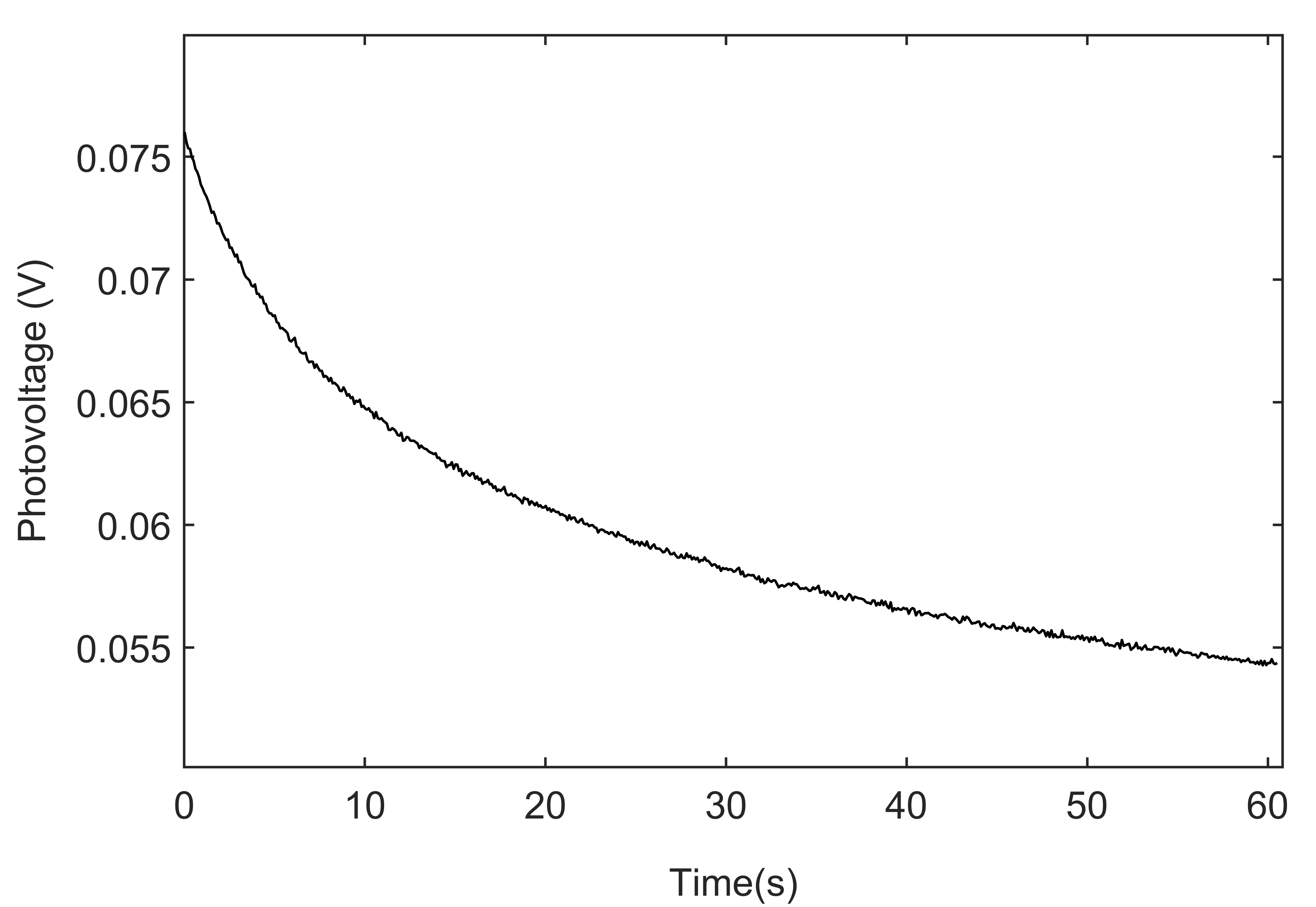}}
  \caption{\T{Autofluorescence} from a ${\sim}100\mu$m region of the PDMS chip without nanodiamond particles measured over 1 minute. Change in autofluorescence over time is at the level of ${\app}1\%$ of typical ND fluorescence from within the droplets. Suppressing this autofluorescence is crucial for high-precision ODMR measurements.}
\zfl{PDMS_autoFlour}
\end{figure}

\begin{figure*}[t]
  \centering
  {\includegraphics[width=0.8 \textwidth]{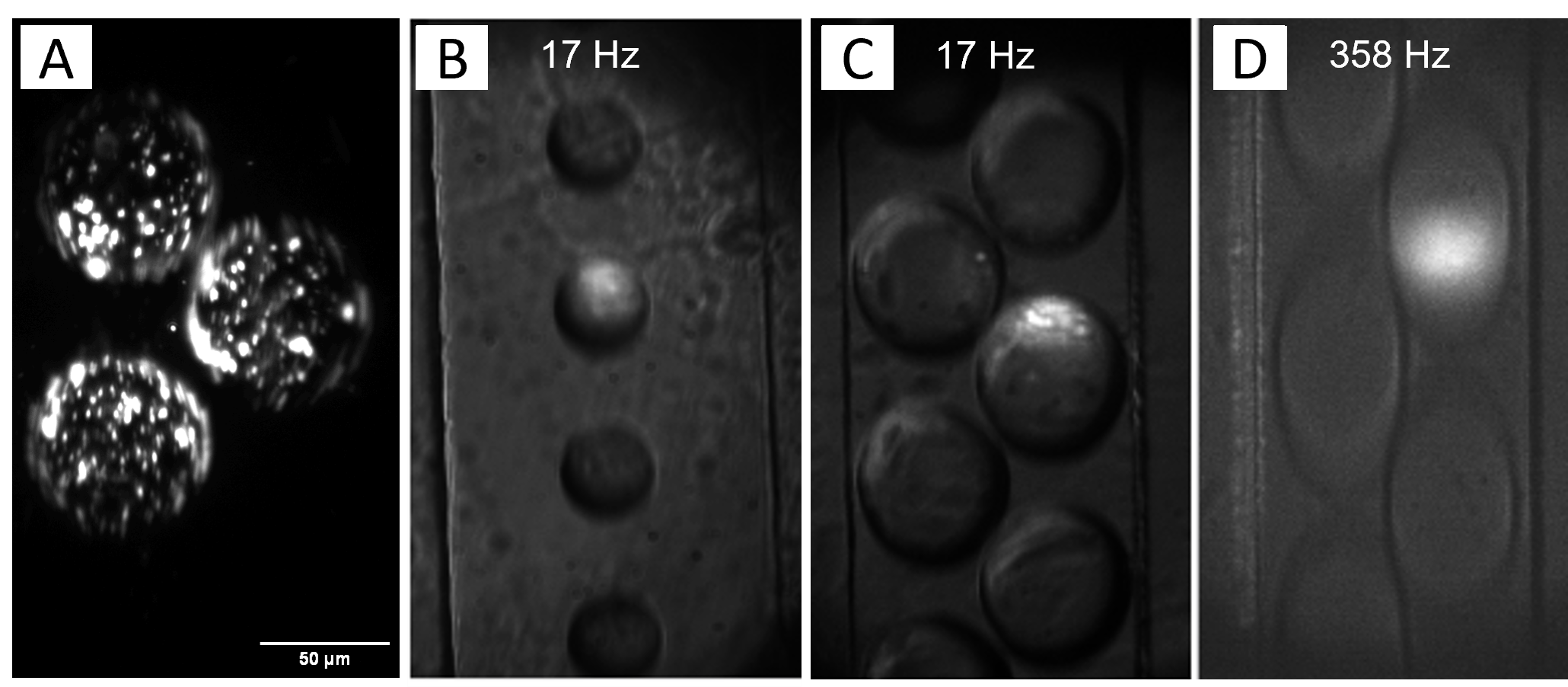}}
  \caption{\T{Movies of droplet flow with NDs.} Still frame grabs are shown, full movies are at Ref. ~\cite{FlowRates_Videos}., featuring 
(A) Fluorescence video of flowing droplets with 750 nm ND particles. (B-D) Fluorescence videos of 100 nm ND particles. Small amount of white light was introduced to make clear the droplet boundaries. Flow is highly regular and orderly. (C-D) At higher speeds, the droplets self-assemble into a double rail configuration, as seen in the last panel. For (D), flow rate exceeds the camera dwell time, resulting in images that appear blurred.}
\zfl{videos}
\end{figure*}

\section{Background fluorescence characterization}
\zsl{autoflour}
The ability to reject background signals is important for increasing the sensitivity of quantum sensing measurements. The measurement of NV center fluorescence changes in the presence of paramagnetic species is a ratiometric measurement in which the drop in fluorescence, measured when MW excitation is on resonance with NV spin transitions, is normalized to total fluoresce emitted from the NV centers. As this total fluorescence value is affected both by signal (NV fluorescence) and by background (e.g. scattering, autofluorescence), we characterized the signal from a new PDMS microfluidic device in the absence of NV diamond. 

The trace in \zfr{PDMS_autoFlour} shows detected photovoltage from PDMS autofluorescence measured using the same detector and under comparable conditions as the data in the main text. The signal begins at approximately 75 mV and decays over the course of the 1 min measurement window to approximately 55mV. Compared to the photovoltages typically measured from passing droplets (~1.2 V for 100 nm NDs), this represents a background signal that can be as much as 6 \% of the total and, importantly, can change by as much as 1-2 \% over the measurement time. While these exact values will depend on many experimental factors, we note that given that the ODMR percent changes are on the order of 1-5 \% fluctuating backgrounds can represent a significant impediment to improving sensitivity. These data demonstrate the need for excluding static and varying backgrounds that are not at the droplet frequency using our Dual Lock-in method.

\section{Particle tracking within droplets}
\zsl{particle_tracking}
In this section, the tracking of Brownian motion of 100 nm ND particles within a single droplet, as shown in \zfr{mfig1}, is described. A movie capturing a droplet with a sparse distribution of ND particles was analyzed using the trackMate plugin for FIJI software~\cite{TrackMate,TrackMate2}. Object size was chosen to be 10 pixels, larger than the actual diameter but allowing some flexability in sizes and objects drift in and out of focus. We then excluded objects below an empirically determined quality threshold. We chose to present trajectories that were monitored for the longest portion of the movie without being interrupted in an attempt to avoid biasing short trajectories. The analysis, spanning a 30 s period, focuses on the motion of individual particles. It is assumed that the particles are dispersed enough to prevent significant overlap in their trajectories of motion, assessed by frame-to-frame movement comparisons within the movie.

The results, depicted in \zfr{mfig1}E, reveal trajectories of the particles spanning ${\sim}$5 $\mu$m, far larger than the diameter of the ND particles. This analysis, when extended to a larger set of droplets, facilitates the creation of a histogram of droplet motion, shown in \zfr{mfig1}Eii. We note that the tracking is limited to two dimensions. Movements in the third dimension cause blurring of the particles as they exit the imaging focal plane, and occasionally results in losing a track. However, the focal region is wide enough to capture most trajectories during the 30s interval.

\section{Videos of flowing droplets}
\zsl{videos}
\zfr{videos} presents videos illustrating the motion of 3 $\mu$m  particles (\zfr{videos}A) and 100 nm ND-filled droplets flowing at various controlled speeds (\zfr{videos}B-D).  

\zfr{videos} shows still frame grabs from these videos, while the videos themselves are available at ~\cite{FlowRates_Videos}. In \zfr{videos}B-D, a small amount of white light is introduced into the imaging system to outline the droplets, and the imaging optical spot is set to be slightly smaller than the droplet diameter. The bright signal then observed is almost exclusively from the ND fluorescence. 

\zfr{videos}A demonstrates the uniform distribution of NDs within a droplet, which can be seen moving and sampling the droplet volume in movie ~\cite{FlowRates_Videos}. \zfr{videos}B-D feature movies showing droplets at varying flow rates: slow, medium, and fast. The droplets form a regular array-like pattern, and stroboscopic sampling gives the impression of each droplet sequentially replacing the next. In \zfr{videos}B, we form a single stream of droplets, comparable to the experiments in the main text. To achieve faster flow rates we use a device that generates an array of droplets 2-wide that is able to achieve higher flow rates (\zfr{videos}C is measured at the same rate as \zfr{videos}A, because our spot size only samples one row of droplets). In faster flows (\zfr{videos}C-D), the droplet movement outpaces the camera's exposure time, causing blurred images. Importantly, the measured signal from all three of these videos yields a stable oscillating signal (main text) highlighting the precision and versatility of droplet-based approaches, particularly in consistently loading droplets with ND particles.

\section{Particle number variation per droplet}

By analyzing fluorescence modulation from flowing droplets, we approximate the number variation of ND particles per droplet. The method involves collecting isolating the 1 KHz band of the Fourier spectrum in \zfr{mfig4}, which contains information about the number of NDs in the field of view at any given time. By plotting a histogram of the variations in intensity of the Fourier band, we estimate an upper bound for the percentage variation in ND loading over time. The data shown in the histogram in \zfr{mfig4}D represent this variability in the Fourier spectrum generated using time bins corresponding to $\sim$ 200 droplets. However, longer time bins will reduce the calculated $\sigma$ by $1/\sqrt{t}$ until the value is no longer limited by the time-bin size. A reasonable estimate of the maximum time bin is given by the Allan deviation in \zfr{mfig4}C. Based on this approach, we calculate the variation between droplets to be \Q{0.23}\%. This variation corresponds to an estimated 2,300 particles per droplet. Despite the potential overestimation, this finding underscores the high degree of control attainable in ND loading within the droplets, highlighting the precision of our experimental setup.

\section{Allan variance analysis} 

We describe the details of Allan variance analysis of the time-dependent droplet PL time series. The PL time series (see main text, \zr{PL}) for the flowing droplets has been divided into smaller time bins of size $\Delta t$ for observing variations in amplitude of the Fourier spectrum in the relevant frequency bands. This time-averaged normalized contrast has been evaluated as 
\beq
\mC(t_i)= \frac{\mF(f_{\mw} \pm f_D,t_i)}{2\mF(f_D,t_i)} 
\zl{contrast_allan}
\eeq
where $\mF \left(f,\Delta t_{i}\right)$ term describes the amplitude of Fourier frequency $f$ in time between $t_i$ and $t-\Delta t_{i}$, extracted through the \textit{fft} numerical algorithm in MATLAB. The DC component of the PL time trace in each time bin $\Delta t$ has been subtracted and therefore, contribution of minor fluctuations from droplet rate is minimized in the contrast calculations. Allan variance is evaluated from the above $\mC(t_i)$ time-series metric. In context of clock stability literature \cite{allan1987time}, we assume our ODMR contrast timepoints to be fractional frequency deviations. The governing exponent $\beta$ of the Allan variance $\sigma\left(\tau\right)$ is related to the Fourier frequency exponent $\alpha$ as $\beta=\alpha-1$. For uncorrelated white Gaussian time-domain noise, the logarithmic scale Allan deviation curve shows $-\frac{1}{2}$ linear slope. Our droplet flow contrast time series attains stable Gaussian noise scaling for several minutes, as shown in \zfr{mfig4}C. A MATLAB repository code \textit{allan} \cite{hopcroftCode} was used to evaluate Allan deviations. 

\begin{figure*}[t]
  \centering
  {\includegraphics[width=0.95\textwidth]{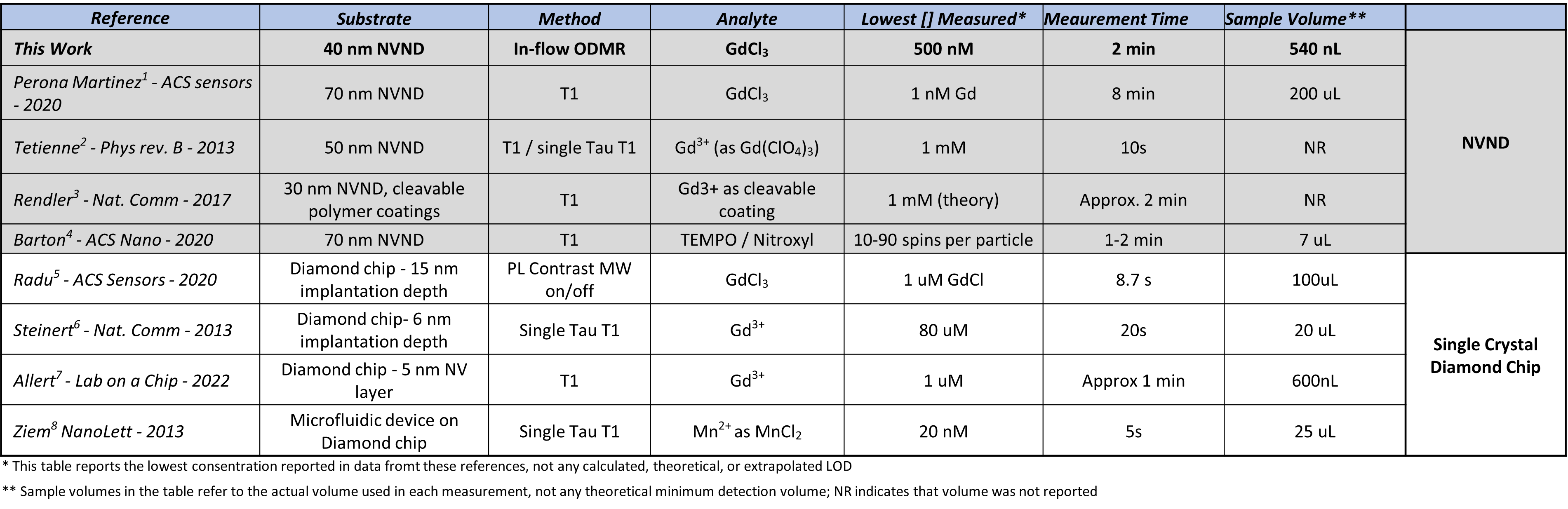}}
  \caption{Table S1: comparison of some related works from literature. Table is organized separating nanodiamond-based methods (grey boxes) and single crystal diamond-based methods (white boxes). $T_1$- based measurements (method column) collect a full $T_1$ curve and fit to determine $T_1$ relaxation times while single Tau $T_1$ methods monitor PL changes at a single point along the $T_1$ relaxation curve. Analyte column refers to the analyte measured in “Lowest [] measured” column. References from top to bottom: \footnotesize{[1] Perona Martinez, F.; et. al. ACS Sensors 2020, 5 (12), 3862–3869. [2] Tetienne, J. P.; et. al. Phys. Rev. B - Condens. Matter Mater. Phys. 2013, 87 (23), 235436. [3] Rendler, T.; et. al. Nat. Commun. 2017, 8. [4] Barton, J.; et. al. ACS Nano 2020, 14 (10), 12938–12950. [5] Radu, V.; et. al. ACS Sensors 2020, 5 (3), 703–710. [6] Steinert, S.; et. al. Nat. Commun. 2013, 4 (1), 1–6. [7] Allert, R. D.; et. al. Lab Chip 2022, 22 (24), 4831–4840. [8] Ziem, F. C.; et. al. Nano Lett. 2013, 13 (9), 4093–4098.}}
  \ztl{table}
\end{figure*}

\section{Comparison of related methods of paramagnetic ion sensing}
\zsl{table}
We note before expanding in detail that the methods compiled here are not exhaustive of the literature, but rather those that are the highest performing or most relevant comparisons to the goals of our work. It should be noted that while many groups aim to detect paramagnetic species, the methods used, the conditions of the experiments, and often the values and units chosen to report are highly variable. As such, a direct comparison between measurements can be dubious, especially across different physical substrates/sensing platforms and across different optical experiment conditions. Indeed, while some groups aim to accelerate measurement times, others aim for superlative sensitivity, while still others aim to decrease sample volume required. As a result, one reference (e.g. Ref.~\cite{Barton20}) may report a sensitivity of 1 $\mu$M in the absence of attempts to measure lower while others (e.g. Ref.~\cite{Perona20}) report high sensitivity while not indicating an intention to miniaturize.

With those caveats in place, we highlight a few relevant works in \ztr{table}, and compare results on axes of parameters that we feel are important to the measurement. We make note that Ref.~\cite{Ziem13} report very high sensitivity to gadolinium ions; however, they also include a thorough (and quite helpful) discussion of the impact of adsorption on the diamond surface. We believe that this complicates the exact value reported, in that while it is certainly true that gadolinium is detected from the 20 nM solution in flow, whether the steady state of that signal is distinguishable from a 100 nM or 500 nM solution is not immediately apparent. We suspect that this effect is at play to some extent on many of these measurements in this table, especially those that use negatively charged diamond to detect paramagnetic cations (e.g. Ref.~\cite{Perona20}). This complicating factor is one reason we approached TEMPOL measurements, which do not have the same adsorption interactions at the diamond surface.

This table highlights the unique benefits of our measurement geometry. Our microfluidic platform enables the combination of high sensitivity, small sample sizes/volumes and short measurement time. While the methods listed each have their own benefits and merit for a variety of purposes, the aspects of our platform that we highlight make it a powerful measurement tool for high throughput and high sensitivity analyte detection.

\end{document}